\pdfoutput=1
\documentclass[a4paper,fleqn]{cas-sc}
\usepackage[authoryear]{natbib}
\usepackage{amsmath,amssymb}
\usepackage{bm}
\usepackage{booktabs}
\usepackage{graphicx}
\usepackage[section]{placeins}
\usepackage{subcaption}
\usepackage{tikz}
\usetikzlibrary{decorations.pathreplacing}
\usepackage{xcolor}

\graphicspath{{figures/}}

\newcommand{\vect}[1]{\mathbf{#1}}
\newcommand{\Ns}{N_s}
\newcommand{\Tve}{T_{\mathrm{ve}}}

\begin{document}
\let\WriteBookmarks\relax
\def\floatpagepagefraction{1}
\def\textpagefraction{.001}

\shorttitle{Implicit high-order flux reconstruction for hypersonic flows of ionised air in thermochemical nonequilibrium}
\shortauthors{R. Dhib et al.}

\title[mode=title]{An implicit high-order flux reconstruction method with logarithmic variables for hypersonic flows in thermochemical nonequilibrium}

\author[1]{Rayan Dhib}[orcid=0000-0003-3670-4678]
\cormark[1]
\ead{rayan.dhib@kuleuven.be}
\credit{Conceptualization, Methodology, Software, Validation, Writing -- original draft}

\author[2]{Vatsalya Sharma}[orcid=0000-0003-2781-6563]
\cormark[1]
\ead{vatsalya@aero.iitkgp.ac.in}

\author[1]{Vincent F. Giangaspero}[orcid=0000-0003-0623-8602]

\author[3]{Firas {Ben Ameur}}[orcid=0000-0001-5097-2885]

\author[1]{José Miguel Luzia Murteira}[orcid=0009-0002-6076-7169]

\author[1,4]{Stefaan Poedts}[orcid=0000-0002-1743-0651]

\author[1]{Andrea Lani}[orcid=0000-0003-4017-215X]

\affiliation[1]{organization={Centre for Mathematical Plasma-Astrophysics, KU Leuven},
            addressline={Celestijnenlaan 200-B},
            postcode={3001},
            city={Leuven},
            country={Belgium}}

\affiliation[2]{organization={Dept. of Aerospace Engineering, Indian Institute of Technology Kharagpur},
            postcode={720301},
            city={Kharagpur},
            country={India}}

\affiliation[3]{organization={Computer, Electrical and Mathematical Sciences and Engineering Division, King Abdullah University of Science and Technology},
            postcode={23955-6900},
            city={Thuwal},
            country={Saudi Arabia}}

\affiliation[4]{organization={Institute of Physics, Maria Curie-Sk{\l}odowska University},
            addressline={ul. Radziszewskiego 10},
            postcode={PL-20-031},
            city={Lublin},
            country={Poland}}

\cortext[1]{Corresponding authors}

\begin{abstract}
Simulating hypersonic flows in thermochemical nonequilibrium requires solving the Navier--Stokes equations for reacting, possibly ionised mixtures. Partial densities span many orders of magnitude, the chemistry is stiff, and shocks must be captured without corrupting the wall heating. As a result, compact high-order computations of viscous two-temperature air flows on unstructured grids have not gone beyond $P = 2$. We present an implicit Flux Reconstruction (FR) method for such flows, built on three ingredients. First, the numerical solver stores and updates the logarithms of the partial densities and temperatures, making every state admissible without a positivity limiter and, for ionised mixtures, imposing charge neutrality consistently with the implicit update. Second, shocks are captured by subcell order blending, extended to steady implicit computations on curved Gauss--Legendre meshes, with second-order subcell blending in subsonic regions and an automatic freeze of the blending near convergence. Third, a pseudo-transient Newton method with a physicality-limited update and automatic CFL control reaches steady state from the free stream with the same settings at every order. The design order is verified up to sixth order on a manufactured solution for five-species air with finite-rate chemistry and vibrational relaxation. Steady viscous solutions up to sixth order are then obtained for five- and eleven-species air on the High Enthalpy shock tunnel Göttingen (HEG) cylinder and for eleven-species air on the RAM-C~II vehicle. To our knowledge, this work gives the first design-order verification of a high-order method with finite-rate air chemistry and vibrational relaxation, and the first high-order finite-element-type computations of viscous eleven-species, two-temperature air. On the HEG cylinder, the stagnation-point heat flux is within 2\,\% of finite-volume references and within 6\,\% of the measurements, with a comparable number of degrees of freedom to second-order finite-volume computations but on meshes with 20 to 40 times fewer elements.

\end{abstract}


\maketitle

\section{Introduction}
\label{sec:intro}

\begin{figure}[pos=htbp]
  \centering
  \begin{subfigure}[b]{0.467\textwidth}
    \centering
    \includegraphics[width=\linewidth]{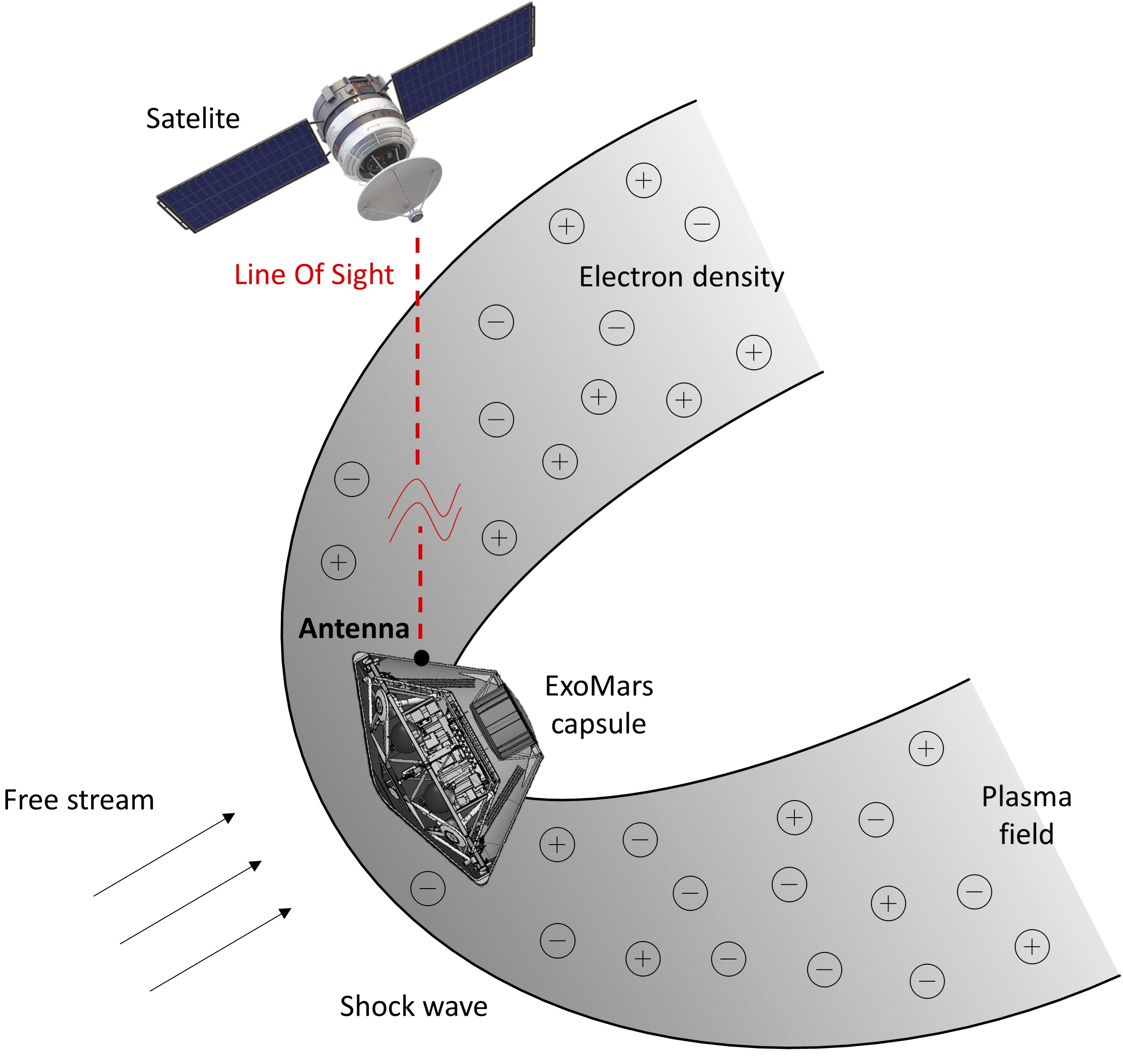}
    \caption{}
    \label{fig:sketch_tcneq1}
  \end{subfigure}\hfill
  \begin{subfigure}[b]{0.492\textwidth}
    \centering
    \includegraphics[width=\linewidth,trim={61bp 39bp 171bp 154bp},clip]{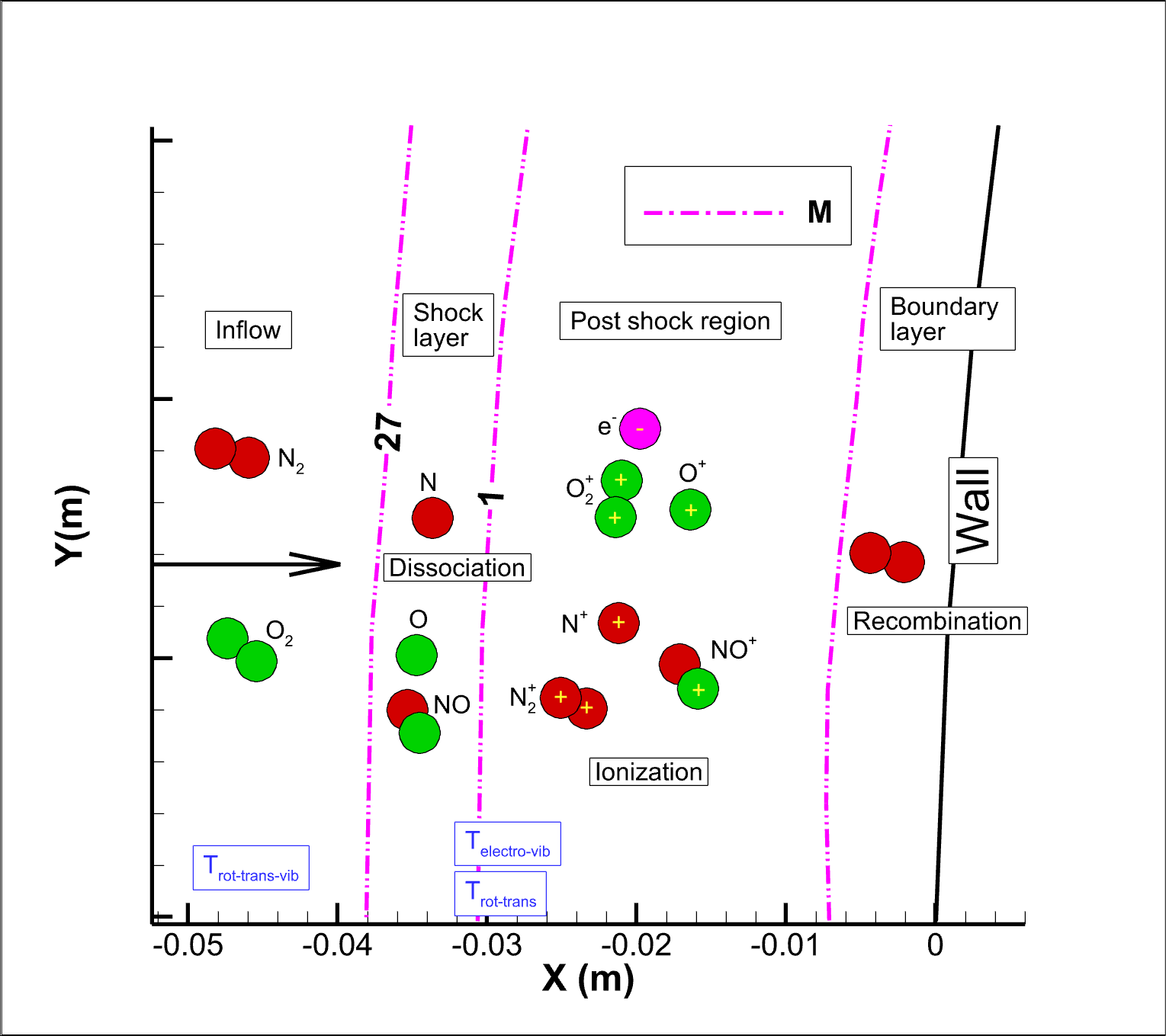}
    \caption{}
    \label{fig:sketch_tcneq2}
  \end{subfigure}
  \caption{Atmospheric entry flows in TCNEQ. (a)~Weakly ionised plasma in the shock layer of an entry capsule, reproduced from \citet{Giangaspero2023} under the CC BY 4.0 license. (b)~TCNEQ effects along the stagnation line of a blunt body as flow decelerates from Mach no. (M) = 27 to 1 over a distance of around 7\,mm during atmospheric entry.}
  \label{fig:sketch_tcneq}
\end{figure}

The design of atmospheric entry vehicles relies on predictions of wall pressure and heat flux, shock stand-off distance, and, at higher speeds, electron density in the shock layer, which causes radio blackout \citep{Grantham1970, Jones1972, Takahashi2014, Giangaspero2023, Giangaspero2026}, as seen in Fig.~\ref{fig:sketch_tcneq1}. Behind the bow shock, the air dissociates and weakly ionises, and its internal-energy modes relax at finite rates, so the flow is in Thermal and Chemical Nonequilibrium (TCNEQ), as seen in Fig.~\ref{fig:sketch_tcneq2}. These predictions have traditionally relied upon finite-volume (FV) codes with mostly second-order accuracy, such as LAURA \citep{Mazaheri2011}, DPLR \citep{Wright1998} and its unstructured counterpart US3D \citep{Candler2015}, which provide the design environments of entry vehicles such as the heat shield of the Mars Science Laboratory \citep{Edquist2013}. Predictive capabilities of such solvers have been assessed against experiments in high-enthalpy wind tunnels \citep{Karl2003, Knight2012, Knight2017, Maier2021} and against flight data, such as the heat flux of FIRE~II \citep{Hash2007} and the electron density of RAM-C~II \citep{Scalabrin2006, Yu2023}, and their uncertainty has been quantified from such comparisons \citep{Bose2013}. Double-cone flows remain a demanding test, where the separation length and the peak heating are sensitive to the thermochemical model \citep{Holloway2020, Holloway2022, Zuo2021, Grover2025}.

With state-of-the-art second-order schemes, the computed wall heat flux is sensitive to shock capturing and to the mesh. The shock-induced instabilities, such as the carbuncle, and errors generated at the shock propagate into the thermal boundary layer \citep{Candler2009}. The heating prediction typically depends on the inviscid flux function, the linear reconstruction of the solution, the limiter \citep{Ching2019} and the treatment of the diffusive fluxes/gradients, and the stagnation-region heating tends to be poorly predicted when cell faces are not properly aligned with the captured shock \citep{Candler2007, Gnoffo2009, Gnoffo2007}.

High-order methods offer several advantages for these flows. In smooth regions, their error decreases as $h^{P+1}$, and they give more accurate heating predictions than low-order methods with the same number of degrees of freedom \citep{Hoskin2024}. 

Their results depend much less on the inviscid flux function and on the mesh alignment with the shock \citep{Ching2019}. Their compact, element-local stencils maintain high order on complex geometries with unstructured and curved meshes and are well suited to modern parallel and Graphics Processing Unit (GPU) architectures \citep{Hoskin2024}. 

The modal content of the solution in each element indicates where it is under-resolved, which can drive shock capturing \citep{Persson2006}. The same properties have motivated high-order Flux Reconstruction (FR) solvers in other fields, such as solar and astrophysical magnetohydrodynamics \citep{Dhib2025, Dhib2026}. Finally, when the discretisation error is reduced by increasing the order, the remaining discrepancy with measurements can be attributed to the physical model, such as surface catalysis or reaction rates, whereas differences between second-order codes for the same case reflect both numerical and modelling choices \citep{Knight2012}. 
TCNEQ models, long used in FV codes, are much less common in higher-order DG methods \citep{Hoskin2024}. 
For two-temperature air, compact high-order discretisations have reached viscous flows only up to third order, with up to five neutral species: Discontinuous Galerkin (DG) at $P\le1$ \citep{Bhatia2013, Bhatia2014, Papoutsakis2014} and DG at $P=1$ with explicit \citep{Absillis2025} or first-order implicit \citep{Luo2026} time stepping, the latter with $P=2$ results shown only along the stagnation line on the HEG cylinder. \citet{Absillis2025} reduce the convective terms to first order with a limiter. 
With a single temperature, HDG reached $P=4$ inviscid and $P=3$ viscous on the HEG cylinder \citep{VanHeyningen2026}, while other DG computations with chemical nonequilibrium were shown at $P=2$ for a one-dimensional inviscid post-shock relaxation \citep{May2021} and for the viscous HEG cylinder \citep{Hoskin2024}. 
Ionised two-temperature air was computed at $P=1$ without viscous terms \citep{Sabo2022}. The design order has not been verified with finite-rate air chemistry and vibrational relaxation. 
The manufactured solutions of \citet{Freno2021} were applied to a second-order FV code, and \citet{Absillis2025} verified their high-order method for a mixture of two fictitious species. To the authors' knowledge, no compact high-order method has therefore computed a steady viscous flow of two-temperature air, neutral or ionised, beyond third order, nor verified its design order with the full thermochemical source terms.

Three difficulties stand in the way of performing higher-order computations of TCNEQ flows. First, the partial densities vary several orders of magnitude, from molecules to atoms, ions, and electrons, and most of them are negligible and non-reacting ahead of the shock, and the high-order polynomials can undershoot to negative partial densities or temperatures. Positivity is usually restored by limiters \citep{Zhang2010, Zhang2011, Du2022, Ching2025}, which modify the solution at every iteration and can stall the convergence of an implicit solver \citep{Venkatakrishnan1995}. 
Moreover, in ionised mixtures, the discrete equations do not preserve charge neutrality. To supplement this, subcell order blending \citep{Hennemann2021, Babbar2024, Babbar2025} has been used with explicit time integration, and, for reacting flows, subcell blending \citep{Peyvan2023, Peyvan2027, Mondal2026} and a posteriori subcell limiting \citep{LiuYuan2023} have been used with explicit transport and without ions, with two temperatures only for an unsteady inviscid blast wave \citep{Mondal2026}. 
Third, the iteration must reach the steady state from a uniform free stream, through a strong transient with stiff chemistry, without a CFL schedule tuned to each case. The pseudo-transient continuation of \citet{Ceze2013, Ceze2015}, which limits the update and automatically adapts the CFL number, was developed for DG discretisations of a perfect gas, but has not been demonstrated for TCNEQ flows using very high-order methods. 

In this work, we present an extended implicit high-order FR method that, for the first time, addresses the three aforementioned difficulties for TCNEQ flows, and we demonstrate its performance on both neutral and ionised air mixtures. The method is implemented in the implicit FR solver \citep{Vandenhoeck2019} of the open-source COOLFluiD platform \citep{lani1, kimpe2, lani13}, with thermodynamic, transport and chemical properties computed by the PLAsmas in Thermodynamic nOn-equilibrium (PLATO) library \citep{Munafo2025}.

The resulting numerical solver stores and updates the logarithms of the partial densities and temperatures, so that every state is evaluated without any limiter. Such variables were previously used for the pressure and the temperature of a perfect gas \citep{Bassi2015, Alberti2024}, but, to the authors' knowledge, not for the partial densities of a multi-species flow. For ionised mixtures, the electron density is set from the ion densities wherever the state is used, as in LAURA \citep{Mazaheri2011}, consistently with the implicit update. Shocks are captured via subcell order blending, which we extend to implicitly compute TCNEQ flows on curved meshes. 

The steady state is reached with the constrained pseudo-transient continuation of \citet{Ceze2013, Ceze2015}, extended with bounds on the logarithmic variables and rules for the electrons, with the same settings at every order and for every case. The design order is verified up to sixth order on the manufactured solution of \citet{Freno2021} for five-species air, with the finite-rate chemistry and vibrational relaxation sources. The method is then validated on the HEG cylinder \citep{Karl2003} with five- and eleven-species air up to $P=5$, and compared against published computations for the same non-catalytic wall on the RAM-C~II flight vehicle \citep{Sharma2024, Yu2023}.

The paper is organised as follows. Section~\ref{sec:thermochem} presents the governing equations, Section~\ref{sec:fr} the FR discretisation, Section~\ref{sec:logvars} the logarithmic variables and the treatment of charge neutrality, Section~\ref{sec:subcell_ob} the subcell order blending, and Section~\ref{sec:implicit} the implicit solver. Section~\ref{sec:verification} verifies the order of accuracy, Section~\ref{sec:validation} validates the method on the HEG cylinder and RAM-C~II, and Section~\ref{sec:conclusions} concludes.

\section{Governing equations}
\label{sec:thermochem}
To account for TCNEQ effects during hypersonic flight, which creates weakly ionised plasma, the governing Navier--Stokes equations need to be modified to include finite-rate chemistry and energy relaxation \citep{Park1989, Park1990}. Behind the bow shock, the energy modes of the air relax at considerably different rates, as seen in Fig.~\ref{fig:sketch_tcneq2}. The rotational energy of the molecules equilibrates with their translational energy within a few collisions, whereas their vibrational energy requires orders of magnitude more collisions \citep {Millikan1963, Park1990, Candler2019}. Therefore, the vibrational energy lags behind over a relaxation zone that covers a significant part of the shock layer. Since vibrationally-excited molecules dissociate more readily, this lag also delays their dissociation. The electronic excitation and the translation of the free electrons, in turn, equilibrate rapidly with the vibrational energy \citep{Park1989}. The two-temperature model of \citet{Park1989, Park1990}, as given by \citet{Gnoffo1989}, retains this separation by adding a single additional energy equation for the vibrational--electronic energy and evaluates the dissociation rates using both temperatures, with the electron-impact rates evaluated at the vibrational--electronic temperature. A single temperature would allow dissociation and ionisation to proceed at the peak translational temperature, whereas multi-temperature and state-to-state models \citep{Munafo2025} add equations and stiffness at a much higher computational cost. 

In this paper, the required thermodynamic, chemical, and transport properties of TCNEQ flows are computed by the PLATO library \citep{Munafo2025}, coupled to the open source COOLFluiD platform (\url{https://github.com/andrealani/COOLFluiD/wiki}) and using a FR-based higher-order discretisation \citep{Dhib2026} instead of the available state-of-the-art FV solver for hypersonic flow applications as in \citep{panesi07, Lani2009, Sharma2024, Sharma2024b, Giangaspero2023, Giangaspero2026}. %
During hypersonic flight, as the gas ionises due to strong shocks, it becomes a mixture of $\Ns$ species, indexed by $s\in\mathcal S=\{1,\dots,\Ns\}$. We write $\mathcal I\subset\mathcal S$ for the positive ions and $e$ for the free electron when the mixture is ionised. Each species is a thermally perfect gas with molar mass $M_s$ and charge number $z_s$ ($z_e=-1$, $z_s>0$ for $s\in\mathcal I$, $z_s=0$ otherwise). With $\rho_s$ the partial density of species $s$, the mixture density and the mass fractions are given as:
\begin{equation}
  \rho=\sum_{s\in\mathcal S}\rho_s,\qquad y_s=\frac{\rho_s}{\rho}.
\end{equation}
For an ionised mixture, the state must be electrically neutral,
\begin{equation}
  \sum_{s\in\mathcal S}\frac{z_s}{M_s}\rho_s=0,
  \label{eq:neutrality}
\end{equation}
which relates the electron density to the ion densities. We report how this constraint is enforced in Sec.~\ref{sec:chargeneutrality}. 
The governing equations in conservative form are given as:
\begin{equation}
  \frac{\partial\vect U}{\partial t}+\nabla\cdot\vect F^{C}(\vect U)=\nabla\cdot\vect F^{D}(\vect U,\nabla\vect U)+\vect S(\vect U,\nabla\vect U),
  \label{eq:cons}
\end{equation}
with the conservative variables
\begin{equation}
  \vect U=\bigl(\rho_1,\dots,\rho_{\Ns},\ \rho\vect V,\ \rho E,\ \rho e^{\mathrm{ve}}\bigr)^{T},
\end{equation}
where $\vect V$ is the mass-averaged velocity, $\rho E$ the total energy per unit volume and $\rho e^{\mathrm{ve}}$ the vibrational--electronic energy per unit volume. The convective and diffusive fluxes and the source are given as:
\begin{equation}
  \vect F^{C}=\begin{pmatrix}
    \rho_s\vect V\\
    \rho\vect V\otimes\vect V+p\,\mathbf I\\
    (\rho E+p)\,\vect V\\
    \rho e^{\mathrm{ve}}\,\vect V
  \end{pmatrix},\qquad
  \vect F^{D}=\begin{pmatrix}
    -\vect J_s\\
    \boldsymbol\tau\\
    \boldsymbol\tau\cdot\vect V-\vect q-\sum_{s\in\mathcal S}h_s\vect J_s\\
    -\vect q^{\mathrm{ve}}-\sum_{s\in\mathcal S}h^{\mathrm{ve}}_s\vect J_s
  \end{pmatrix},\qquad
  \vect S=\begin{pmatrix}
    \dot\omega_s\\
    \vect 0\\
    0\\
    \Omega^{\mathrm{ve}}
  \end{pmatrix},
  \label{eq:fluxes}
\end{equation}
where $\Ns$ is the number of species equations in the first row, which, for an ionised mixture, includes free electrons. In $d$ space dimensions, Eq.~\eqref{eq:cons} has $\Ns+d+2$ components. The pressure $p$ is given by Eq.~\eqref{eq:pressure}, the production rates $\dot\omega_s$ and the source $\Omega^{\mathrm{ve}}$ by Eqs.~\eqref{eq:wdot} and~\eqref{eq:omegave}, and the diffusion fluxes $\vect J_s$, the viscous stress tensor $\boldsymbol\tau$ and the heat fluxes $\vect q$ and $\vect q^{\mathrm{ve}}$ with Eqs.~\eqref{eq:stress} and~\eqref{eq:heatflux}. The specific enthalpy of species $s$ is $h_s$, and its vibrational--electronic part $h^{\mathrm{ve}}_s$ equals $e^{\mathrm{ve}}_s$ for the heavy species and $e^{\mathrm{ve}}_e+p_e/\rho_e$ for the electrons.

As discussed earlier, in the two-temperature model of \citet {Park1989, Park1990}, the translational and rotational modes of the heavy particles are in equilibrium at the temperature $T$, and their vibrational and electronic modes, together with the translational mode of the free electrons, are at the temperature $\Tve$. The mixture energies per unit volume are
\begin{equation}
  \rho E=\sum_{s\in\mathcal S}\rho_se_s(T,\Tve)+\tfrac12\rho|\vect V|^2,\qquad
  \rho e^{\mathrm{ve}}=\sum_{s\in\mathcal S}\rho_se^{\mathrm{ve}}_s(\Tve),
  \label{eq:energies}
\end{equation}
where $e_s$ is the specific internal energy of species $s$, including its formation enthalpy, and $e^{\mathrm{ve}}_s$ its vibrational--electronic part; for the electrons, $e_e=e^{\mathrm{ve}}_e$ depends on $\Tve$ only. The species energies are those of the rigid-rotor harmonic-oscillator model with electronic levels, computed by PLATO from the species partition functions using its spectroscopic database \citep{Munafo2025}: the rotational energy of the molecules is at $T$, and the vibrational energy $e^{\mathrm v}_s$ of the molecules (harmonic oscillator) and the electronic energy $e^{\mathrm{el}}_s$ of the heavy species (Boltzmann distribution over the electronic levels), $e^{\mathrm{ve}}_s=e^{\mathrm v}_s+e^{\mathrm{el}}_s$, are at $\Tve$. With the electrons at $\Tve$, Dalton's law gives the pressure
\begin{equation}
  p=\sum_{s\neq e}\rho_s\frac{R_u}{M_s}T+\rho_e\frac{R_u}{M_e}\Tve,
  \label{eq:pressure}
\end{equation}
where $R_u$ is the universal gas constant and the last term is the electron pressure $p_e$. %

The convective flux $\vect F^{C}$ is hyperbolic. Its largest wave speeds in the direction of a unit vector $\vect n$ are $\vect V\cdot\vect n\pm a_f$ \citep{Gnoffo1989}, where the frozen speed of sound
\begin{equation}
  a_f=\sqrt{\gamma_f\,\frac{p}{\rho}},\qquad
  \gamma_f=1+\frac{\sum_{s\neq e}y_s R_u/M_s}{\sum_{s\neq e}y_s\,c^{\mathrm{tr}}_{v,s}(T)},
  \label{eq:soundspeed}
\end{equation}
involves only the specific heats $c^{\mathrm{tr}}_{v,s}$ of the translational and rotational modes, since the vibrational--electronic energy is transported by its own equation and is frozen for the acoustic waves.

The mixture reacts through $N_r$ reactions,
\begin{equation}
  \sum_{s\in\mathcal S}\nu'_{s,r}\,\mathrm A_s\rightleftharpoons\sum_{s\in\mathcal S}\nu''_{s,r}\,\mathrm A_s,\qquad r=1,\dots,N_r,
\end{equation}
where $\mathrm A_s$ is the chemical symbol of species $s$ and $\nu'_{s,r}$ and $\nu''_{s,r}$ are the stoichiometric coefficients of the reactants and products. The mass production rate of species $s$ is
\begin{equation}
  \dot\omega_s=M_s\sum_{r=1}^{N_r}\bigl(\nu''_{s,r}-\nu'_{s,r}\bigr)\bigl(R^f_r-R^b_r\bigr),\qquad
  R^f_r=k^f_r\prod_{s'\in\mathcal S}\Bigl(\frac{\rho_{s'}}{M_{s'}}\Bigr)^{\nu'_{s',r}},\qquad
  R^b_r=k^b_r\prod_{s'\in\mathcal S}\Bigl(\frac{\rho_{s'}}{M_{s'}}\Bigr)^{\nu''_{s',r}}.
  \label{eq:wdot}
\end{equation}
By this law of mass action, the rate at which a species is destroyed is proportional to its concentration, a property used in Sec.~\ref{sec:logvars_diffusion}. The forward rate coefficients follow the modified Arrhenius law
\begin{equation}
  k^f_r=C_r\,T_r^{\eta_r}\exp\Bigl(-\frac{E_{a,r}}{k_BT_r}\Bigr),\qquad
  T_r=T^{a_r}\,\Tve^{b_r},
  \label{eq:rxn_coeff}
\end{equation}
with the rate-controlling temperature $T_r$ of \citet{Park1989}: $T_r=\sqrt{T\Tve}$ ($a_r=b_r=\tfrac12$) for dissociation by heavy-particle impact, $T_r=\Tve$ for the reactions by electron impact, and $T_r=T$ for the exchange, charge-exchange and associative-ionisation reactions. The backward coefficients follow from detailed balance at the temperature $T_b$ of the reverse process, $k^b_r=k^f_r(T_b)/K^{\mathrm{eq}}_r(T_b)$, where the forward law~\eqref{eq:rxn_coeff} is evaluated at $T_r=T_b$ and the equilibrium constant $K^{\mathrm{eq}}_r$ is computed by PLATO from the partition functions of the species; $T_b=T$ for the heavy-particle reactions, including the recombination that reverses dissociation, and $T_b=\Tve$ for the electron-impact reactions and the dissociative recombination that reverses associative ionisation. The five-species mixture ($\mathrm N_2$, $\mathrm O_2$, NO, N, O) and the eleven-species mixture, which adds $\mathrm N_2^+$, $\mathrm O_2^+$, $\mathrm{NO}^+$, $\mathrm N^+$, $\mathrm O^+$ and $\mathrm e^-$, use the mechanism and the rate coefficients of \citet{Park1993}. The 49 reactions of the eleven-species mechanism and their coefficients are listed in \citet[App.~A]{Sharma2024}. The five-species mechanism has 17 reactions, listed in Table~\ref{tab:air5}: the dissociation of $\mathrm N_2$, $\mathrm O_2$, and NO by each of the five species, and the two Zel'dovich exchange reactions, forming a subset of that list among neutral species. %

The source of the vibrational--electronic energy is
\begin{equation}
  \Omega^{\mathrm{ve}}=\Omega^{\mathrm{VT}}+\Omega^{\mathrm{chem}}+\Omega^{\mathrm{ET}}-p_e\,\nabla\cdot\vect V,
  \qquad
  \Omega^{\mathrm{VT}}=\sum_{s\in\mathcal M}\rho_s\,\frac{e^{\mathrm v}_s(T)-e^{\mathrm v}_s(\Tve)}{\tau_s}.
  \label{eq:omegave}
\end{equation}
The first term, $\Omega^{\mathrm{VT}}$, is the Landau--Teller energy exchange with the translational--rotational modes, summed over the molecules $\mathcal M\subset\mathcal S$. Its relaxation times $\tau_s$ combine the correlation of \citet{Millikan1963}, the coefficients of \citet{Park1993}, and the high-temperature correction of \citet{Park1990}, and are averaged over the heavy collision partners with their mole fractions. The second term, $\Omega^{\mathrm{chem}}$, is the vibrational--electronic energy created or removed by the reactions. In PLATO,
\begin{equation}
  \Omega^{\mathrm{chem}}=\sum_{s\neq e}\dot\omega_s\,e^{\mathrm{ve}}_s(\Tve)-\sum_{r\in\mathcal R_e}\Delta E_r\bigl(R^f_r-R^b_r\bigr),
  \label{eq:omegachem}
\end{equation}
where molecules are created and destroyed with the mean vibrational energy at $\Tve$ (non-preferential dissociation model) and every heavy species with its mean electronic energy at $\Tve$, and where, in the reactions by electron impact $\mathcal R_e$, the free electrons supply the reaction energy $\Delta E_r$: the ionisation energy of N and O and the dissociation energy of $\mathrm N_2$. The third term, present for an ionised mixture, is the elastic energy exchange between the free electrons and the heavy particles, $\Omega^{\mathrm{ET}}=\tfrac32k_B\,n_e\,\nu_{eh}\,(T-\Tve)$, with the electron number density $n_e$ and the electron--heavy-particle collision frequency $\nu_{eh}$. The last term is the work of the electron pressure. Since it involves the divergence of the velocity, the source depends on $\nabla\vect U$ in Eq.~\eqref{eq:cons}.
The viscous stress tensor of the Newtonian mixture, with Stokes' hypothesis, is
\begin{equation}
  \boldsymbol\tau=\mu\Bigl(\nabla\vect V+(\nabla\vect V)^{T}-\tfrac23(\nabla\cdot\vect V)\,\mathbf I\Bigr),
  \label{eq:stress}
\end{equation}
and the heat fluxes follow Fourier's law for each temperature,
\begin{equation}
  \vect q=\vect q^{\mathrm{tr}}+\vect q^{\mathrm{ve}},\qquad
  \vect q^{\mathrm{tr}}=-\lambda^{\mathrm{tr}}\nabla T,\qquad
  \vect q^{\mathrm{ve}}=-\lambda^{\mathrm{ve}}\nabla\Tve.
  \label{eq:heatflux}
\end{equation}
The diffusion fluxes $\vect J_s=\rho_s\vect u^{d}_s$, with the diffusion velocities $\vect u^{d}_s$, are obtained from the Stefan--Maxwell equations, whose driving forces are computed from the gradients of the mass fractions. They satisfy $\sum_{s\in\mathcal S}\vect J_s=\vect 0$ and, for an ionised mixture, carry no net electric current, $\sum_{s\in\mathcal S}z_s\vect J_s/M_s=\vect 0$ (ambipolar diffusion). The driving forces are the gradients of the mole fractions, computed from those of the mass fractions; pressure diffusion and thermal diffusion are neglected. For an ionised mixture, the ambipolar electric field is an additional unknown of the Stefan--Maxwell system, determined by the zero-current condition, and the electron equation is written at $\Tve$. The viscosity $\mu$, the thermal conductivities $\lambda^{\mathrm{tr}}$ and $\lambda^{\mathrm{ve}}$, and the diffusion coefficients are computed by PLATO from the first-order Chapman--Enskog solution of the Boltzmann equation, with Laguerre--Sonine polynomial expansions and collision integrals for each pair of species; we refer to \citet{Munafo2025} and \citet{Sharma2024} for the details.

Equation~\eqref{eq:cons} is written for a planar flow. For an axisymmetric flow, it is solved in a meridian plane, with $x$ along the axis of symmetry and $y$ the distance from it, and the geometric source
\begin{equation}
  \vect S^{\mathrm{axi}}=-\frac{1}{y}\bigl(\vect F^{C}-\vect F^{D}\bigr)\cdot\vect e_y+\frac{p-\tau_{\theta\theta}}{y}\,\vect e_{\rho v},
  \qquad
  \tau_{\theta\theta}=2\mu\,\frac{v}{y}-\tfrac23\mu\,\nabla\cdot\vect V,
  \label{eq:axi}
\end{equation}
is added to $\vect S$, where $\vect e_y$ is the unit vector along $y$, $v$ the velocity component along $y$, and $\vect e_{\rho v}$ the unit vector that selects the momentum equation along $y$. The divergence of the velocity in $\tau_{\theta\theta}$ in the stress tensor~\eqref{eq:stress} and in the work of the electron pressure~\eqref{eq:omegave} then includes the term $v/y$, which takes the limit $\partial v/\partial y$ on the axis.

The discretisation of the following sections thus relies on three properties of Eq.~\eqref{eq:cons}: the convective flux is hyperbolic, with the wave speeds $|\vect V\cdot\vect n|+a_f$ of Eq.~\eqref{eq:soundspeed}; the diffusive flux depends on the gradients of the mass fractions, of the velocity and of the two temperatures and, through the work of the electron pressure, depends on the divergence of the velocity. The TCNEQ governing equations (Equation~\eqref{eq:cons}) introduced here are discretised in space with the FR method presented in the next section. 

\section{FR discretisation}
\label{sec:fr}

This section presents the FR discretisation used in this work, based on \citet{Huynh2007}, \citet{Vincent2011}, and our previous implicit FR solvers for high-speed flows \citep{Vandenhoeck2019,Dhib2024} and for magnetohydrodynamics \citep{Dhib2026}, where the method is described in detail. All computations of this paper use quadrilateral elements, and the presentation is restricted to them. For generality, the formulation is written for a vector of \emph{update variables} $\vect W$, the quantities stored at the solution points, from which the conservative variables $\vect U(\vect W)$ are computed, and for a vector of \emph{gradient variables} $\vect G(\vect W)$, whose gradients enter the diffusive fluxes. The choice of $\vect W$ and $\vect G$ made in this work is the subject of Sec.~\ref{sec:logvars}.

\subsection{Domain discretisation and coordinate transformation}
\label{sec:fr_mapping}

The computational domain is partitioned into non-overlapping, conforming quadrilateral elements $\Omega_K$. Following the finite-element paradigm, the solution within each element is approximated by a polynomial of degree $P$ defined exclusively within $\Omega_K$. To facilitate implementation, each element is mapped to a reference square $[-1,1]^2$, with coordinates $\boldsymbol\xi=(\xi,\eta)$, by the isoparametric transformation $\vect x=\boldsymbol\Theta_K(\boldsymbol\xi)$ of a quadratic (Q2) element. The conservation laws~\eqref{eq:cons} transform to the reference domain as
\begin{equation}
  \frac{\partial(J_K\vect U)}{\partial t}+\hat\nabla\cdot\hat{\vect f}=J_K\vect S,
  \qquad
  \hat{\vect f}=\Bigl(\frac{\partial y}{\partial\eta}\vect F_x-\frac{\partial x}{\partial\eta}\vect F_y,\ -\frac{\partial y}{\partial\xi}\vect F_x+\frac{\partial x}{\partial\xi}\vect F_y\Bigr),
  \label{eq:transformed}
\end{equation}
where $J_K$ is the Jacobian determinant of $\boldsymbol\Theta_K$, $\hat\nabla$ the differentiation in the reference coordinates, and $\vect F_x$, $\vect F_y$ are the Cartesian components of $\vect F=\vect F^{C}-\vect F^{D}$.

\subsection{Solution representation}
\label{sec:fr_elements}
\label{sec:fr_traces}

The approximate solution within $\Omega_K$, marked by the superscript $\delta$, is the tensor-product polynomial of degree $P$
\begin{equation}
  \vect W^{\delta}(\boldsymbol\xi)=\sum_{i=0}^{P}\sum_{j=0}^{P}\vect W_{ij}\,l_i(\xi)\,l_j(\eta),
  \label{eq:solpoly}
\end{equation}
defined by its values $\vect W_{ij}$ at the $(P+1)^2$ solution points $\boldsymbol\xi_{ij}=(\xi_i,\eta_j)$, where $\xi_0<\dots<\xi_P$ are the Gauss--Legendre (GL) points and $l_i$ is the one-dimensional Lagrange polynomial associated with $\xi_i$. Furthermore, $P+1$ flux points $\boldsymbol\xi_{f,q}$, $q=0,\dots,P$, are placed at the GL points along each face $f=1,\dots,4$, with outward unit normal $\vect n_{f,q}$. At each flux point, the solution is evaluated on both sides of the face. The inner state $\vect W^{-}_{f,q}=\vect W^{\delta}(\boldsymbol\xi_{f,q})$ is the solution polynomial~\eqref{eq:solpoly} of the local element at the flux point. The outer state $\vect W^{+}_{f,q}$ is that of the neighbouring element or, at a boundary, the ghost state of Sec.~\ref{sec:fr_bc}.

\subsection{Flux correction and residual}
\label{sec:fr_conv}
\label{sec:fr_residual}
\label{sec:fr_correction}

The flux built from the solution polynomial is discontinuous between elements. FR corrects it so that its normal component at every flux point equals a common flux shared by the two elements, and takes the divergence of the corrected flux as the residual.

We write $\vect F^{C}(\vect W)$ and $\vect F^{D}(\vect W,\vect Q)$ for the fluxes evaluated with the state $\vect W$ and the gradient $\vect Q$ of $\vect G$, and $\vect S(\vect W,\vect Q)$ for the source. Within each element, the discontinuous flux $\hat{\vect f}^{\delta D}$ is the polynomial of degree $P$ interpolating the transformed flux of $\vect F^{C}(\vect W_{ij})-\vect F^{D}(\vect W_{ij},\vect Q^{\delta D}_{ij})$, where $\vect Q^{\delta D}_{ij}$ are the gradients at the solution points (Sec.~\ref{sec:fr_diff}); the superscript $\delta D$ marks an element-wise approximation, discontinuous between elements.

At each flux point, the common normal flux, marked by the superscript $\delta I$, is computed from the inner and outer states,
\begin{equation}
  (\vect F\cdot\vect n)^{\delta I}_{f,q}
  =\vect F^{*}\bigl(\vect W^{-}_{f,q},\vect W^{+}_{f,q},\vect n_{f,q}\bigr)
  -\vect F^{D}\bigl(\tfrac12(\vect W^{-}_{f,q}+\vect W^{+}_{f,q}),\,\tfrac12(\vect Q^{-}_{f,q}+\vect Q^{+}_{f,q})\bigr)\cdot\vect n_{f,q},
  \label{eq:commonflux}
\end{equation}
where $\vect F^{*}$ is a Riemann flux; in this work, a local Lax--Friedrichs flux \citep{Rusanov1962}, and $\vect Q^{\mp}_{f,q}$ are the face gradients of the two elements, given by the second scheme of \citet{Bassi2000} (BR2). At a boundary face, $\vect F^{D}$ is evaluated with the boundary state and gradient of Sec.~\ref{sec:fr_bc}.

The discontinuous flux is corrected with one correction function per flux point,
\begin{equation}
  \hat{\vect f}^{\delta}=\hat{\vect f}^{\delta D}+\hat{\vect f}^{\delta C},\qquad
  \hat{\vect f}^{\delta C}=\sum_{f=1}^{4}\sum_{q=0}^{P}\Delta_{f,q}\,\vect h_{f,q},\qquad
  \Delta_{f,q}=J_{f,q}\,(\vect F\cdot\vect n)^{\delta I}_{f,q}-\hat{\vect f}^{\delta D}(\boldsymbol\xi_{f,q})\cdot\hat{\vect n},
  \label{eq:fluxcorr}
\end{equation}
where $\hat{\vect f}^{\delta C}$ is the correction flux and $\Delta_{f,q}$ is the difference between the common normal flux and the discontinuous one extrapolated to the flux point, both in the reference element. $J_{f,q}$ is the face Jacobian at the flux point: the ratio of physical to reference length along the face, and $\hat{\vect n}$ is the outward unit normal of the reference element. The vector correction function $\vect h_{f,q}$ has a normal component $\vect h_{f,q}\cdot\hat{\vect n}$ equal to one at flux point $q$ of face $f$ and zero at the other flux points, so that the normal component of the corrected flux $\hat{\vect f}^{\delta}$ equals the common flux at every flux point. We use the VCJH correction functions \citep{Vincent2011}, a one-parameter family parameterised by $c_{\mathrm{VCJH}}$. On quadrilaterals, they are built from the one-dimensional functions $h_L$ and $h_R$ of the left and right faces.

The residual $\vect R_{\mathrm{FR},ij}$ is the divergence of the corrected flux at the solution points $\boldsymbol{\xi}_{ij}$, which reads:
\begin{equation}
  \vect R_{\mathrm{FR},ij}
  =-\bigl(\hat\nabla\cdot\hat{\vect f}^{\delta}\bigr)\big|_{\boldsymbol\xi_{ij}}
  =-\bigl(\hat\nabla\cdot\hat{\vect f}^{\delta D}\bigr)\big|_{\boldsymbol\xi_{ij}}
    -\sum_{f=1}^{4}\sum_{q=0}^{P}\Delta_{f,q}\,\phi_{f,q}(\boldsymbol\xi_{ij}),
  \label{eq:sob_RFR}
\end{equation}
where $\phi_{f,q}=\hat\nabla\cdot\vect h_{f,q}$ is the correction field of the flux point. We write $\vect R^{C}_{\mathrm{FR},ij}$ and $\vect R^{D}_{\mathrm{FR},ij}$ for the parts of $\vect R_{\mathrm{FR},ij}$ that come from $\vect F^{C}$ and from $\vect F^{D}$.

With the source $\vect S_{ij}=\vect S(\vect W_{ij},\vect Q^{\delta D}_{ij})$, the solution evolves as
\begin{equation}
  J_{K,ij}\,\frac{d\vect U_{ij}}{dt}=\vect R_{ij},\qquad
  \vect R_{ij}=\vect R^{C}_{\mathrm{FR},ij}+\vect R^{D}_{\mathrm{FR},ij}+J_{K,ij}\,\vect S_{ij},
  \label{eq:semidiscrete}
\end{equation}
where $J_{K,ij}$ is the Jacobian determinant at $\boldsymbol\xi_{ij}$. We denote $\vect R(\vect W)$ as the vector of all $\vect R_{ij}$ in the mesh. With shock capturing, $\vect R^{C}_{\mathrm{FR},ij}$ is replaced by the blended residual of Sec.~\ref{sec:subcell_ob}.

\subsection{Solution gradients}
\label{sec:fr_diff}

The diffusive flux needs the gradients of $\vect G$. As the flux, it is corrected at the element faces with the same correction fields,
\begin{equation}
  \vect Q^{\delta D}=\nabla\vect G^{\delta}+\nabla\vect G^{\delta C},
  \label{eq:gradient}
\end{equation}
where $\vect G^{\delta}$ is the polynomial of degree $P$ interpolating $\vect G$ at the solution points and $\nabla\vect G^{\delta C}$ is the correction gradient. It lifts into the element the differences $\boldsymbol\Pi_{f,q}=\vect G^{\delta I}_{f,q}-\vect G^{\delta}(\boldsymbol\xi_{f,q})$ between the common and the local values of $\vect G$ at the flux points,
\begin{equation}
  \nabla\vect G^{\delta C}(\boldsymbol\xi)=\frac{1}{J_K(\boldsymbol\xi)}\sum_{f=1}^{4}\sum_{q=0}^{P}\phi_{f,q}(\boldsymbol\xi)\,J_{f,q}\;\boldsymbol\Pi_{f,q}\otimes\vect n_{f,q},
  \label{eq:lifting}
\end{equation}
where $\otimes$ is the outer product, $(\boldsymbol\Pi\otimes\vect n)_{kl}=\Pi_k\,n_l$, so that component $k$ of $\vect G$ receives the gradient $\Pi_k\,\vect n$. As in BR2, at an interior face the common value $\vect G^{\delta I}_{f,q}$ is the average of the local values of the two elements at the flux point; at a boundary face it is the boundary value of Sec.~\ref{sec:fr_bc}. Like the solution, $\vect Q^{\delta D}$ is represented by its values $\vect Q^{\delta D}_{ij}$ at the solution points; they enter the diffusive part of the discontinuous flux and the source.

At the flux points, the diffusive part of the common flux~\eqref{eq:commonflux} uses the face gradients
\begin{equation}
  \vect Q^{-}_{f,q}=\bigl(\nabla\vect G^{\delta}+\eta_{\mathrm{BR2}}\,\nabla\vect G^{\delta C}_{f}\bigr)\big|_{\boldsymbol\xi_{f,q}},
  \label{eq:br2face}
\end{equation}
where $\nabla\vect G^{\delta C}_{f}$ is the term of face $f$ in Eq.~\eqref{eq:lifting}, $\eta_{\mathrm{BR2}}=5$ a penalty factor~\citep{Brezzi2000,Hartmann2008}. Similarly, we define $\vect Q^{+}_{f,q}$ as the associated face gradient in the neighbouring element. As $\vect Q^{\delta D}$, the sum $\nabla\vect G^{\delta}+\eta_{\mathrm{BR2}}\nabla\vect G^{\delta C}_{f}$ is computed at the solution points and interpolated to the flux point. Since each face gradient uses only the correction of its own face, the residual of an element depends only on the element and its face neighbours, which is the compactness of BR2.

\subsection{Boundary conditions}
\label{sec:fr_bc}

At a flux point of a boundary face, only the inner state $\vect W^-$ is available. Each boundary condition prescribes a boundary state $\vect W^{b}$ and a boundary gradient $\vect Q^{b}$, which are imposed weakly. In the convective part of the common flux, the outer state is the ghost state
\begin{equation}
  \vect W^{g}=2\vect W^{b}-\vect W^-,
  \label{eq:ghost}
\end{equation}
so that the average of the inner and ghost states is the boundary state. The diffusive part of the common flux is $\vect F^{D}(\vect W^{b},\vect Q^{b})\cdot\vect n$. In the correction gradient~\eqref{eq:lifting}, the common value of $\vect G$ is the boundary value $\vect G^{b}$. The components of $\vect Q^{b}$ and $\vect G^{b}$ that a condition does not prescribe keep the inner values $\vect Q^{-}_{f,q}$ and $\vect G^{\delta}(\boldsymbol\xi_{f,q})$, and these components of $\vect G$ receive no correction.
\begin{itemize}
  \item \emph{Supersonic inflow}: $\vect W^{b}$ is the free-stream state $\vect W_\infty$, and $\vect G^{b}=\vect G(\vect W_\infty)$.
  \item \emph{Supersonic outflow}: $\vect W^{g}=\vect W^-$, thus $\vect W^{b}=\vect W^-$.
  \item \emph{Slip wall}: the ghost state is the inner state with the normal velocity reversed,
  \begin{equation}
    \vect V^{g}=\vect V^- -2(\vect V^-\cdot\vect n)\,\vect n,
    \label{eq:slipghost}
  \end{equation}
  so that the zero normal velocity is imposed weakly. In $\vect W^{b}$ and $\vect G^{b}$, the normal velocity is set to zero.
  \item \emph{Isothermal non-catalytic wall} at the temperature $T_w$: the boundary state is
  \begin{equation}
    \rho_s^{b}=\rho_s^-\,\frac{T^-}{T_w},\qquad \vect V^{b}=\vect 0,\qquad T^{b}=\Tve^{b}=T_w,
    \label{eq:wallbnd}
  \end{equation}
  and $\vect G^{b}$ has the same velocity and temperatures. In $\vect Q^{b}$, the normal gradients of the mass fractions are set to zero, so that $\vect J_s\cdot\vect n=0$. Instead of Eq.~\eqref{eq:ghost}, the ghost state is
  \begin{equation}
    \rho_s^{g}=\rho_s^-,\qquad \vect V^{g}=-\vect V^-,\qquad T^{g}=2T_w-T^-,\qquad \Tve^{g}=2T_w-\Tve^-.
    \label{eq:wallghost}
  \end{equation}
\end{itemize}

\section{Logarithmic variables and charge neutrality}
\label{sec:logvars}

In hypersonic nonequilibrium flows, partial densities vary over many orders of magnitude, since atoms and ions are produced behind shocks and may be present only as traces in the free-stream cold gas. Near shocks and in boundary layers, high-order polynomials of these partial densities can undershoot, potentially resulting in negative partial densities or temperatures and causing the computation to fail. Positivity is usually restored with limiters \citep{Zhang2010,Zhang2011}, which for multi-species reacting flows also act on the partial densities \citep{Du2022,Ching2025}. These limiters are typically designed for explicit time integration and do not extend easily to implicit time integration \citep{vanderVegt2019}. Ionised mixtures add an additional challenge, as the charge-neutrality constraint~\eqref{eq:neutrality} is not preserved exactly by the discrete equations.

In this work, we still solve the conservation laws for $\vect U$, but the scheme stores, interpolates and updates the logarithms of the partial densities and temperatures instead (Sec.~\ref{sec:logvars_def}), which, to the authors' knowledge, has never been done before for multi-species nonequilibrium flows. Every state evaluated by the scheme is then admissible without any limiter (Sec.~\ref{sec:logvars_positivity}), with major and trace species being treated alike by the implicit scheme (Secs.~\ref{sec:impl_jacobian} and~\ref{sec:impl_update}). For ionised mixtures, the electron density is set from the ion densities, imposing charge neutrality wherever the state is used: in property evaluation, in the stored state, and in the implicit update (Secs.~\ref{sec:chargeneutrality} and~\ref{sec:impl_update}).

\subsection{Update variables}
\label{sec:logvars_def}

While the equations are formulated in conservative variables, we define the variables in which the solution is actually stored at each degree of freedom and updated (at each pseudo time-step in steady simulations) as \emph{update variables}. For this work, the chosen update variables are the logarithms of the partial densities and of the two temperatures, together with the velocity:
\begin{equation}
  \vect W=\bigl(\ln\rho_1,\dots,\ln\rho_{\Ns},\ \vect V,\ \ln T,\ \ln\Tve\bigr)^{T},
  \label{eq:logvars}
\end{equation}
with the components $w_s=\ln\rho_s$, $w_T=\ln T$ and $w_{\mathrm{ve}}=\ln\Tve$. The velocity vector has $d$ components, where $d$ is the number of space dimensions ($d=2$ in this work). The conservative variables follow explicitly from $\rho_s=\exp(w_s)$, $T=\exp(w_T)$ and $\Tve=\exp(w_{\mathrm{ve}})$,
\begin{equation}
  \rho\vect V=\Bigl(\sum_s \rho_s\Bigr)\vect V,\qquad
  \rho E=\sum_s \rho_s\,e_s(T,\Tve)+\tfrac12\Bigl(\sum_s \rho_s\Bigr)|\vect V|^2,\qquad
  \rho e^{\mathrm{ve}}=\sum_s \rho_s\,e^{\mathrm{ve}}_s(\Tve),
  \label{eq:w_of_v}
\end{equation}
and since the energies increase strictly with the temperatures, the map $\vect W\mapsto\vect U$ is a bijection between $\mathbb R^{\Ns+d+2}$ and the set of states with positive partial densities and temperatures. Unlike the inverse map, it requires no iterative recovery of the temperatures. The residual is still evaluated for the conservation laws~\eqref{eq:cons}, from the states $\vect U(\vect W)$, so the update variables only change the parametrisation of the discrete solution and the scheme remains conservative.

Since the logarithm of a vanishing partial density is undefined, species absent from the free stream, such as the atoms, ions, and electrons of cold air, are given small inflow densities. This is common practice in nonequilibrium flow solvers, which seed the inflow with a small degree of ionisation for stable computations \citep{Takahashi2014} or bound the species densities from below by a small floor \citep{Mazaheri2011}.

Logarithms have been used as unknowns before, in the Newton iteration of NASA's Chemical Equilibrium with Applications program (CEA), which corrects the logarithms of the species amounts and of the temperature of a single state \citep{Gordon1994}, and as the working variables of implicit DG schemes for a perfect gas, in the form of the logarithms of the pressure and the temperature \citep{Bassi2015,Alberti2024}. Neither applies them to the partial densities, which in nonequilibrium flows span many orders of magnitude.

\subsection{Positivity by construction}
\label{sec:logvars_positivity}

In each element the solution is a polynomial in the update variables (Sec.~\ref{sec:fr}), that is, in the logarithms of the partial densities and temperatures. Whatever values these polynomials take, and whatever update the implicit scheme applies, the partial densities $\rho_s=\exp(w_s)$ and the temperatures $T=\exp(w_T)$ and $\Tve=\exp(w_{\mathrm{ve}})$ are positive, as is the pressure~\eqref{eq:pressure}. Every state the scheme evaluates is therefore admissible by construction.

Positivity by construction does not exclude non-finite states, which arise from an overflow of the exponential or from a logarithm tending to $-\infty$; the limited update of Sec.~\ref{sec:impl_update} prevents them, since it bounds the change of the logarithms in each iteration and requires the density and pressure of the updated state to stay finite.

\subsection{Gradients and diffusion fluxes}
\label{sec:logvars_diffusion}

The diffusive fluxes~\eqref{eq:fluxes} require the gradients of the mass fractions, through the diffusion driving forces, of the velocity and of the temperatures. The gradients are computed for the \emph{gradient variables}:
\begin{equation}
  \vect G=\bigl(\ln y_1,\dots,\ln y_{\Ns},\ \vect V,\ \ln T,\ \ln\Tve\bigr)^{T},\qquad
  \ln y_s=w_s-\ln\sum_{s'\in\mathcal S}\exp(w_{s'}),
  \label{eq:gradvars}
\end{equation}
which, like $\vect W$, are finite for every admissible state. The correction gradient~\eqref{eq:lifting} lifts the differences of $\vect G$ at the flux points, and the common diffusive flux is evaluated at the average of the inner and outer states, that is, at the geometric means of their partial densities and temperatures. The physical gradients are recovered at each evaluation point from the gradients of $\vect G$,
\begin{equation}
  \nabla y_s=y_s\,\nabla\ln y_s,\qquad \nabla T=T\,\nabla\ln T,\qquad \nabla\Tve=\Tve\,\nabla\ln\Tve,
  \label{eq:gradrecovery}
\end{equation}
with $y_s$, $T$ and $\Tve$ taken at the same point.
This choice is motivated by a property of the continuous equations that concerns the positivity of the species densities, which becomes visible when the species equation \(\partial_t\rho_s+\nabla\cdot(\rho_s\vect V+\vect J_s)=\dot\omega_s\) is divided by \(\rho_s\) to give, with \(D(\cdot)/Dt=\partial_t(\cdot)+\vect V\cdot\nabla(\cdot)\), the evolution equation of the logarithm of the partial density,
\begin{equation}
  \frac{D\ln\rho_s}{Dt}
  = -\nabla\cdot\vect V+\frac{\dot\omega_s}{\rho_s}-\frac{\nabla\cdot\vect J_s}{\rho_s}.
\end{equation}
Since production can only increase \(\rho_s\), only the loss terms of this equation can drive the density to zero, and each of them remains bounded as \(\rho_s\to0\): the convective term is bounded for a regular velocity field, and the destruction rate is proportional to the concentration of \(s\) by the law of mass action, so that its contribution to \(\dot\omega_s/\rho_s\) stays bounded as long as the other concentrations and the rate coefficients do. If the diffusive term is bounded as well, then \(D\ln\rho_s/Dt\ge -C\) along fluid paths for some finite constant \(C\), hence \(\rho_s(t)\ge\rho_s(0)\,e^{-Ct}>0\), and the density cannot vanish in finite time; conversely, a loss term that did not vanish with \(\rho_s\) would drive \(\ln\rho_s\) to \(-\infty\) in finite time.

For the diffusive term, the corresponding requirement is that the flux \(\vect J_s=\rho_s\vect u^d_s\) vanish with \(\rho_s\), which holds in the continuous model because the Stefan--Maxwell driving forces of Sec.~\ref{sec:thermochem} are the mole-fraction gradients \(\nabla x_s=x_s\nabla\ln x_s\), each proportional to the concentration of the species it acts on, so that the diffusion velocity \(\vect u^d_s\) remains bounded as \(x_s\to0\) whenever the logarithmic gradients do. The discretisation preserves this behaviour by computing the gradients for the variables \(\ln y_s\) and recovering the physical gradients as Eq.~\eqref{eq:gradrecovery}, so that the driving force of species \(s\), and hence the discrete flux \(\vect J_s\), vanishes with \(y_s\) provided the discrete logarithmic gradients remain bounded.

\subsection{Charge neutrality}
\label{sec:chargeneutrality}

For an ionised mixture, the neutrality constraint~\eqref{eq:neutrality} with $z_e=-1$ determines the electron density from the ion densities,
\begin{equation}
  \rho_e^{\star}=\sum_{s\in\mathcal I}c_s\,\rho_s,\qquad c_s=\frac{M_e\,z_s}{M_s}.
  \label{eq:rhoe_neutral}
\end{equation}
In the continuous model the constraint is preserved, since every reaction conserves charge and the ambipolar diffusion fluxes carry no net current; the electron continuity equation is then the combination of the ion equations with the weights $c_s$. In the discrete model, this no longer holds exactly. The electron diffusion flux is discretised separately, the truncation errors of the electron and ion equations are not related by the weights $c_s$, and the discrete equations are solved only approximately. Because $M_e/M_s$ is of order $10^{-5}$, an imbalance that is small in terms of ion mass is a large relative error in $\rho_e$. The electron density enters the electron-impact reaction rates, the electron pressure in Eqs.~\eqref{eq:pressure} and~\eqref{eq:omegave}, and the transport properties. We therefore impose the constraint in the evaluation of the properties and in the stored state, and make both consistent with the implicit update (Sec.~\ref{sec:impl_update}).

Every evaluation of thermodynamic, kinetic and transport properties, including the pressure~\eqref{eq:pressure} and the energies of the map~\eqref{eq:w_of_v}, uses $\rho_e^{\star}$ of Eq.~\eqref{eq:rhoe_neutral} in place of the electron density of the state (neutral evaluation), and the electron diffusion driving force is computed from the ion gradients, $\nabla y_e^{\star}=\sum_{s\in\mathcal I}c_s\nabla y_s$. The stored electron density then enters only the quantities computed directly from the state, namely the mixture density $\rho$, in which its mass is negligible, the electron pressure work in Eq.~\eqref{eq:omegave}, and the output fields.

So that these quantities are consistent with the neutral evaluation, a state filter sets the electron entry of the state at every solution point to its neutral value after each update of the solution,
\begin{equation}
  w_e\leftarrow\ln\sum_{s\in\mathcal I}c_s\,\exp(w_s),
  \label{eq:filter}
\end{equation}
and the stored state then satisfies the constraint~\eqref{eq:neutrality} exactly at every solution point. This reset is the default of LAURA \citep{Mazaheri2011}.

The electron equation is kept in the discrete system, so that the data layout is the same for all mixtures, but the electron entry it yields is overwritten by the filter (Sec.~\ref{sec:impl_update}). At convergence, the ion and neutral equations, the momentum and the energy equations are satisfied, and neutrality holds pointwise, while the residual of the electron equation is not driven to zero and is excluded from the convergence monitoring.

\section{Subcell order blending}
\label{sec:subcell_ob}

High-order finite-element-type methods, FR included, produce spurious oscillations near shocks, a manifestation of the Gibbs phenomenon. With the logarithmic variables of Sec.~\ref{sec:logvars}, these oscillations no longer lead to negative densities or temperatures, but they still pollute the solution near the shock, and in hypersonic flows the wall heat flux is sensitive to how the shock is captured and to any numerical dissipation that reaches the boundary layer \citep{Ching2019}.

In this work, shocks are captured by subcell order blending: drawing on the subcell FV scheme of \citet{Sonntag2017} and on the convex blending of \citet{Hennemann2021}, the FR residual of each element is blended with a low-order FV residual computed on a subcell grid, with a coefficient given by an element-wise smoothness indicator. The closest work is the subcell blending of \citet{Babbar2024} for Lax--Wendroff FR, extended to curvilinear meshes \citep{Babbar2025} and to multiderivative Runge--Kutta FR \citep{Babbar2026}, all for the Euler equations with explicit time integration. The subcell methods applied to reacting flows advance the flow equations explicitly, with one temperature and no ions \citep{Peyvan2023,LiuYuan2023,Peyvan2027}; \citet{Mondal2026} use two temperatures, once again with explicit time integration, for an unsteady inviscid blast wave of five-species air. To the authors' knowledge, subcell blending is used here for the first time with a fully implicit time integration of reacting flows, and for an ionised mixture with two temperatures.

Steady implicit computations of viscous TCNEQ flows on curved meshes with GL points require three new components that were not considered in these previous works. First, the subcell face normals are built from the FR derivative of the metric, so that the blended scheme preserves a uniform state on curved elements whenever FR does (Sec.~\ref{sec:sob_normals}). Second, the low-order scheme is second order in subsonic elements and first order elsewhere, providing the desired accuracy where the flow is smooth and keeping the limiter away from shocks, where it would contribute to stalling convergence (Sec.~\ref{sec:sob_selective}). Third, the blending coefficient is frozen automatically once the implicit iteration stalls (Sec.~\ref{sec:sob_freeze}).

\subsection{Blended residual and subcell grid}

The semi-discrete update at every solution point of element $K$ reads:
\begin{equation}
  \frac{d\hat{\mathbf U}_{ij}}{dt}
  =\underbrace{\mathbf R^{D}_{\mathrm{FR},ij}}_{\text{diffusive}}
  +(1-\alpha_K)\,\underbrace{\mathbf R^{C}_{\mathrm{FR},ij}}_{\text{high order}}
  +\alpha_K\,\underbrace{\mathbf R^{C}_{\mathrm{SC},ij}}_{\text{low order}}
  +J_{K,ij}\,\mathbf S_{ij},
  \label{eq:sob_blend}
\end{equation}
where $\alpha_K\in[0,1]$ is the blending coefficient of the element, computed from a smoothness indicator (Sec.~\ref{sec:sob_alpha}), $\hat{\mathbf U}_{ij}=J_{K,ij}\mathbf U(\mathbf W_{ij})$ is the transformed conservative state and $\mathbf R^{C}_{\mathrm{SC}}$ is an FV residual on a subcell grid that has one subcell per solution point. With $\alpha_K=0$ the element uses the FR scheme of Sec.~\ref{sec:fr}; with $\alpha_K=1$ its convective terms are discretised by the subcell FV scheme alone.

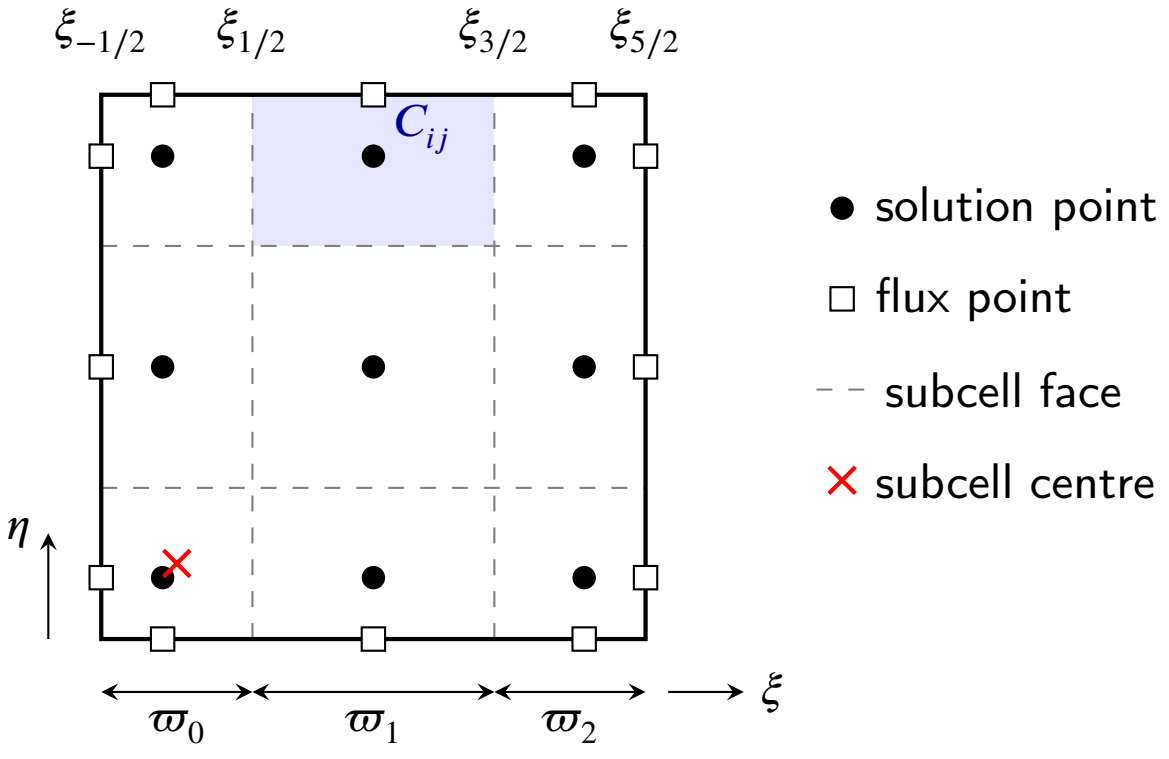
\begin{figure}[pos=htbp]
  \centering
\begin{tikzpicture}[>=stealth,font=\small]
\fill[blue!10] (-0.800,0.800) rectangle (0.800,1.800);
\node[blue!60!black] at (0.330,1.574) {$C_{ij}$};
\draw[dashed,gray] (-0.800,-1.800) -- (-0.800,1.800);
\draw[dashed,gray] (0.800,-1.800) -- (0.800,1.800);
\draw[dashed,gray] (-1.800,-0.800) -- (1.800,-0.800);
\draw[dashed,gray] (-1.800,0.800) -- (1.800,0.800);
\draw[thick] (-1.800,-1.800) rectangle (1.800,1.800);
\fill (-1.394,-1.394) circle (2.2pt);
\fill (-1.394,0.000) circle (2.2pt);
\fill (-1.394,1.394) circle (2.2pt);
\fill (0.000,-1.394) circle (2.2pt);
\fill (0.000,0.000) circle (2.2pt);
\fill (0.000,1.394) circle (2.2pt);
\fill (1.394,-1.394) circle (2.2pt);
\fill (1.394,0.000) circle (2.2pt);
\fill (1.394,1.394) circle (2.2pt);
\draw[fill=white] (-1.394,-1.800) +(-2.2pt,-2.2pt) rectangle +(2.2pt,2.2pt);
\draw[fill=white] (0.000,-1.800) +(-2.2pt,-2.2pt) rectangle +(2.2pt,2.2pt);
\draw[fill=white] (1.394,-1.800) +(-2.2pt,-2.2pt) rectangle +(2.2pt,2.2pt);
\draw[fill=white] (-1.394,1.800) +(-2.2pt,-2.2pt) rectangle +(2.2pt,2.2pt);
\draw[fill=white] (0.000,1.800) +(-2.2pt,-2.2pt) rectangle +(2.2pt,2.2pt);
\draw[fill=white] (1.394,1.800) +(-2.2pt,-2.2pt) rectangle +(2.2pt,2.2pt);
\draw[fill=white] (-1.800,-1.394) +(-2.2pt,-2.2pt) rectangle +(2.2pt,2.2pt);
\draw[fill=white] (-1.800,0.000) +(-2.2pt,-2.2pt) rectangle +(2.2pt,2.2pt);
\draw[fill=white] (-1.800,1.394) +(-2.2pt,-2.2pt) rectangle +(2.2pt,2.2pt);
\draw[fill=white] (1.800,-1.394) +(-2.2pt,-2.2pt) rectangle +(2.2pt,2.2pt);
\draw[fill=white] (1.800,0.000) +(-2.2pt,-2.2pt) rectangle +(2.2pt,2.2pt);
\draw[fill=white] (1.800,1.394) +(-2.2pt,-2.2pt) rectangle +(2.2pt,2.2pt);
\draw[red,thick] (-1.300,-1.300) +(-2.5pt,-2.5pt) -- +(2.5pt,2.5pt) +(-2.5pt,2.5pt) -- +(2.5pt,-2.5pt);
\draw[<->] (-1.800,-2.150) -- node[below]{$\varpi_0$} (-0.800,-2.150);
\draw[<->] (-0.800,-2.150) -- node[below]{$\varpi_1$} (0.800,-2.150);
\draw[<->] (0.800,-2.150) -- node[below]{$\varpi_2$} (1.800,-2.150);
\node[above] at (-1.800,1.920) {$\xi_{-1/2}$};
\node[above] at (-0.800,1.920) {$\xi_{1/2}$};
\node[above] at (0.800,1.920) {$\xi_{3/2}$};
\node[above] at (1.800,1.920) {$\xi_{5/2}$};
\draw[->] (-2.150,-1.800) -- (-2.150,-1.100) node[left]{$\eta$};
\draw[->] (1.950,-2.150) -- (2.450,-2.150) node[right]{$\xi$};
\fill (3.100,1.050) circle (2.2pt) node[right=3pt]{solution point};
\draw[fill=white] (3.100,0.450) +(-2.2pt,-2.2pt) rectangle +(2.2pt,2.2pt);
\node[right=3pt] at (3.100,0.450) {flux point};
\draw[dashed,gray] (2.930,-0.150) -- +(0.34,0) node[right,black]{subcell face};
\draw[red,thick] (3.100,-0.750) +(-2.5pt,-2.5pt) -- +(2.5pt,2.5pt) +(-2.5pt,2.5pt) -- +(2.5pt,-2.5pt);
\node[right=3pt] at (3.100,-0.750) {subcell centre};
\end{tikzpicture}
  \caption{Subcell grid in the reference element for $P=2$.}
  \label{fig:sob_grid}
\end{figure}

As in \citet{Sonntag2017}, each element is divided into $(P+1)^2$ subcells, one per solution point, whose sizes in the reference element are the GL quadrature weights $\varpi_0,\dots,\varpi_P$ ($\sum_i\varpi_i=2$). Along $\xi$ the subcell faces are therefore
\begin{equation}
  \xi_{-1/2}=-1,\qquad \xi_{i+1/2}=\xi_{i-1/2}+\varpi_i,\qquad i=0,\dots,P,
  \label{eq:sob_faces}
\end{equation}
so that $\xi_{P+1/2}=1$, and identically along $\eta$. A half-integer index denotes a subcell face: $i+1/2$ is the face at $\xi=\xi_{i+1/2}$, between the points $i$ and $i+1$. Subcell $C_{ij}=[\xi_{i-1/2},\xi_{i+1/2}]\times[\eta_{j-1/2},\eta_{j+1/2}]$ contains the solution point $\boldsymbol\xi_{ij}$ and has reference area $\varpi_i\varpi_j$ (Fig.~\ref{fig:sob_grid}). The nodal value $\hat{\mathbf U}_{ij}$ is interpreted as the subcell value of $C_{ij}$ \citep{Sonntag2017,Hennemann2021}. Since $\sum_{ij}\varpi_i\varpi_j\hat{\mathbf U}_{ij}$ is the GL quadrature of the element integral, the FR and the subcell representations carry the same element mean.

\subsection{Subcell fluxes}
\label{sec:sob_fluxes}

In reference coordinates, the low-order residual at the solution point $\boldsymbol\xi_{ij}$ is the FV balance of its subcell $C_{ij}$:
\begin{equation}
  \mathbf R^{C}_{\mathrm{SC},ij}
  =-\frac{1}{\varpi_i}\Bigl(\hat{\mathbf F}_{i+1/2,j}-\hat{\mathbf F}_{i-1/2,j}\Bigr)
   -\frac{1}{\varpi_j}\Bigl(\hat{\mathbf F}_{i,j+1/2}-\hat{\mathbf F}_{i,j-1/2}\Bigr),
  \label{eq:sob_RSC}
\end{equation}
where $\hat{\mathbf F}_{i+1/2,j}$ is the transformed numerical flux through the subcell face at $\xi=\xi_{i+1/2}$ of the row $j$, and $\hat{\mathbf F}_{i,j+1/2}$ the one through the subcell face at $\eta=\eta_{j+1/2}$ of the column $i$. In what follows, the fluxes, face normals and reconstruction are written for the faces $\xi=\xi_{i+1/2}$ of the row $j$, the faces $\eta=\eta_{j+1/2}$ of the column $i$ being treated in the same way.

\subsubsection{Internal and element faces}

Through an internal subcell face, $0\le i<P$, the flux has the same form as the element-face flux $J_{f,q}\mathbf F^{*}$ of FR, with the face Jacobian and normal of the subcell face,
\begin{equation}
  \hat{\mathbf F}_{i+1/2,j}=J_{i+1/2,j}\,
  \mathbf F^{*}\bigl(\mathbf W^{-}_{i+1/2,j},\,\mathbf W^{+}_{i+1/2,j},\,\mathbf n_{i+1/2,j}\bigr),
  \label{eq:sob_intflux}
\end{equation}
where $\mathbf W^{\mp}_{i+1/2,j}$ are the reconstructed states on the two sides of the subcell face, $\mathbf n_{i+1/2,j}$ its unit normal and $J_{i+1/2,j}$ its face Jacobian, the ratio of physical to reference length along it (both defined below), and $\mathbf F^{*}$ is the Riemann flux of Eq.~\eqref{eq:commonflux}.

The subcells along the boundary of the element share one or more of their faces with the element faces. Since the flux points lie at the same 1D GL coordinates as the solution points, each shared face contains exactly one flux point (Fig.~\ref{fig:sob_grid}). The flux used at that flux point is
\begin{equation}
  \mathbf F^{*}_{f,q}
  =(1-\alpha_f)\,\mathbf F^{*}\bigl(\mathbf W^{-}_{f,q},\mathbf W^{+}_{f,q},\mathbf n_{f,q}\bigr)
  +\alpha_f\,\mathbf F^{*}\bigl(\mathbf W^{\mathrm{SC},-}_{f,q},\mathbf W^{\mathrm{SC},+}_{f,q},\mathbf n_{f,q}\bigr),
  \qquad \alpha_f=\max(\alpha_K,\alpha_{K'}),
  \label{eq:sob_faceflux}
\end{equation}
where $\mathbf W^{\mp}_{f,q}$ are the inner and outer states of Sec.~\ref{sec:fr_traces}, $K'$ is the element on the other side of the face, $\alpha_K,\alpha_{K'}$ are the blending coefficients of the two elements, and $\mathbf W^{\mathrm{SC},\mp}_{f,q}$ are the reconstructed face states of the two adjacent subcells. This flux replaces $\mathbf F^{*}(\mathbf W^{-}_{f,q},\mathbf W^{+}_{f,q},\mathbf n_{f,q})$ in the common normal flux~\eqref{eq:commonflux}. The single value~\eqref{eq:sob_faceflux}, transformed to the reference element, is used both in the FR correction of Eq.~\eqref{eq:sob_RFR} and as the outer subcell flux in Eq.~\eqref{eq:sob_RSC} by \emph{both} elements sharing the face. With $\alpha_K=1$, the element runs a genuine subcell FV scheme, with no contribution of the inner and outer FR states. At physical boundaries, the boundary flux, computed from the inner and ghost states, is used for both parts.

Since the FR correction and the subcell FV integrate exactly to the face fluxes over the element, and both elements of a face use the same value~\eqref{eq:sob_faceflux}, the blended scheme~\eqref{eq:sob_blend} is conservative for any blending coefficients. \citet{Babbar2024} start from the same construction, replacing the maximum with the mean of the two coefficients, and then correct the flux for admissibility.

\subsubsection{Face normals}
\label{sec:sob_normals}

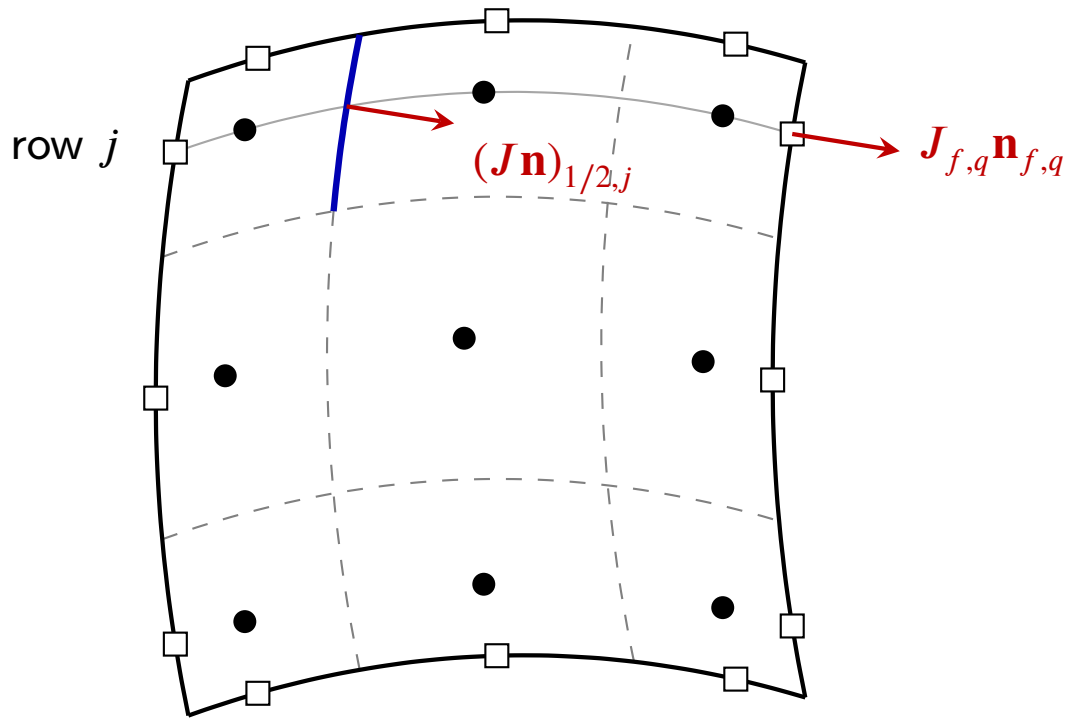
\begin{figure}[pos=htbp]
  \centering
\begin{tikzpicture}[>=stealth,font=\small,scale=1.2]
\draw[dashed,gray] plot[smooth] coordinates {(-0.576,-1.548) (-0.595,-1.450) (-0.613,-1.353) (-0.631,-1.256) (-0.647,-1.159) (-0.662,-1.061) (-0.676,-0.964) (-0.688,-0.867) (-0.700,-0.770) (-0.711,-0.673) (-0.720,-0.575) (-0.728,-0.478) (-0.736,-0.381) (-0.742,-0.284) (-0.747,-0.186) (-0.751,-0.089) (-0.753,0.008) (-0.755,0.105) (-0.756,0.202) (-0.755,0.300) (-0.753,0.397) (-0.751,0.494) (-0.747,0.591) (-0.742,0.689) (-0.736,0.786) (-0.728,0.883) (-0.720,0.980) (-0.711,1.077) (-0.700,1.175) (-0.688,1.272) (-0.676,1.369) (-0.662,1.466) (-0.647,1.564) (-0.631,1.661) (-0.613,1.758) (-0.595,1.855) (-0.576,1.952)};
\draw[dashed,gray] plot[smooth] coordinates {(-1.664,-0.828) (-1.570,-0.795) (-1.476,-0.763) (-1.381,-0.734) (-1.287,-0.706) (-1.192,-0.680) (-1.098,-0.656) (-1.003,-0.633) (-0.909,-0.612) (-0.814,-0.593) (-0.720,-0.575) (-0.626,-0.560) (-0.531,-0.546) (-0.437,-0.533) (-0.342,-0.523) (-0.248,-0.514) (-0.153,-0.507) (-0.059,-0.501) (0.036,-0.498) (0.130,-0.496) (0.224,-0.496) (0.319,-0.497) (0.413,-0.500) (0.508,-0.505) (0.602,-0.512) (0.697,-0.521) (0.791,-0.531) (0.886,-0.543) (0.980,-0.556) (1.074,-0.572) (1.169,-0.589) (1.263,-0.608) (1.358,-0.628) (1.452,-0.651) (1.547,-0.675) (1.641,-0.700) (1.736,-0.728)};
\draw[dashed,gray] plot[smooth] coordinates {(0.936,-1.503) (0.916,-1.406) (0.898,-1.309) (0.881,-1.211) (0.864,-1.114) (0.849,-1.017) (0.836,-0.920) (0.823,-0.823) (0.811,-0.725) (0.801,-0.628) (0.791,-0.531) (0.783,-0.434) (0.776,-0.336) (0.769,-0.239) (0.764,-0.142) (0.761,-0.045) (0.758,0.052) (0.756,0.150) (0.756,0.247) (0.756,0.344) (0.758,0.441) (0.761,0.539) (0.764,0.636) (0.769,0.733) (0.776,0.830) (0.783,0.927) (0.791,1.025) (0.801,1.122) (0.811,1.219) (0.823,1.316) (0.836,1.414) (0.849,1.511) (0.864,1.608) (0.881,1.705) (0.898,1.802) (0.916,1.900) (0.936,1.997)};
\draw[dashed,gray] plot[smooth] coordinates {(-1.664,0.728) (-1.570,0.761) (-1.476,0.792) (-1.381,0.822) (-1.287,0.850) (-1.192,0.876) (-1.098,0.900) (-1.003,0.923) (-0.909,0.944) (-0.814,0.963) (-0.720,0.980) (-0.626,0.996) (-0.531,1.010) (-0.437,1.022) (-0.342,1.033) (-0.248,1.042) (-0.153,1.049) (-0.059,1.054) (0.036,1.058) (0.130,1.060) (0.224,1.060) (0.319,1.058) (0.413,1.055) (0.508,1.050) (0.602,1.043) (0.697,1.035) (0.791,1.025) (0.886,1.013) (0.980,0.999) (1.074,0.984) (1.169,0.967) (1.263,0.948) (1.358,0.927) (1.452,0.905) (1.547,0.881) (1.641,0.855) (1.736,0.828)};
\draw[gray!70,thin] plot[smooth] coordinates {(-1.592,1.306) (-1.498,1.339) (-1.403,1.370) (-1.309,1.399) (-1.214,1.427) (-1.120,1.453) (-1.025,1.478) (-0.931,1.500) (-0.836,1.521) (-0.742,1.541) (-0.648,1.558) (-0.553,1.574) (-0.459,1.588) (-0.364,1.600) (-0.270,1.611) (-0.175,1.619) (-0.081,1.627) (0.014,1.632) (0.108,1.636) (0.202,1.637) (0.297,1.638) (0.391,1.636) (0.486,1.633) (0.580,1.628) (0.675,1.621) (0.769,1.613) (0.864,1.602) (0.958,1.591) (1.052,1.577) (1.147,1.562) (1.241,1.544) (1.336,1.526) (1.430,1.505) (1.525,1.483) (1.619,1.459) (1.714,1.433) (1.808,1.406)};
\draw[thick] plot[smooth] coordinates {(-1.520,-1.800) (-1.539,-1.703) (-1.558,-1.606) (-1.575,-1.508) (-1.591,-1.411) (-1.606,-1.314) (-1.620,-1.217) (-1.633,-1.119) (-1.644,-1.022) (-1.655,-0.925) (-1.664,-0.828) (-1.673,-0.731) (-1.680,-0.633) (-1.686,-0.536) (-1.691,-0.439) (-1.695,-0.342) (-1.698,-0.244) (-1.699,-0.147) (-1.700,-0.050) (-1.699,0.047) (-1.698,0.144) (-1.695,0.242) (-1.691,0.339) (-1.686,0.436) (-1.680,0.533) (-1.673,0.631) (-1.664,0.728) (-1.655,0.825) (-1.644,0.922) (-1.633,1.019) (-1.620,1.117) (-1.606,1.214) (-1.591,1.311) (-1.575,1.408) (-1.558,1.506) (-1.539,1.603) (-1.520,1.700)};
\draw[thick] plot[smooth] coordinates {(-1.520,-1.800) (-1.426,-1.767) (-1.331,-1.736) (-1.237,-1.706) (-1.142,-1.678) (-1.048,-1.652) (-0.953,-1.628) (-0.859,-1.605) (-0.764,-1.584) (-0.670,-1.565) (-0.576,-1.548) (-0.481,-1.532) (-0.387,-1.518) (-0.292,-1.505) (-0.198,-1.495) (-0.103,-1.486) (-0.009,-1.479) (0.086,-1.474) (0.180,-1.470) (0.274,-1.468) (0.369,-1.468) (0.463,-1.469) (0.558,-1.473) (0.652,-1.478) (0.747,-1.484) (0.841,-1.493) (0.936,-1.503) (1.030,-1.515) (1.124,-1.529) (1.219,-1.544) (1.313,-1.561) (1.408,-1.580) (1.502,-1.600) (1.597,-1.623) (1.691,-1.647) (1.786,-1.673) (1.880,-1.700)};
\draw[thick] plot[smooth] coordinates {(1.880,-1.700) (1.861,-1.603) (1.842,-1.506) (1.825,-1.408) (1.809,-1.311) (1.794,-1.214) (1.780,-1.117) (1.767,-1.019) (1.756,-0.922) (1.745,-0.825) (1.736,-0.728) (1.727,-0.631) (1.720,-0.533) (1.714,-0.436) (1.709,-0.339) (1.705,-0.242) (1.702,-0.144) (1.701,-0.047) (1.700,0.050) (1.701,0.147) (1.702,0.244) (1.705,0.342) (1.709,0.439) (1.714,0.536) (1.720,0.633) (1.727,0.731) (1.736,0.828) (1.745,0.925) (1.756,1.022) (1.767,1.119) (1.780,1.217) (1.794,1.314) (1.809,1.411) (1.825,1.508) (1.842,1.606) (1.861,1.703) (1.880,1.800)};
\draw[thick] plot[smooth] coordinates {(-1.520,1.700) (-1.426,1.733) (-1.331,1.764) (-1.237,1.794) (-1.142,1.822) (-1.048,1.848) (-0.953,1.872) (-0.859,1.895) (-0.764,1.916) (-0.670,1.935) (-0.576,1.952) (-0.481,1.968) (-0.387,1.982) (-0.292,1.995) (-0.198,2.005) (-0.103,2.014) (-0.009,2.021) (0.086,2.026) (0.180,2.030) (0.274,2.032) (0.369,2.032) (0.463,2.031) (0.558,2.027) (0.652,2.022) (0.747,2.016) (0.841,2.007) (0.936,1.997) (1.030,1.985) (1.124,1.971) (1.219,1.956) (1.313,1.939) (1.408,1.920) (1.502,1.900) (1.597,1.877) (1.691,1.853) (1.786,1.827) (1.880,1.800)};
\draw[blue!70!black,very thick] plot[smooth] coordinates {(-0.720,0.980) (-0.714,1.041) (-0.708,1.102) (-0.701,1.163) (-0.694,1.223) (-0.687,1.284) (-0.679,1.345) (-0.670,1.406) (-0.662,1.466) (-0.652,1.527) (-0.643,1.588) (-0.633,1.649) (-0.622,1.709) (-0.611,1.770) (-0.600,1.831) (-0.588,1.892) (-0.576,1.952)};
\fill (-1.209,-1.282) circle (1.8pt);
\fill (-1.317,0.073) circle (1.8pt);
\fill (-1.209,1.429) circle (1.8pt);
\fill (0.108,-1.076) circle (1.8pt);
\fill (0.000,0.280) circle (1.8pt);
\fill (0.108,1.636) circle (1.8pt);
\fill (1.425,-1.205) circle (1.8pt);
\fill (1.317,0.151) circle (1.8pt);
\fill (1.425,1.506) circle (1.8pt);
\draw[fill=white] (-1.592,-1.406) +(-1.8pt,-1.8pt) rectangle +(1.8pt,1.8pt);
\draw[fill=white] (1.808,-1.306) +(-1.8pt,-1.8pt) rectangle +(1.8pt,1.8pt);
\draw[fill=white] (-1.137,-1.677) +(-1.8pt,-1.8pt) rectangle +(1.8pt,1.8pt);
\draw[fill=white] (-1.137,1.823) +(-1.8pt,-1.8pt) rectangle +(1.8pt,1.8pt);
\draw[fill=white] (-1.700,-0.050) +(-1.8pt,-1.8pt) rectangle +(1.8pt,1.8pt);
\draw[fill=white] (1.700,0.050) +(-1.8pt,-1.8pt) rectangle +(1.8pt,1.8pt);
\draw[fill=white] (0.180,-1.470) +(-1.8pt,-1.8pt) rectangle +(1.8pt,1.8pt);
\draw[fill=white] (0.180,2.030) +(-1.8pt,-1.8pt) rectangle +(1.8pt,1.8pt);
\draw[fill=white] (-1.592,1.306) +(-1.8pt,-1.8pt) rectangle +(1.8pt,1.8pt);
\draw[fill=white] (1.808,1.406) +(-1.8pt,-1.8pt) rectangle +(1.8pt,1.8pt);
\draw[fill=white] (1.497,-1.599) +(-1.8pt,-1.8pt) rectangle +(1.8pt,1.8pt);
\draw[fill=white] (1.497,1.901) +(-1.8pt,-1.8pt) rectangle +(1.8pt,1.8pt);
\draw[->,thick,red!75!black] (-0.648,1.558) -- (-0.055,1.464) node[below right]{$(J\mathbf n)_{1/2,j}$};
\draw[->,thick,red!75!black] (1.808,1.406) -- (2.401,1.311) node[right]{$J_{f,q}\mathbf n_{f,q}$};
\node[left] at (-1.792,1.326) {row $j$};
\end{tikzpicture}
  \caption{A curved mesh element for $P=2$, with its subcell faces (dashed) and the row $j$ of solution points (grey); markers as in Fig.~\ref{fig:sob_grid}. Arrows: the scaled normal $(J\mathbf n)_{1/2,j}$ of the subcell face drawn in blue, and $J_{f,q}\mathbf n_{f,q}$ at the flux point that closes the row.}
  \label{fig:sob_normals}
\end{figure}

The face Jacobian $J_{i+1/2,j}$ and the unit normal $\mathbf n_{i+1/2,j}$ in Eq.~\eqref{eq:sob_intflux} are the length and the direction of the scaled normal $(J\mathbf n)_{i+1/2,j}$ of the subcell face. Like the fluxes of Eq.~\eqref{eq:sob_RSC}, it points towards increasing $\xi$ (Fig.~\ref{fig:sob_normals}). The two end faces of the row $j$, at $\xi=-1$ and $\xi=1$, coincide with element faces. They take the scaled normal $J_{f,q}\mathbf n_{f,q}$ that FR already uses at the flux point there, oriented towards increasing $\xi$, and we write these end values as $(J\mathbf n)_{-1/2,j}$ and $(J\mathbf n)_{P+1/2,j}$.

For the internal faces, the obvious choice for $(J\mathbf n)_{i+1/2,j}$ is the scaled normal given by the mapping,
\begin{equation}
  (J\mathbf n)^{\xi}=\Bigl(\frac{\partial y}{\partial\eta},\,-\frac{\partial x}{\partial\eta}\Bigr),
  \label{eq:sob_Jn}
\end{equation}
evaluated at $(\xi_{i+1/2},\eta_j)$. Its direction is normal to the curve $\xi=\xi_{i+1/2}$, and its length is the ratio of physical to reference length along that curve. These are the metric terms of the transformed flux~\eqref{eq:transformed}, $\hat{\mathbf f}=\bigl(\mathbf F\cdot(J\mathbf n)^{\xi},\,\mathbf F\cdot(J\mathbf n)^{\eta}\bigr)$, with $(J\mathbf n)^{\eta}=(-\partial y/\partial\xi,\,\partial x/\partial\xi)$.

This choice, however, does not preserve a uniform state on curved elements. For a uniform state, every numerical flux is exact, so the subcell residual vanishes only if, for every subcell, the outward scaled normals of its four faces, each multiplied by the reference length of the face, sum to zero. Normals evaluated from the mapping do not satisfy this condition on curved elements (Appendix~\ref{app:subcell_metrics}).

We therefore build the internal face normals from the derivative that FR itself applies to the metric. For a uniform state, FR differentiates along the row $j$ the polynomial of $\mathbf F^{C}\cdot(J\mathbf n)^{\xi}$, corrected at both ends of the row. This gives $\mathbf F^{C}\cdot\boldsymbol\delta^{\xi}_{ij}$, with
\begin{equation}
  \boldsymbol\delta^{\xi}_{ij}
  =\sum_{m=0}^{P}D_{im}\,(J\mathbf n)^{\xi}_{mj}
  +h_L'(\xi_i)\Bigl((J\mathbf n)_{-1/2,j}-\sum_{m=0}^{P}l_m(-1)\,(J\mathbf n)^{\xi}_{mj}\Bigr)
  +h_R'(\xi_i)\Bigl((J\mathbf n)_{P+1/2,j}-\sum_{m=0}^{P}l_m(1)\,(J\mathbf n)^{\xi}_{mj}\Bigr).
  \label{eq:sob_metricderiv}
\end{equation}
Here $(J\mathbf n)^{\xi}_{mj}$ is the scaled normal~\eqref{eq:sob_Jn} at the solution point $\boldsymbol\xi_{mj}$, $l_m$ the one-dimensional Lagrange polynomial of the point $\xi_m$, $D_{im}=l_m'(\xi_i)$ the one-dimensional derivative matrix, and $h_L$, $h_R$ the left and right one-dimensional correction functions of Sec.~\ref{sec:fr_correction}, with $h_L(-1)=1$, $h_L(1)=0$, $h_R(-1)=0$, $h_R(1)=1$. The first term is the derivative of the interpolated scaled normal. The other two correct its values at the ends of the row to the element-face values, as FR corrects the flux. The vector $\boldsymbol\delta^{\eta}_{ij}$ is defined in the same way along the column $i$.

The internal face normals then follow by telescoping along the row, starting from the end value $(J\mathbf n)_{-1/2,j}$ at $\xi=-1$,
\begin{equation}
  (J\mathbf n)_{i+1/2,j}=(J\mathbf n)_{i-1/2,j}+\varpi_i\,\boldsymbol\delta^{\xi}_{ij},\qquad i=0,\dots,P-1.
  \label{eq:sob_telescope}
\end{equation}
The recursion lands exactly on the element-face value at the other end of the row. For a uniform state, the subcell residual then equals the FR residual at every solution point (Appendix~\ref{app:subcell_metrics}), so the blended residual~\eqref{eq:sob_blend} preserves a uniform state for any $\alpha_K$ whenever FR does.

This is the construction of \citet[App.~B.4]{Hennemann2021} for Lobatto points, also used by \citet{Babbar2025} on curvilinear meshes. There, the solution points include the element faces, and the correction terms of Eq.~\eqref{eq:sob_metricderiv} vanish. On GL points, the correction terms are what make the recursion land on the element face.

\subsection{Subcell reconstruction}
\label{sec:sob_subcell_options}

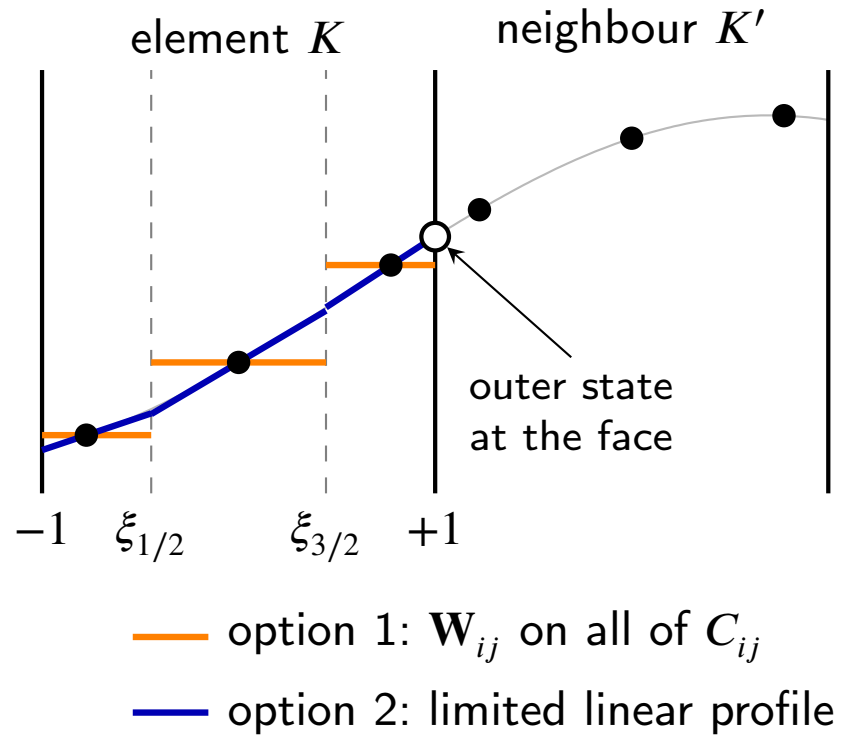
\begin{figure}[pos=htbp]
  \centering
\begin{tikzpicture}[>=stealth,font=\small]
\draw[gray!55,thin] plot[smooth] coordinates {(0.000,0.400) (0.026,0.406) (0.052,0.413) (0.078,0.420) (0.105,0.427) (0.131,0.434) (0.157,0.442) (0.183,0.449) (0.209,0.457) (0.235,0.466) (0.261,0.474) (0.287,0.483) (0.314,0.492) (0.340,0.501) (0.366,0.510) (0.392,0.520) (0.418,0.529) (0.444,0.539) (0.470,0.549) (0.496,0.560) (0.523,0.570) (0.549,0.581) (0.575,0.592) (0.601,0.603) (0.627,0.615) (0.653,0.626) (0.679,0.638) (0.706,0.650) (0.732,0.662) (0.758,0.675) (0.784,0.687) (0.810,0.700) (0.836,0.713) (0.862,0.726) (0.888,0.739) (0.915,0.752) (0.941,0.766) (0.967,0.779) (0.993,0.793) (1.019,0.807) (1.045,0.821) (1.071,0.835) (1.097,0.850) (1.124,0.864) (1.150,0.879) (1.176,0.894) (1.202,0.909) (1.228,0.924) (1.254,0.939) (1.280,0.954) (1.307,0.970) (1.333,0.985) (1.359,1.001) (1.385,1.017) (1.411,1.033) (1.437,1.049) (1.463,1.065) (1.489,1.081) (1.516,1.097) (1.542,1.113) (1.568,1.130) (1.594,1.146) (1.620,1.163) (1.646,1.179) (1.672,1.196) (1.698,1.213) (1.725,1.229) (1.751,1.246) (1.777,1.263) (1.803,1.280) (1.829,1.297) (1.855,1.314) (1.881,1.331) (1.908,1.348) (1.934,1.365) (1.960,1.382) (1.986,1.400) (2.012,1.417) (2.038,1.434) (2.064,1.451) (2.090,1.468) (2.117,1.485) (2.143,1.502) (2.169,1.520) (2.195,1.537) (2.221,1.554) (2.247,1.571) (2.273,1.588) (2.299,1.605) (2.326,1.622) (2.352,1.639) (2.378,1.656) (2.404,1.673) (2.430,1.690) (2.456,1.707) (2.482,1.723) (2.509,1.740) (2.535,1.757) (2.561,1.773) (2.587,1.790) (2.613,1.806) (2.639,1.822) (2.665,1.839) (2.691,1.855) (2.718,1.871) (2.744,1.887) (2.770,1.903) (2.796,1.918) (2.822,1.934) (2.848,1.950) (2.874,1.965) (2.901,1.980) (2.927,1.996) (2.953,2.011) (2.979,2.026) (3.005,2.041) (3.031,2.055) (3.057,2.070) (3.083,2.084) (3.110,2.099) (3.136,2.113) (3.162,2.127) (3.188,2.141) (3.214,2.154) (3.240,2.168) (3.266,2.181) (3.292,2.194) (3.319,2.207) (3.345,2.220) (3.371,2.233) (3.397,2.246) (3.423,2.258) (3.449,2.270) (3.475,2.282) (3.502,2.294) (3.528,2.306) (3.554,2.317) (3.580,2.328) (3.606,2.339) (3.632,2.350) (3.658,2.361) (3.684,2.371) (3.711,2.381) (3.737,2.391) (3.763,2.401) (3.789,2.411) (3.815,2.420) (3.841,2.429) (3.867,2.438) (3.893,2.447) (3.920,2.455) (3.946,2.464) (3.972,2.472) (3.998,2.479) (4.024,2.487) (4.050,2.494) (4.076,2.501) (4.103,2.508) (4.129,2.515) (4.155,2.521) (4.181,2.528) (4.207,2.533) (4.233,2.539) (4.259,2.545) (4.285,2.550) (4.312,2.555) (4.338,2.559) (4.364,2.564) (4.390,2.568) (4.416,2.572) (4.442,2.575) (4.468,2.579) (4.494,2.582) (4.521,2.585) (4.547,2.588) (4.573,2.590) (4.599,2.592) (4.625,2.594) (4.651,2.596) (4.677,2.597) (4.704,2.598) (4.730,2.599) (4.756,2.600) (4.782,2.600) (4.808,2.600) (4.834,2.600) (4.860,2.599) (4.886,2.599) (4.913,2.598) (4.939,2.597) (4.965,2.595) (4.991,2.593) (5.017,2.591) (5.043,2.589) (5.069,2.587) (5.095,2.584) (5.122,2.581) (5.148,2.578) (5.174,2.574) (5.200,2.570)};
\draw[thick] (0.000,0.100) -- (0.000,2.900);
\draw[dashed,gray] (0.722,0.100) -- (0.722,2.900);
\draw[dashed,gray] (1.878,0.100) -- (1.878,2.900);
\draw[thick] (2.600,0.100) -- (2.600,2.900);
\draw[thick] (5.200,0.100) -- (5.200,2.900);
\node[above] at (1.300,2.900) {element $K$};
\node[above] at (3.900,2.900) {neighbour $K'$};
\draw[orange,very thick] (0.000,0.485) -- (0.722,0.485);
\draw[blue!70!black,very thick] (0.000,0.386) -- (0.722,0.629);
\fill (0.293,0.485) circle (2.2pt);
\draw[orange,very thick] (0.722,0.966) -- (1.878,0.966);
\draw[blue!70!black,very thick] (0.722,0.624) -- (1.878,1.308);
\fill (1.300,0.966) circle (2.2pt);
\draw[orange,very thick] (1.878,1.610) -- (2.600,1.610);
\draw[blue!70!black,very thick] (1.878,1.330) -- (2.600,1.801);
\fill (2.307,1.610) circle (2.2pt);
\fill (2.893,1.976) circle (2.2pt);
\fill (3.900,2.449) circle (2.2pt);
\fill (4.907,2.598) circle (2.2pt);
\draw[thick,fill=white] (2.600,1.798) circle (2.6pt);
\draw[<-,shorten <=3pt] (2.600,1.798) -- (3.500,1.000) node[below,align=center,font=\footnotesize]{outer state\\at the face};
\node[below] at (0.000,0.100) {$-1$};
\node[below] at (0.722,0.100) {$\xi_{1/2}$};
\node[below] at (1.878,0.100) {$\xi_{3/2}$};
\node[below] at (2.600,0.100) {$+1$};
\draw[orange,very thick] (0.600,-0.850) -- +(0.5,0) node[right,black]{option 1: $\mathbf W_{ij}$ on all of $C_{ij}$};
\draw[blue!70!black,very thick] (0.600,-1.350) -- +(0.5,0) node[right,black]{option 2: limited linear profile};
\end{tikzpicture}
  \caption{Subcell reconstruction for $P=2$: a smooth field (grey) along one row of solution points across the face shared by elements $K$ and $K'$, with its Option~1 and Option~2 representations. The outer state at the face is that of the neighbour $K'$.}
  \label{fig:sob_recon}
\end{figure}

The face states $\mathbf W^{\mp}_{i+1/2,j}$ of Eq.~\eqref{eq:sob_intflux} and $\mathbf W^{\mathrm{SC},\mp}_{f,q}$ of Eq.~\eqref{eq:sob_faceflux} are obtained from the subcell values in one of two ways (Fig.~\ref{fig:sob_recon}). Option~1 takes the subcell value over the whole subcell, giving a first-order FV scheme; Option~2 reconstructs a limited linear profile in each subcell, giving a second-order FV scheme. In this work, we propose to combine the two options element by element (Sec.~\ref{sec:sob_selective}).

\subsubsection{Option 1: first-order subcells}
\label{sec:sob_option1}

With Option~1, the two states at an internal subcell face are the values at the solution points on either side of it,
\begin{equation}
  \mathbf W^{-}_{i+1/2,j}=\mathbf W_{ij},\qquad \mathbf W^{+}_{i+1/2,j}=\mathbf W_{i+1,j},
  \label{eq:sob_fo}
\end{equation}
and the states $\mathbf W^{\mathrm{SC},\mp}_{f,q}$ at an element face are the values at the solution points of the two elements closest to the flux point (Fig.~\ref{fig:sob_recon}), which gives the first-order subcell scheme of \citet{Sonntag2017} and \citet{Hennemann2021}.

In the elements where this first-order scheme is used, the diffusive terms are blended with the same coefficient, so the low-order scheme covers both the convective and diffusive terms, as in \citet{Krais2021} and \citet{Lin2023}. The update~\eqref{eq:sob_blend} then becomes
\begin{equation}
  \frac{d\hat{\mathbf U}_{ij}}{dt}
  =(1-\alpha_K)\,\underbrace{\bigl(\mathbf R^{C}_{\mathrm{FR},ij}+\mathbf R^{D}_{\mathrm{FR},ij}\bigr)}_{\text{high order}}
  +\alpha_K\,\underbrace{\bigl(\mathbf R^{C}_{\mathrm{SC},ij}+\mathbf R^{D}_{\mathrm{SC},ij}\bigr)}_{\text{low order}}
  +J_{K,ij}\,\mathbf S_{ij},
  \label{eq:sob_blend_fo}
\end{equation}
where $\mathbf R^{D}_{\mathrm{SC}}$ is an FV residual of the diffusive terms on the subcell grid, built from two-point fluxes between neighbouring solution points (Appendix~\ref{app:subcell_diffusion}).

However, applied to the GL subcell grid, the first-order scheme presents a drawback that should be noted. Its face flux is built from the two neighbouring point values, whereas the GL points are neither evenly spaced nor at the centres of their subcells (Fig.~\ref{fig:sob_grid}). For a smooth field, the discrete derivative at every solution point therefore carries an error that does not shrink with the mesh size (Appendix~\ref{app:pointwise}). Although these errors cancel in the element mean, on a smooth flow they leave a small oscillation that repeats in every blended element and, in the blunt-body test of Appendix~\ref{app:gnoffo}, reaches the stagnation heat flux. This is why the second-order option is used where the flow is smooth.

\subsubsection{Option 2: limited second-order subcells}

Following the second-order variants of \citet{Sonntag2017}, \citet{RuedaRamirez2022} and \citet{Babbar2024}, the subcell values are reconstructed linearly up to the subcell faces. The reconstruction is done direction by direction, in reconstruction variables $\mathbf Z=\mathcal T(\mathbf W)$, where $\mathcal T$ is an invertible variable transformation. For the thermochemical nonequilibrium model, $\mathcal T$ is the identity: the reconstruction is done in the logarithmic update variables~\eqref{eq:logvars}, and every reconstructed face state is admissible (Sec.~\ref{sec:logvars_positivity}). For the perfect gas of Appendix~\ref{app:gnoffo}, $\mathbf Z=(p,u,v,T)$.

Along the row $j$, the subcell $C_{ij}$ carries the linear profile
\begin{equation}
  \mathbf Z_{ij}(\xi)=\mathbf Z_{ij}+\mathbf L_{ij}\,(\xi-\xi_i),\qquad \xi\in[\xi_{i-1/2},\xi_{i+1/2}],
  \label{eq:sob_profile}
\end{equation}
where $\mathbf L_{ij}$ is a limited slope along $\xi$, computed from the neighbouring values (Fig.~\ref{fig:sob_recon}). As in \citet{Babbar2024}, the profile takes the subcell value at the solution point, not at the subcell centre, which is what makes the reconstruction exact for linear fields: a profile taking the subcell value at the centre would be off by $\mathcal O(h)$ at the faces, the order of the Option~1 error that the reconstruction is meant to remove. The internal face states are
\begin{equation}
  \mathbf W^{-}_{i+1/2,j}=\mathcal T^{-1}\!\bigl(\mathbf Z_{ij}+\mathbf L_{ij}(\xi_{i+1/2}-\xi_i)\bigr),\qquad
  \mathbf W^{+}_{i+1/2,j}=\mathcal T^{-1}\!\bigl(\mathbf Z_{i+1,j}+\mathbf L_{i+1,j}(\xi_{i+1/2}-\xi_{i+1})\bigr).
\end{equation}
At the two ends of the row, $\xi_{-1/2}=-1$ and $\xi_{P+1/2}=1$, the same expressions give the state of each element on its side of the element face, $\mathbf W^{\mathrm{SC},\mp}_{f,q}$ in Eq.~\eqref{eq:sob_faceflux}.

The slope $\mathbf L_{ij}$ is built from the values on either side of the point: with the index $j$ dropped, let $\mathbf Z_{i-1}$, $\mathbf Z_i$, $\mathbf Z_{i+1}$ be three consecutive values along a row, at the reference positions $\tilde\xi_{i-1}<\xi_i<\tilde\xi_{i+1}$, and
\begin{equation}
  \mathbf D^{-}_i=\frac{\mathbf Z_i-\mathbf Z_{i-1}}{\xi_i-\tilde\xi_{i-1}},\qquad
  \mathbf D^{+}_i=\frac{\mathbf Z_{i+1}-\mathbf Z_i}{\tilde\xi_{i+1}-\xi_i},
\end{equation}
the backward and forward secants. For an interior point, $0<i<P$, the neighbours are the adjacent solution points, $\tilde\xi_{i\pm1}=\xi_{i\pm1}$. For an end point, one neighbour lies outside the element and is replaced by a value sampled \emph{on the element face}, at $\tilde\xi=\pm1$: the outer state $\mathbf W^{+}$, from the neighbouring element, at the flux point that ends the row (open circle in Fig.~\ref{fig:sob_recon}), or, at a physical boundary, the average $\tfrac12(\mathbf W^{-}+\mathbf W^{g})$ of the inner and ghost states of Sec.~\ref{sec:fr_bc}. The three values then lie on one row of the element, so both secants are taken along the same reference direction and the slope is exact for linear fields on any affine element, skewed or not. Unlike \citet{Babbar2024}, who take the outer value from the solution points of the neighbouring element, the slope thus involves only the element's reference coordinate, even on curved meshes.

These two secants are combined, component by component, through a limiter $\Psi$, for which the minmod limiter used by \citet{Sonntag2017} and \citet{RuedaRamirez2022} is not suited to a steady implicit computation. Minmod is not differentiable where a secant changes sign, which happens at every smooth extremum, and limiters that switch in this way are known to stall the convergence to a steady state, whereas a Newton method requires a differentiable limiter \citep{Venkatakrishnan1995}. We therefore use a form of the limiter of \citet{VanAlbada1982},
\begin{equation}
  \mathbf L_i=\Psi(\mathbf D^{-}_i,\mathbf D^{+}_i),\qquad
  \Psi(D^-,D^+)=\frac{\bigl(\max(D^-D^+,0)+\vartheta^2\bigr)(D^-+D^+)}{(D^-)^2+(D^+)^2+2\vartheta^2},
  \qquad
  \vartheta=\vartheta_r\max\bigl(|Z_{i-1}|,|Z_i|,|Z_{i+1}|\bigr),
  \label{eq:sob_va}
\end{equation}
where $D^-$ and $D^+$ stand for one component of $\mathbf D^{-}_i$ and $\mathbf D^{+}_i$, respectively, and $Z_{i-1}$, $Z_i$, $Z_{i+1}$ for the same component of $\mathbf Z_{i-1}$, $\mathbf Z_i$, $\mathbf Z_{i+1}$. The threshold $\vartheta$, with $\vartheta_r=10^{-3}$, plays the role of the small parameter of \citet{VanAlbada1982} and \citet{Venkatakrishnan1995}. For $D^-D^+>0$, $\Psi$ is the smooth average of \citet[Eq.~37]{VanAlbada1982}, which for secants much larger than $\vartheta$ reduces to the classical van Albada slope $D^-D^+(D^-+D^+)/\bigl((D^-)^2+(D^+)^2\bigr)$. Where the secants have opposite signs and are much larger than $\vartheta$ (an overshoot), the term $\max(D^-D^+,0)$ gives $\Psi\approx0$. Secants much smaller than $\vartheta$ (a smooth extremum or a near-constant region) give their average $\tfrac12(D^-+D^+)$: smooth extrema are not clipped, and the limiter does not switch there.

\subsubsection{Selective reconstruction}
\label{sec:sob_selective}

Selective reconstruction uses Option~2 where the flow is smooth, for accuracy, and Option~1 at shocks, for robustness. Near a shock, the secants are much larger than $\vartheta$, so the smoothing of Eq.~\eqref{eq:sob_va} has no effect there, and a limiter that acts at a shock keeps switching and stalls the convergence to a steady state, while the first-order scheme converges \citep{Venkatakrishnan1995}. Since the flow ahead of a shock is supersonic, the choice is made from the local Mach number, through a flag $\chi_K\in\{0,1\}$ of each element,
\begin{equation}
  \chi_K=1\quad\Longleftrightarrow\quad
  \max_{ij}\mathit{Ma}_{ij}<\mathit{Ma}_{\mathrm{rec}},
\end{equation}
where $\mathit{Ma}_{ij}=|\mathbf V_{ij}|/a_{f,ij}$ is the local Mach number at the solution points, computed with the frozen speed of sound, and $\mathit{Ma}_{\mathrm{rec}}$ is a Mach limit. An element with $\chi_K=1$ uses Option~2 and blends only its convective terms, as in Eq.~\eqref{eq:sob_blend}, while its diffusive terms keep their FR form. An element with $\chi_K=0$ uses Option~1 and blends both its convective and diffusive terms, as in Eq.~\eqref{eq:sob_blend_fo}. Each element uses its option also on its side of the element faces. With $\mathit{Ma}_{\mathrm{rec}}=1$, the low-order operator is second order in the subsonic regions, where the flow behind a shock is smooth, and Option~1 would leave its oscillation, and first order in every element that contains supersonic flow, including the shocks. The price is a less sharp shock profile \citep{RuedaRamirez2022}, but not a lower order of accuracy, since any shock-capturing scheme is first order at a shock.

\subsection{Blending coefficient}
\label{sec:sob_alpha}

The blending coefficient $\alpha_K$ of Eq.~\eqref{eq:sob_blend} is computed from a modal smoothness indicator of the pressure. Let $\hat p_k$, $k=1,\dots,(P+1)^2$, be the coefficients of the pressure of element $K$ in the tensor-product Legendre basis, obtained from its values at the solution points with the inverse Vandermonde matrix. Mode $k$ is the product of Legendre polynomials of degrees $k_\xi$ and $k_\eta$, and its degree is $\deg(k)=\max(k_\xi,k_\eta)$. With $\mathcal E_\ell=\sum_{\deg(k)=\ell}\hat p_k^2$ the energy of the modes of degree $\ell$, the indicator of \citet{Hennemann2021}, which follows \citet{Persson2006}, is
\begin{equation}
  \sigma_K=\max\Bigl(\frac{\mathcal E_P}{\sum_{\ell=0}^{P}\mathcal E_\ell},\;\frac{\mathcal E_{P-1}}{\sum_{\ell=0}^{P-1}\mathcal E_\ell}\Bigr),
\end{equation}
and the smoothness of element $K$ is $\zeta_K=\log_{10}\sigma_K$.

The smoothness is mapped to a coefficient with the sine ramp of \citet{Persson2006},
\begin{equation}
  \mathcal A(\zeta)=\begin{cases}
    0, & \zeta<\zeta_0-\kappa,\\
    \tfrac12\Bigl(1+\sin\dfrac{\pi(\zeta-\zeta_0)}{2\kappa}\Bigr), & |\zeta-\zeta_0|\le\kappa,\\
    1, & \zeta>\zeta_0+\kappa,
  \end{cases}
\end{equation}
centred on the reference smoothness $\zeta_0=-c_0\log_{10}(P+1)$, with the parameter $c_0$ and the half-width $\kappa$. As in \citet{Hennemann2021}, values below $\alpha_{\min}$ are set to zero and values above $\alpha_{\max}$ are limited to $\alpha_{\max}$, which gives the coefficient requested by the indicator,
\begin{equation}
  \hat\alpha_K=\begin{cases}
    0, & \mathcal A(\zeta_K)<\alpha_{\min},\\
    \min\bigl(\mathcal A(\zeta_K),\alpha_{\max}\bigr), & \text{otherwise}.
  \end{cases}
\end{equation}
Following the same work, the requested coefficient is then spread to the neighbouring elements, here the elements sharing a node with $K$, denoted $\mathcal N(K)$, by two Jacobi passes of
\begin{equation}
  \hat\alpha_K\leftarrow\max\Bigl(\hat\alpha_K,\;\tfrac12\max_{K'\in\mathcal N(K)}\hat\alpha_{K'}\Bigr).
\end{equation}

In the implicit iteration, the applied coefficient $\alpha_K$ follows $\hat\alpha_K$ with a relaxation and is frozen once the iteration stalls (Sec.~\ref{sec:sob_freeze}).

\section{Implicit time marching}
\label{sec:implicit}

The steady solutions are obtained by marching the semi-discrete equations~\eqref{eq:semidiscrete} in pseudo-time with a linearised backward Euler scheme, as in our previous implicit FR solvers for high-speed flows \citep{Vandenhoeck2019,Dhib2024}. Starting from a uniform free stream, the iteration goes through a strong transient, in which shocks form and move to their positions. The chemistry and the vibrational relaxation are stiff, and the partial densities span many orders of magnitude. A step acceptable for the mixture can then be far outside the range of validity of its linearisation for a trace species or temperature, and a CFL schedule tuned for one case does not carry over to the next.

In this work, the update is limited and the CFL number adapted automatically, following the pseudo-transient continuation of \citet{Ceze2013,Ceze2015}, developed for DG discretisations of the RANS equations of a perfect gas. We extend this approach to high-order FR discretisations of TCNEQ flows. The update is bounded in the logarithmic partial densities and temperatures, and two rules handle the electrons of ionised mixtures (Sec.~\ref{sec:impl_update}). To the authors' knowledge, such a limited update has never been applied before to high-order simulations of nonequilibrium flows. No CFL schedule is prescribed. Each computation from the free stream or from a lower order starts with a small CFL number, and the CFL controller of Sec.~\ref{sec:impl_cfl}, together with the automatic freeze of the blending coefficient (Sec.~\ref{sec:sob_freeze}), brings the computation to the steady state. With the same settings at every order and independently of the flow conditions, all the computations of Sec.~\ref{sec:validation} reach the maximum CFL number, $\mathrm{CFL}_{\max}=1000$, and converge in fewer than 3800 iterations.

\subsection{Linearised backward Euler scheme}
\label{sec:impl_newton}
\label{sec:impl_dt}
\label{sec:impl_linear}

In this section, the indices $i$ and $j$ number the solution points of the whole mesh. For implicit time stepping, consider the backward Euler scheme applied to the semi-discrete equations~\eqref{eq:semidiscrete} at solution point $i$, written in the update variables:
\begin{equation}
  J_i\Bigl(\frac{\partial\vect U}{\partial\vect W}\Bigr)_i\frac{\vect W_i^{n+1}-\vect W_i^{n}}{\Delta t_K}=\vect R_i(\vect W^{n+1}),
  \label{eq:beuler}
\end{equation}
where the superscript $n$ is the pseudo-time step, $J_i$ the Jacobian determinant of the mapping at the point, $(\partial\vect U/\partial\vect W)_i$ the Jacobian of the map~\eqref{eq:w_of_v} at $\vect W_i$, and $\Delta t_K$ the local pseudo-time step of the element $K$ that contains point $i$, obtained from the CFL number and the largest wave speeds on the faces of the element. Since $\vect W^{n+1}$ is unknown, this constitutes a nonlinear system that must be linearised. Using a first-order Taylor expansion of the residual,
\begin{equation}
  \vect R_i(\vect W^{n+1})\approx\vect R_i(\vect W^{n})+\sum_{j\in\mathcal D(i)}\Bigl(\frac{\partial\vect R_i}{\partial\vect W_j}\Bigr)^{n}\bigl(\vect W_j^{n+1}-\vect W_j^{n}\bigr),
  \label{eq:taylor}
\end{equation}
where the stencil $\mathcal D(i)$ contains the solution points of the element of point $i$ and of its face neighbours (Sec.~\ref{sec:fr_diff}), the linearised system for the change $\Delta\vect W^{n}=\vect W^{n+1}-\vect W^{n}$ becomes
\begin{equation}
  \Bigl(\mathbf M^{n}-\mathbf A^{n}\Bigr)\,\Delta\vect W^{n}=\vect R(\vect W^{n}),
  \label{eq:newton_v}
\end{equation}
where $\mathbf A^{n}$ is the Jacobian matrix, with the blocks $\partial\vect R_i/\partial\vect W_j$, and $\mathbf M^{n}$ is the block-diagonal matrix with the blocks $J_i(\partial\vect U/\partial\vect W)_i/\Delta t_K$. The system is linearised once per pseudo-time step and solved using the GMRES method \citep{Saad1986} of PETSc \citep{Balay2018} with an additive Schwarz preconditioner. The new state is
\begin{equation}
  \vect W^{n+1}=\vect W^{n}+\omega^{n}\,\Delta\vect W^{n},
  \label{eq:update}
\end{equation}
where $\omega^{n}\in(0,1]$ is the relaxation factor of the limited update of Sec.~\ref{sec:impl_update}. For ionised mixtures, the state filter~\eqref{eq:filter} is applied to $\vect W^{n+1}$. As the pseudo-time step grows, $\mathbf M^{n}$ vanishes and the scheme becomes Newton's method for the steady equations $\vect R(\vect W)=\vect 0$.

The linearised step does not depend on the choice of update variables, since $\mathbf M^{n}$ and $\mathbf A^{n}$ both carry the factor $\partial\vect U/\partial\vect W$; the system~\eqref{eq:newton_v} gives the same change of the conservative variables, $(\partial\vect U/\partial\vect W)\Delta\vect W^{n}$, as with $\vect U$ as unknowns. The logarithmic variables of Sec.~\ref{sec:logvars} change how the Jacobian is approximated (Sec.~\ref{sec:impl_jacobian}) and how the step is applied (Sec.~\ref{sec:impl_update}).

\subsection{Jacobian matrix computation}
\label{sec:impl_jacobian}

The Jacobian matrix is computed element by element using numerical differentiation with a first-order finite difference scheme. For component $l$ of the update state at solution point $j$,
\begin{equation}
  \frac{\partial\vect R_i}{\partial w_{j,l}}\approx\frac{\vect R_i(\vect W^{(j,l)})-\vect R_i(\vect W)}{\varepsilon_l},\qquad
  \varepsilon_l=\delta_J\,\operatorname{sign}(w_{j,l})\max\bigl(|w_{j,l}|,\,|w_{l,\mathrm{ref}}|\bigr),
  \label{eq:fdeps}
\end{equation}
where $\vect W^{(j,l)}$ is $\vect W$ with $w_{j,l}$ increased by $\varepsilon_l$, $w_{l,\mathrm{ref}}$ a reference value (one for the logarithms, the free-stream speed for the velocity components) and $\delta_J=10^{-8}$. For a logarithmic partial density, $\varepsilon_l$ is a relative perturbation of $\rho_s$, so the Jacobian columns of all species, whether major components or traces many orders of magnitude smaller, are computed with the same relative accuracy, without tuning the perturbation species by species.

The perturbed residuals are those of the element at point $j$ and its face neighbours, with the gradients recomputed from the perturbed state; for ionised mixtures, they also include the dependence on the ion densities through the neutral electron density~\eqref{eq:rhoe_neutral}. With shock capturing, the blending coefficient is computed once per iteration and held fixed in the finite differences (Sec.~\ref{sec:subcell_ob}). Option~2 of the subcell scheme widens the dependence of the residual, since the slope at an endpoint depends on the outer state at the face (Sec.~\ref{sec:sob_subcell_options}). These neighbour samples are kept frozen while the local states are perturbed, and so are the point gradients of the subcell diffusive flux (Appendix~\ref{app:subcell_diffusion}). Only the derivatives through the end-point slopes and through these point gradients are therefore omitted, and all other dependences, including those through the element-face fluxes~\eqref{eq:sob_faceflux}, are exact.

\subsection{Limited update for TCNEQ}
\label{sec:impl_update}

The step $\Delta\vect W^{n}$ can leave the range in which the linearisation~\eqref{eq:taylor} is valid, in particular during the start-up transient. It is therefore scaled by the relaxation factor $\omega^{n}$ of Eq.~\eqref{eq:update}, computed with the physical update limiter of \citet{Ceze2015}, without their residual penalties and residual line search. Each solution point $i$ yields the largest factor $\omega_i\le1$ for which its updated state satisfies a set of constraints, and the smallest is applied to the whole mesh, $\omega^{n}=\min_i\omega_i$. The constraint of \citet{Ceze2015} bounds the relative decrease of density and pressure. In FV codes for nonequilibrium flows, LAURA limits the relative change of the solution in one update, by default to 10\,\% \citep{Mazaheri2011}.

With the logarithmic variables, admissibility itself never limits the update. Let $\Delta w_s$, $\Delta w_T$ and $\Delta w_{\mathrm{ve}}$ be the components of $\Delta\vect W^{n}_i$. The linear model predicts the change $\exp(w_s)\,\Delta w_s$ of a partial density, and a negative density for $\Delta w_s<-1$, whereas the update gives $\exp(w_s+\Delta w_s)>0$ for any $\Delta w_s$. The two agree to $\mathcal O(\Delta w_s^2)$ for small $|\Delta w_s|$. No step has to be cut to remain admissible, and the bounds limit only the change per iteration, in the same way for major and trace species. To handle TCNEQ cases, we add two bounds on the logarithmic variables, of the same kind as those of the Newton iteration of chemical-equilibrium solvers \citep[Sec.~3.3]{Gordon1994}. The factor $\omega_i$ is the largest for which:
\begin{itemize}
  \item $\omega_i\,|\Delta w_s|\le\Delta_{\max}$ for all species, with $\Delta_{\max}=2.3$, so that every partial density changes by at most a factor of about ten. A bound on the mixture density instead would let a trace species change by orders of magnitude.
  \item $\omega_i\,|\Delta w_T|\le\ln(1+c_T)$ and $\omega_i\,|\Delta w_{\mathrm{ve}}|\le\ln(1+c_T)$, with $c_T=0.5$, so that every temperature changes by at most a factor $1+c_T$. The density and pressure constraint below would not catch such steps, since it limits only decreases and the pressure~\eqref{eq:pressure} depends on $\Tve$ only through the electrons.
  \item the mixture density and pressure of the updated state stay finite and decrease by at most a fraction $c_\rho=0.1$ of their current values.
\end{itemize}
The first two bounds are linear in $\omega_i$ and give it explicitly. The third is tested on the state they allow; if it fails, the largest admissible factor is found by bisection between the current state and that state.

If $\omega^{n}<\omega_{\min}$, with $\omega_{\min}=10^{-2}$, the update is rejected. The state is then kept, the CFL number is reduced (Sec.~\ref{sec:impl_cfl}) and the iteration is repeated. The computation stops after ten consecutive rejections.

For ionised mixtures, the neutral evaluation and the state filter of Sec.~\ref{sec:chargeneutrality} must be made consistent with the limited update. The linear system~\eqref{eq:newton_v} still returns an electron component $\Delta w_e$, but it is never applied, since the filter~\eqref{eq:filter} overwrites the electron entry after the update. The change of the electron entry applied in an iteration is therefore the change produced by the filter. Let the state satisfy Eq.~\eqref{eq:rhoe_neutral} before the update, in which $\ln\rho_s$ of the ion $s$ changes by $\omega^{n}\Delta w_s$. With the weights $\beta_s=c_s\rho_s/\rho_e^{\star}\ge0$, which sum to one,
\begin{equation}
  \ln\frac{\rho_e^{\star,\mathrm{new}}}{\rho_e^{\star}}=\ln\sum_{s\in\mathcal I}\beta_s\,\exp(\omega^{n}\Delta w_s)\in\Bigl[\min_{s\in\mathcal I}\omega^{n}\Delta w_s,\ \max_{s\in\mathcal I}\omega^{n}\Delta w_s\Bigr],
  \label{eq:electron_bound}
\end{equation}
since the argument of the logarithm is a convex combination of the $\exp(\omega^{n}\Delta w_s)$. The applied electron change is thus bounded by the ion changes, which the bounds already limit. Two rules follow. First, the electron component of $\Delta\vect W^{n}$ is excluded from the bounds, since the change requested by the linearised electron equation bears no relation to the change the filter produces and, if included, could set the global relaxation factor of every update and stall the convergence. Second, every trial state is filtered before the density and pressure constraint is tested, so that the test sees the state that will actually be accepted, and not an electron entry moved by the unbounded electron component.

\subsection{CFL control}
\label{sec:impl_cfl}

The CFL number is adapted to the outcome of each update with the law of \citet[Eq.~12]{Ceze2015}, to which we add an upper bound and no growth in an iteration with a rejected update:
\begin{equation}
  \mathrm{CFL}^{n+1}=
  \begin{cases}
    \min\bigl(c_g\,\mathrm{CFL}^{n},\ \mathrm{CFL}_{\max}\bigr), & \omega^{n}=1\ \text{and no rejection in iteration } n,\\
    \mathrm{CFL}^{n}, & \omega_{\min}\le\omega^{n}<1,\ \text{or}\ \omega^{n}=1\ \text{after a rejection in iteration } n,\\
    c_b\,\mathrm{CFL}^{n}, & \omega^{n}<\omega_{\min}\ \text{(rejected update)},
  \end{cases}
  \label{eq:cfl}
\end{equation}
with the growth factor $c_g=1.025$, the reduction factor $c_b=0.1$ and the maximum $\mathrm{CFL}_{\max}=1000$. The CFL number therefore grows only while the full step is taken, and it is held as long as the step has to be limited. The automatic freeze of the blending coefficient (Sec.~\ref{sec:sob_freeze}) uses the same CFL history. A computation is considered converged when the norm of the $\ln T$ component of the update over all solution points, $r^{n}=\bigl(\sum_i(\Delta w_{T,i}^{n})^2\bigr)^{1/2}$, falls below $10^{-6}$.

\subsection{Relaxation and freezing of the blending coefficient}
\label{sec:impl_blending}
\label{sec:sob_freeze}

The blending coefficient of Sec.~\ref{sec:subcell_ob} also evolves with the iteration. The coefficient applied at iteration $n$ is relaxed towards the coefficient $\hat\alpha_K$ requested by the indicator (Sec.~\ref{sec:sob_alpha}),
\begin{equation}
  \alpha^{n}_K=\alpha^{n-1}_K+\gamma_r\bigl(\hat\alpha_K-\alpha^{n-1}_K\bigr),\qquad \gamma_r\in(0,1],
\end{equation}
with the relaxation factor $\gamma_r$. Relaxation damps the variations of the coefficient during the start-up transient, when shocks move through the mesh.

The coefficient $\alpha_K$, the selection $\chi_K$ and the limited slopes depend on the solution through thresholds, in $\hat\alpha_K$ ($\alpha_{\min}$ and $\alpha_{\max}$), in the Mach selection and in the limiter. Near a steady state, a small change of the solution can carry an element or a point across one of these thresholds. The residual then changes discontinuously between iterations, and the iteration stalls. \citet{Venkatakrishnan1995} observed this with limiters and froze the limiter to restore convergence. \citet{LiuYuan2023} met the same difficulty with a posteriori subcell limiting in FR, where cells alternating between the FR and subcell schemes prevented convergence, and resolved it by marking the troubled cells consistently over the iterations.

We therefore freeze the three quantities once, when the iteration stalls. Iteration $n$ is a stall if, over the previous $N_W$ iterations, the CFL number has stayed at the maximum $\mathrm{CFL}_{\max}$ of the CFL law~\eqref{eq:cfl} and the update norm $r^{n}$ of Sec.~\ref{sec:impl_cfl} has decreased by less than a tenth of a decade:
\begin{equation}
  \min_{n-N_W\le m\le n}\mathrm{CFL}^{m}=\mathrm{CFL}_{\max}
  \quad\text{and}\quad
  \log_{10}\frac{r^{n-N_W}}{\min_{n-N_W<m\le n}r^{m}}<0.1,
\end{equation}
with $N_W=50$. The condition on the CFL number avoids freezing a transient solution, in which shocks still move. At the first stall, $n=n_f$, the coefficient is set to the requested one without relaxation, $\alpha^{n_f}_K=\hat\alpha_K$, and kept for the rest of the computation, as is $\chi_K$. Each limited slope keeps its ratio to the average of its two secants, per component and direction: $\mathbf L_i=\boldsymbol\Lambda_i\,\tfrac12(\mathbf D^{-}_i+\mathbf D^{+}_i)$ for $n\ge n_f$, with the ratio $\boldsymbol\Lambda_i$ of iteration $n_f$. After the freeze, none of the thresholds act, and the residual changes continuously with the solution.

\section{Verification}
\label{sec:verification}

To verify the order of accuracy, we use the manufactured solution of \citet[Sec.~4.2.2]{Freno2021} for five-species air ($\mathrm N_2$, $\mathrm O_2$, NO, N, O) in TCNEQ, designed to verify the code for hypersonic reacting flows. To the authors' knowledge, the design order of a high-order method has not been verified before with finite-rate air chemistry and vibrational relaxation (Sec.~\ref{sec:intro}).

On the square $(x,y)\in[0,1]^2$\,m, the fields $\rho_s$, $u$, $v$, $T$ and $\Tve$ are the trigonometric functions of \citet[Eq.~19]{Freno2021}, with the amplitude factor 0.05 and the reference values $\bar\rho_{\mathrm N_2}=7.7\times10^{-3}$, $\bar\rho_{\mathrm O_2}=2.0\times10^{-3}$, $\bar\rho_{\mathrm{NO}}=\bar\rho_{\mathrm N}=\bar\rho_{\mathrm O}=10^{-4}$\,kg\,m$^{-3}$, $\bar u=\bar v=11484.10$\,m\,s$^{-1}$ (reference Mach number 8), $\bar T=5000$\,K and $\bar T_{\mathrm{ve}}=1000$\,K. The flow is supersonic throughout the domain, and $v=0$ on $y=0$ and $y=1$. The manufactured state is imposed at the inflow $x=0$; $x=1$ is a supersonic outflow, and $y=0$ and $y=1$ are slip walls, with the ghost state of the slip wall of Sec.~\ref{sec:fr_bc}. The manufactured state $\vect U^{\mathrm{ms}}$ is the steady solution of:
\begin{equation}
  \frac{\partial\vect U}{\partial t}+\nabla\cdot\vect F^{C}(\vect U)=\vect S(\vect U)+\vect S^{\mathrm{ms}},\qquad
  \vect S^{\mathrm{ms}}=\nabla\cdot\vect F^{C}(\vect U^{\mathrm{ms}})-\vect S(\vect U^{\mathrm{ms}}),
  \label{eq:mms}
\end{equation}
where the manufactured source $\vect S^{\mathrm{ms}}$ is evaluated at the solution points. As in \citet{Freno2021}, $\vect S^{\mathrm{ms}}$ is computed from the flux and source functions of the solver, but the divergence is approximated by eighth-order central differences with a step of $10^{-3}$\,m instead of automatic differentiation; halving or doubling the step changes the error norms by about $10^{-4}$ in relative terms. In this study, we employ the thermochemical model of Sec.~\ref{sec:thermochem}, which differs from that of \citet{Freno2021}, so only the orders are comparable between the two works.

\begin{figure}[pos=htbp]
  \centering
  \includegraphics[width=90mm]{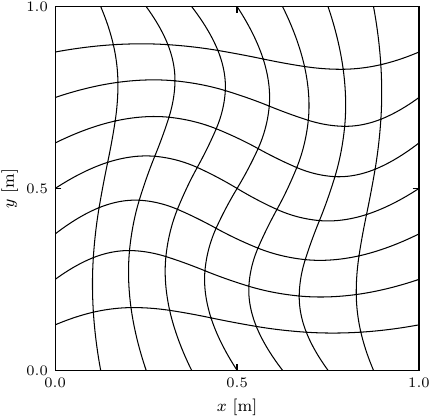}
  \caption{Manufactured solution: curved mesh with $N=8$. Inflow at $x=0$, outflow at $x=1$, slip walls at $y=0$ and $y=1$.}
  \label{fig:mms_mesh}
\end{figure}

The problem is solved on meshes of $N\times N$ curved Q2 elements, obtained by evaluating the nonuniform map of \citet[App.~B]{Freno2021} at the nine nodes of each element of a uniform grid (Fig.~\ref{fig:mms_mesh}). The values of $N$ are listed in Table~\ref{tab:mms}. The discretisation uses the logarithmic update variables of Sec.~\ref{sec:logvars}, the local Lax--Friedrichs flux, GL points and $c_{\mathrm{VCJH}}=0$, without blending. The implicit solver, with the limited update and the CFL controller of Sec.~\ref{sec:implicit}, starts from the manufactured solution with the CFL number at its maximum, $10^{3}$, and stays there throughout each run. After 15 iterations, the relative change in the error norms is less than $3\times10^{-5}$ between successive iterations.

For each field $\psi\in\{\rho_s,u,v,T,\Tve\}$ the error is measured by the norms $\|\psi_h-\psi^{\mathrm{ms}}\|/\bar\psi$, where $\psi_h$ is obtained from the solution polynomial of the update variables and $\psi^{\mathrm{ms}}$ is the manufactured field. The norms $\|\cdot\|_1$ and $\|\cdot\|_2$ are computed with $(P+3)^2$ Gauss--Legendre quadrature points per element, and $\|\cdot\|_\infty$ is the maximum over the same points. The observed order between two meshes is the logarithm of the ratio of their errors divided by the logarithm of the ratio of their $N$ values.

\begin{figure}[pos=htbp]
  \centering
  \includegraphics[width=\textwidth]{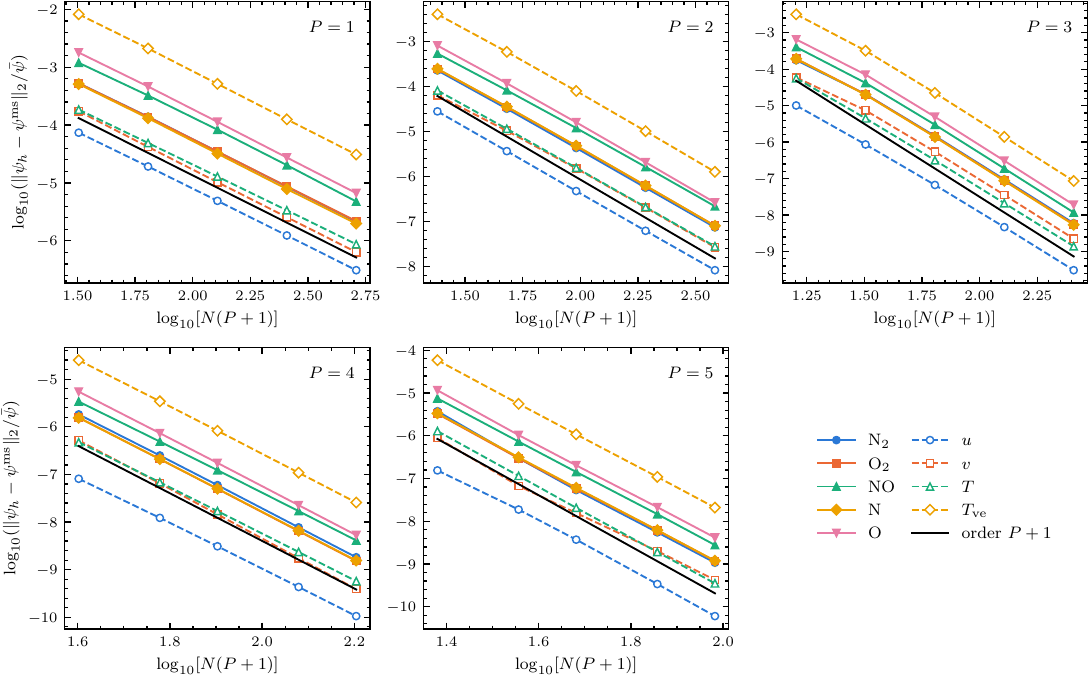}
  \caption{Manufactured solution: $\|\cdot\|_2$ error of the nine fields against $N(P+1)$, the number of solution points per direction, for $P=1$ to $5$. Black lines: order $P+1$.}
  \label{fig:mms}
\end{figure}

The errors of the nine fields decrease at the design rate $P+1$ for $P=1$ to $5$ (Fig.~\ref{fig:mms}): between the two finest meshes, the observed orders in the $\|\cdot\|_1$ and $\|\cdot\|_2$ norms are within 0.2 of $P+1$ up to $P=4$ and between 5.4 and 6.0 at $P=5$, and scatter more in the $\|\cdot\|_\infty$ norm (Table~\ref{tab:mms}). With their second-order FV scheme, \citet{Freno2021} observed order two on this problem. The present results extend this verification to orders up to six.

\begin{table}[pos=htbp]
  \centering
  \caption{Manufactured solution: observed order between the two finest meshes, smallest and largest values over the nine fields.}
  \label{tab:mms}
  \begin{tabular}{clccc}
    \toprule
    $P$ & $N$ & $\|\cdot\|_1$ & $\|\cdot\|_2$ & $\|\cdot\|_\infty$ \\
    \midrule
    1 & 16, 32, 64, 128, 256 & 1.95--2.07 & 1.95--2.08 & 1.95--2.15 \\
    2 & 8, 16, 32, 64, 128   & 2.92--2.99 & 2.90--3.01 & 2.88--3.20 \\
    3 & 4, 8, 16, 32, 64     & 3.93--4.01 & 3.92--4.01 & 3.74--4.16 \\
    4 & 8, 12, 16, 24, 32    & 4.90--5.20 & 4.90--5.19 & 4.46--5.04 \\
    5 & 4, 6, 8, 12, 16      & 5.37--6.02 & 5.40--6.01 & 4.99--6.23 \\
    \bottomrule
  \end{tabular}
\end{table}

\section{Validation}
\label{sec:validation}

To validate the proposed methodology on TCNEQ flows, we consider two blunt bodies: the cylinder tested by \citet{Karl2003} in the High Enthalpy Shock Tunnel G\"ottingen (HEG), computed with five- and eleven-species air, and the RAM-C~II flight vehicle at 61\,km of \citet{Grantham1970}, whose shock layer is ionised, with eleven-species air.

\subsection{HEG cylinder, five-species air}
\label{sec:val_heg}

\begin{figure}[pos=htbp]
  \centering
  \includegraphics[width=\textwidth]{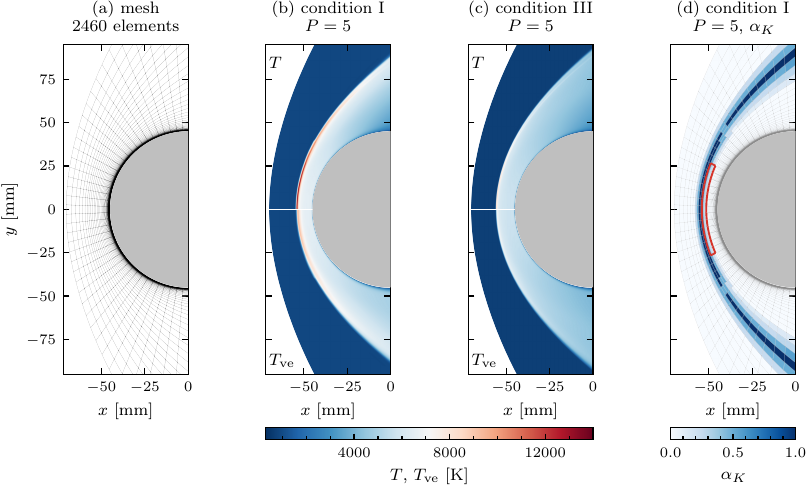}
  \caption{HEG cylinder, five-species air at $P=5$: (a) mesh. (b) Condition I and (c) condition III, translational temperature $T$ (upper half) and vibrational--electronic temperature $\Tve$ (lower half). (d) Blending coefficient distribution (condition I); the elements with second-order subcells are outlined in red.}
  \label{fig:heg_fields}
\end{figure}

The cylinder has a radius of 45\,mm, and the tested conditions, I and III, with the free streams given by \citet{Knight2012}, are shown in Table~\ref{tab:heg_cond}. Available high-order computations of this cylinder used DG at $P=1$, with the convective terms computed from the element means in every element \citep{Absillis2025} or with $P=2$ results shown only along the stagnation line \citep{Luo2026}, DG at $P=2$ with chemical nonequilibrium \citep{Hoskin2024}, or HDG with a single temperature \citep{VanHeyningen2026}; here both conditions are computed with two temperatures and up to $P=5$. Both free streams contain atomic oxygen and nitric oxide from the nozzle expansion. The nozzle airflow is close to thermal equilibrium \citep{Karl2003}, and the free-stream vibrational--electronic temperature is set to $T_\infty$. The air is modelled with five species; N, absent from the free-stream condition III, is given a density of $10^{-9}\rho_\infty$ (Sec.~\ref{sec:logvars_def}). The wall is isothermal at $T_w=300$\,K and non-catalytic, as in the 3D calculations by Lani (using the COOLFluiD's FV solver) reported by \citet{Knight2012}. The mesh has 41 Q2 elements around the half cylinder and 60 in the normal direction, with a wall-normal element size of 0.9\,$\mu$m at the wall (Fig.~\ref{fig:heg_fields}a).

\begin{table}[pos=htbp]
  \centering
  \caption{HEG cylinder: free-stream conditions \citep[Table~2]{Knight2012}. The pressure is computed from $\rho_\infty$, $T_\infty$ and the mass fractions through Dalton's law (484 and 698\,Pa, 1.7 and 1.6\,\% above the tabulated values).}
  \label{tab:heg_cond}
  \begin{tabular}{lcc}
    \toprule
     & condition I & condition III \\
    \midrule
    $V_\infty$ [m\,s$^{-1}$] & 5956 & 4776 \\
    $\rho_\infty$ [kg\,m$^{-3}$] & $1.547\times10^{-3}$ & $3.26\times10^{-3}$ \\
    $T_\infty$ [K] & 901 & 694 \\
    $y_{\mathrm N_2}$, $y_{\mathrm O_2}$, $y_{\mathrm{NO}}$ & 0.7543, 0.00713, 0.01026 & 0.7356, 0.1340, 0.0509 \\
    $y_{\mathrm N}$, $y_{\mathrm O}$ & $6.5\times10^{-7}$, 0.2283 & 0, 0.07955 \\
    \bottomrule
  \end{tabular}
\end{table}

The same figure shows the solution at $P=5$ for conditions I (b) and III (c). The translational temperature peaks behind the shock and relaxes to $\Tve$ before the wall. The shock stand-off, measured at the half rise of $T$ on the stagnation line, is 9.4\,mm (condition I) and 10.8\,mm (condition III), 9\,\% below the measured 10.3 and 11.9\,mm \citep[Fig.~11]{Karl2003}. The stand-off is sensitive to the chemistry model: in the computations of \citet{Karl2003}, changing the reaction-rate model (Park, Dunn and Kang, or Gupta) in both the nozzle and the cylinder flow changes it by up to 10\,\% (condition I) and 23\,\% (condition III). In condition I (d), the blending coefficient is largest in the elements that contain the shock and decreases over a few elements on both sides; it is zero in every element within 5\,mm of the wall, at every order and in both conditions, so the boundary layer is discretised by FR alone. The second-order subcells are used in the subsonic region behind the shock, around the stagnation line.

\begin{figure}[pos=htbp]
  \centering
  \includegraphics[width=\textwidth]{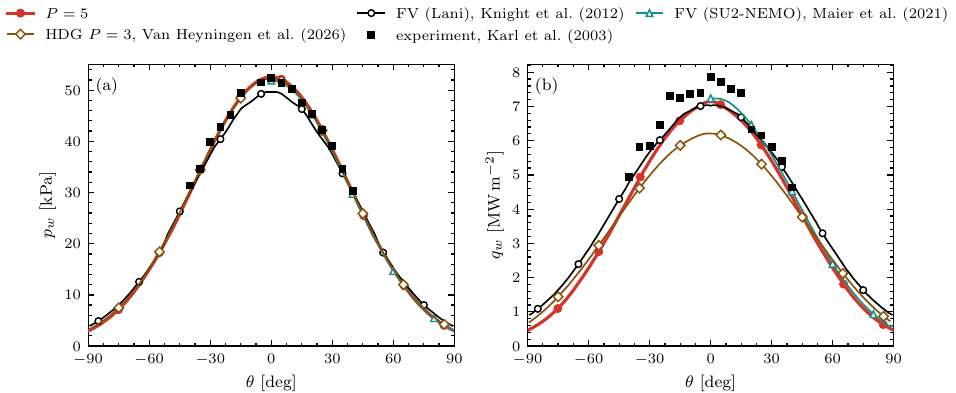}
  \caption{HEG cylinder, condition I: (a) wall pressure and (b) heat flux against the angle from the stagnation point, five-species air at $P=5$. Measurements of \citet[Fig.~10 and Table~4]{Karl2003}; FV computations of Lani \citep[Figs.~12 and~13]{Knight2012} and with SU2-NEMO \citep[Fig.~7]{Maier2021}; HDG computation at $P=3$ \citep[Fig.~6]{VanHeyningen2026}.}
  \label{fig:heg_wall_I}
\end{figure}

\begin{figure}[pos=htbp]
  \centering
  \includegraphics[width=\textwidth]{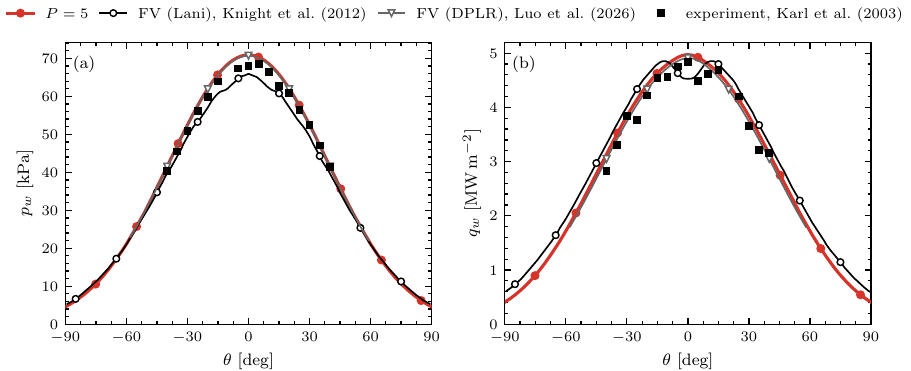}
  \caption{HEG cylinder, condition III: (a) wall pressure and (b) heat flux against the angle from the stagnation point, five-species air at $P=5$. Measurements of \citet[Fig.~10 and Table~4]{Karl2003}; FV computations of Lani \citep[Figs.~15 and~16]{Knight2012} and with DPLR \citep[Fig.~6]{Luo2026}.}
  \label{fig:heg_wall_III}
\end{figure}

\begin{figure}[pos=htbp]
  \centering
  \includegraphics[width=\textwidth]{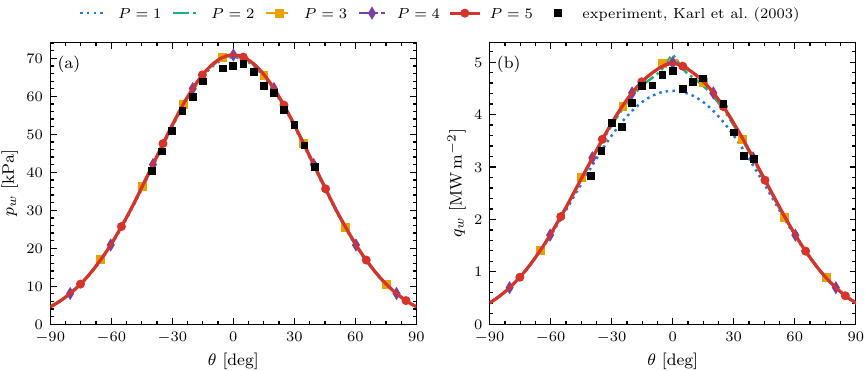}
  \caption{HEG cylinder, condition III: (a) wall pressure and (b) heat flux against the angle from the stagnation point, five-species air from $P=1$ to $P=5$.}
  \label{fig:heg_conv_heat1}
\end{figure}

At the wall, Figures~\ref{fig:heg_wall_I} and~\ref{fig:heg_wall_III} compare the pressure and heat flux with the measurements of \citet[Fig.~10 and Table~4]{Karl2003} and with other computations with a non-catalytic wall at 300\,K, and Table~\ref{tab:heg_stag} lists the stagnation values. Figure~\ref{fig:heg_conv_heat1} shows the evolution of the wall quantities with the solution order for condition III. The stagnation pressure changes by less than 0.3\,\% from $P=2$ to $P=5$, and the heat flux along the wall by less than 0.8\,\% (condition I) and 0.4\,\% (condition III) of its stagnation value from $P=4$ to $P=5$. At $P=2$ the heat flux has a narrow peak at the stagnation point, 1.8 and 2.2\,\% above $P=3$, as on the cylinder of Appendix~\ref{app:gnoffo} at the same order. The pressure agrees with the measurements in condition I and is 4\,\% above the measured stagnation value in condition III. The heat flux is 6\,\% below the measured stagnation value of \citet[Table~4]{Karl2003} in condition I and 3\,\% above it in condition III. \citet{Knight2012} report this underprediction in condition I for all computations except one with a catalytic wall, and attribute it to surface catalysis.

\begin{table}[pos=htbp]
  \centering
  \caption{HEG cylinder: stagnation-point wall pressure $p_0$ [kPa] and heat flux $q_0$ [MW\,m$^{-2}$]. The values of the other computations are read from their wall distributions (Figs.~\ref{fig:heg_wall_I} and~\ref{fig:heg_wall_III}).}
  \label{tab:heg_stag}
  \begin{tabular}{lrrcrr}
    \toprule
     & \multicolumn{2}{c}{condition I} & & \multicolumn{2}{c}{condition III} \\
    \cmidrule(lr){2-3}\cmidrule(lr){5-6}
     & $p_0$ & $q_0$ & & $p_0$ & $q_0$ \\
    \midrule
    $P=5$, five species & 52.60 & 7.15 & & 70.91 & 4.97 \\
    $P=5$, eleven species & 52.59 & 7.16 & & & \\
    \addlinespace
    experiment \citep[Table~4]{Karl2003} & 52.2 & 7.60 & & 68.0 & 4.83 \\
    \addlinespace
    FV, Lani \citep{Knight2012} & 49.6 & 7.02 & & 65.9 & 4.53 \\
    FV, SU2-NEMO \citep{Maier2021} & 51.9 & 7.24 & & & \\
    HDG, $P=3$ \citep{VanHeyningen2026} & 52.3 & 6.21 & & & \\
    FV, DPLR \citep{Luo2026} & & & & 70.7 & 4.91 \\
    \bottomrule
  \end{tabular}
\end{table}

At the stagnation point, the heat flux at $P=5$ is within 2\,\% of the FV computations of Lani \citep{Knight2012} and SU2-NEMO \citep{Maier2021} in condition I, and of DPLR \citep{Luo2026} in condition III. The HDG computation of \citet{VanHeyningen2026} with a single temperature and Fick diffusion gives 13\,\% less. The DG computation of \citet[Sec.~5.2]{Absillis2025} in condition III, with $P=1$ diffusive terms and the convective terms computed from the element means, gives 4 to 6\,\% less, depending on the gradient formulation. The pressures of the other computations are within 1.3\,\% of $P=5$, except that of Lani, which is 6 and 7\,\% lower; Lani's heat flux in condition III has a local minimum at the stagnation point. For this computation, SU2-NEMO uses 68\,000 quadrilateral cells, and the HDG computation uses approximately 19\,200 elements at $P=3$ (about $3.1\times10^{5}$ solution points), compared with 2460 elements here, with 88\,560 solution points at $P=5$. Since the HDG study combines uniform refinement with $r$-adaptive mesh movement, a direct element-to-element comparison is not meaningful; nevertheless, the present computation reaches comparable surface pressure and heat flux with a number of unknowns similar to the second-order SU2-NEMO grid and roughly 3.5 times fewer than the HDG solution.

\begin{figure}[pos=htbp]
  \centering
  \includegraphics[width=\textwidth]{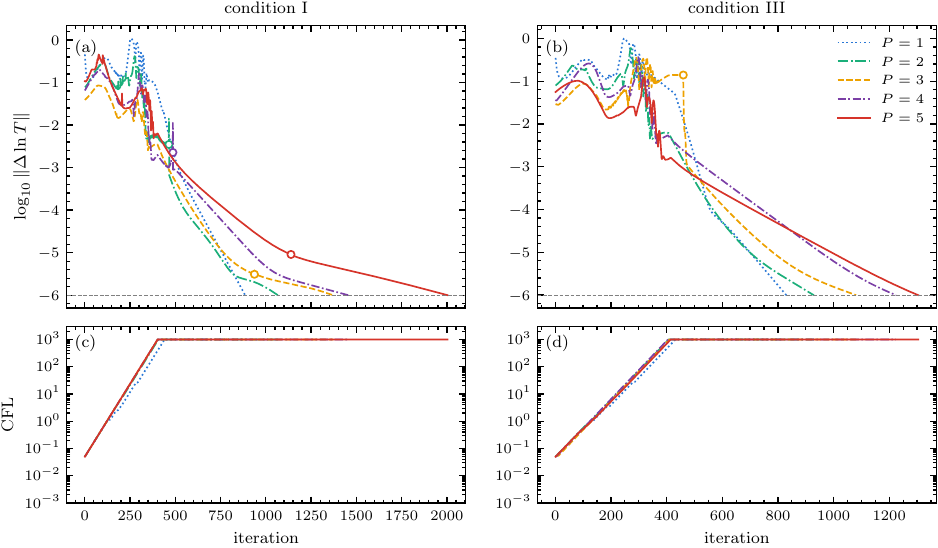}
  \caption{HEG cylinder: (a, b) norm of the $\ln T$ update and (c, d) CFL number against the iteration, for $P=1$ to $5$. The circles mark the automatic freeze of the blending coefficient; the dashed line is the stopping threshold.}
  \label{fig:heg_conv}
\end{figure}

\begin{figure}[pos=htbp]
  \centering
  \includegraphics[width=0.5\textwidth]{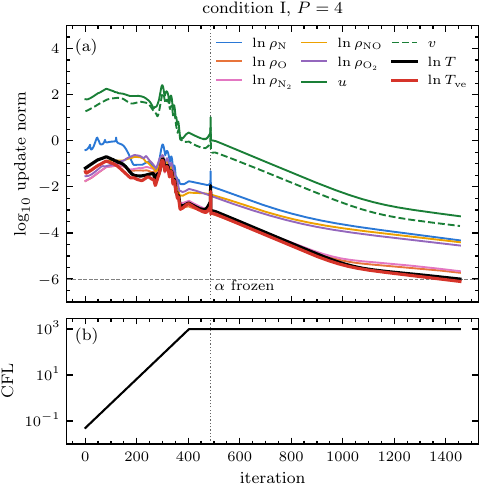}
  \caption{HEG cylinder, condition I, five-species air at $P=4$: (a) norm of the update of each variable and (b) CFL number against the iteration. The dotted line marks the automatic freeze of the blending coefficient; the dashed line is the stopping threshold of the $\ln T$ update.}
  \label{fig:heg_conv_cond1}
\end{figure}

These solutions are obtained order by order: each order starts from the converged solution of the order below, and $P=0$ from the free stream. From $P=1$ to $5$, both conditions use the same settings: the blending of Sec.~\ref{sec:subcell_ob} with $c_0=3$, $\kappa=0.5$, $\alpha_{\min}=0.005$, $\alpha_{\max}=1$ and $\gamma_r=0.25$, first-order subcells at $P=1$ and the selective reconstruction of Sec.~\ref{sec:sob_selective} from $P=2$, together with the implicit scheme of Sec.~\ref{sec:implicit} started from $\mathrm{CFL}^0=0.05$. Figure~\ref{fig:heg_conv} shows the convergence histories for each order, for conditions I (a, c) and III (b, d). Every computation from $P=1$ reaches $\mathrm{CFL}_{\max}=1000$ and converges in 830 to 2003 iterations. Since the CFL number grows by at most a factor $c_g=1.025$ per iteration, it reaches its maximum at iteration 403 at the earliest. No update is rejected, and limited updates mainly hold the growth back at $P=1$, where the maximum is reached at iteration 441 (condition I) and 432 (condition III), compared with 403 to 416 at higher orders. In condition I at $P=2$ and $4$ and in condition III at $P=3$, the update norm $r^n$ of Sec.~\ref{sec:impl_cfl} stalls between about $10^{-3}$ and $10^{-1}$ until the blending coefficient is frozen (Sec.~\ref{sec:sob_freeze}), and then converges; in condition I at $P=3$ and $5$, the freeze occurs late, at about $10^{-5}$, while the update norm still decreases. Figure~\ref{fig:heg_conv_cond1} shows the history of every variable in condition I at $P=4$, where the freeze ends the stall.

\subsection{HEG cylinder, eleven-species air}
\label{sec:val_heg_air11}

\begin{figure}[pos=htbp]
  \centering
  \includegraphics[width=0.48\textwidth]{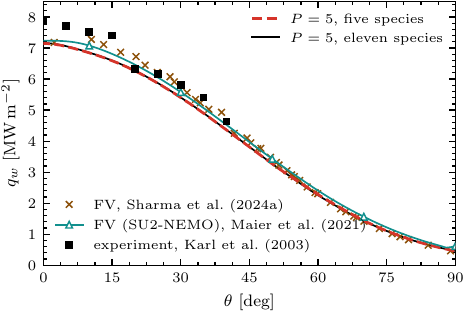}
  \caption{HEG cylinder, condition I: wall heat flux at $P=5$ with five- and eleven-species air, the FV computations of \citet[Fig.~4]{Sharma2024} with eleven species and with SU2-NEMO \citep[Fig.~7b]{Maier2021}, and the measurements of \citet[Fig.~10 and Table~4]{Karl2003}.}
  \label{fig:heg_air11}
\end{figure}

To test the ionised model on the same flow, we repeat the computation of condition I from $P=1$ to $P=5$ with eleven-species air and the same numerical parameters as the five-species computations: only the mixture changes, with the charge-neutrality treatment of Sec.~\ref{sec:chargeneutrality}. The ions enter with an ionisation degree, the total mole fraction of the ions, of $10^{-10}$, as in \citet{Takahashi2014}, and the electron density follows from charge neutrality. The iteration counts stay within 1.5\,\% of the five-species ones at every order. At $P=5$, the stagnation pressure and heat flux differ from the five-species values by less than 0.2\,\% (Table~\ref{tab:heg_stag}), and the wall heat flux by less than 0.4\,\% of its stagnation value (Fig.~\ref{fig:heg_air11}). 
The COOLFluiD FV solver for TCNEQ flows described in \citet{Sharma2024} uses the same thermochemical module, coupled to PLATO, as the present FR solver. Its computation with eleven species gives a heat flux within 1\,\% of $P=5$ near the stagnation point and 6 to 8\,\% above $P=5$ between $10^\circ$ and $30^\circ$, on a mesh of 107\,256 cells, against the 88\,560 solution points used here at $P=5$.
To the authors' knowledge, these are the first high-order finite-element-type computations of a viscous flow of eleven-species, two-temperature air at up to sixth order.

\subsection{RAM-C II, eleven-species air}
\label{sec:val_ramc}

\begin{figure}[pos=htbp]
  \centering
  \includegraphics[width=0.75\textwidth]{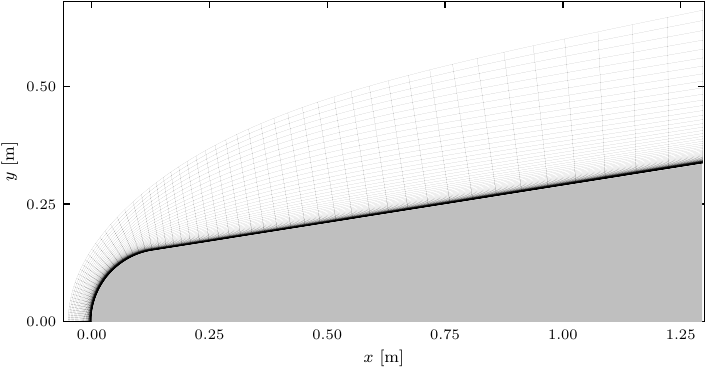}
  \caption{RAM-C~II: mesh. $x$ is the distance from the nose along the axis.}
  \label{fig:ramc_mesh}
\end{figure}

On the HEG cylinder, the ions barely change the wall quantities; in the RAM-C~II flight experiment at 61\,km, the electron density itself is the quantity of interest. Its payload has a hemispherical nose of radius 15.24\,cm, followed by a cone with a $9^\circ$ half-angle \citep{Grantham1970}. As in the computation of \citet{Sharma2024}, with the same eleven-species, two-temperature model and thermochemical library, the free stream has the velocity $V_\infty=7636$\,m\,s$^{-1}$ (Mach 23.9), the density $\rho_\infty=2.720\times10^{-4}$\,kg\,m$^{-3}$, the temperature $T_\infty=254$\,K and the mass fractions 0.79 of $\mathrm N_2$ and 0.21 of $\mathrm O_2$, and the wall is isothermal at $T_w=1500$\,K and non-catalytic. N, O and NO enter with the density $10^{-9}\rho_\infty$ (Sec.~\ref{sec:logvars_def}) and the ions with the ionisation degree of Sec.~\ref{sec:val_heg_air11}. The flow is axisymmetric and is computed with the geometric source~\eqref{eq:axi}. The mesh has 60 Q2 elements along the body, 28 of them on the nose, and 72 in the normal direction; the wall-normal element size at the wall is 1.4\,$\mu$m at the stagnation point and 31\,$\mu$m at the end of the domain, 1.3\,m from the nose (Fig.~\ref{fig:ramc_mesh}).

\begin{figure}[pos=htbp]
  \centering
  \includegraphics[width=\textwidth]{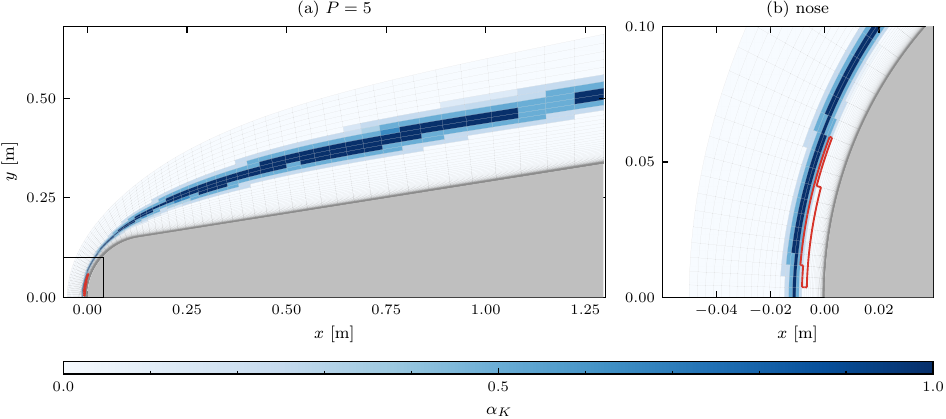}
  \caption{RAM-C~II at $P=5$: blending coefficient distribution, (a) along the body and (b) at the nose (frame in (a)); the elements with second-order subcells are outlined in red.}
  \label{fig:ramc_alpha}
\end{figure}

\begin{figure}[pos=htbp]
  \centering
  \includegraphics[width=\textwidth]{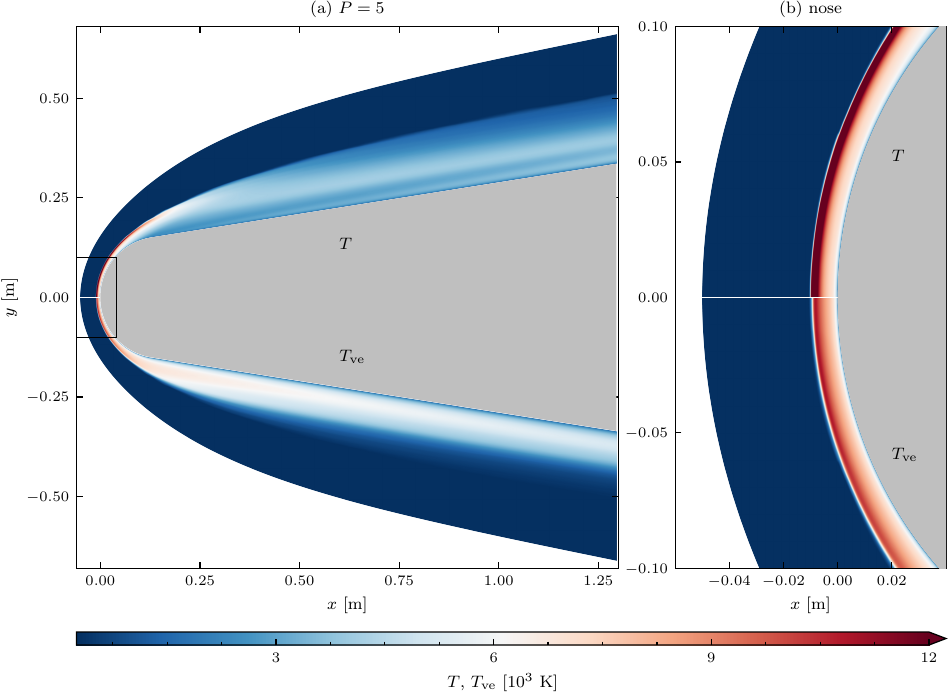}
  \caption{RAM-C~II at $P=5$: translational temperature $T$ (upper half) and vibrational--electronic temperature $\Tve$ (lower half), (a) along the body and (b) at the nose (frame in (a)). The colour scale is limited to 12\,000\,K, so that the shock layer along the cone stays visible; behind the shock at the nose, $T$ exceeds this limit.}
  \label{fig:ramc_temperatures}
\end{figure}

\begin{figure}[pos=htbp]
  \centering
  \includegraphics[width=\textwidth]{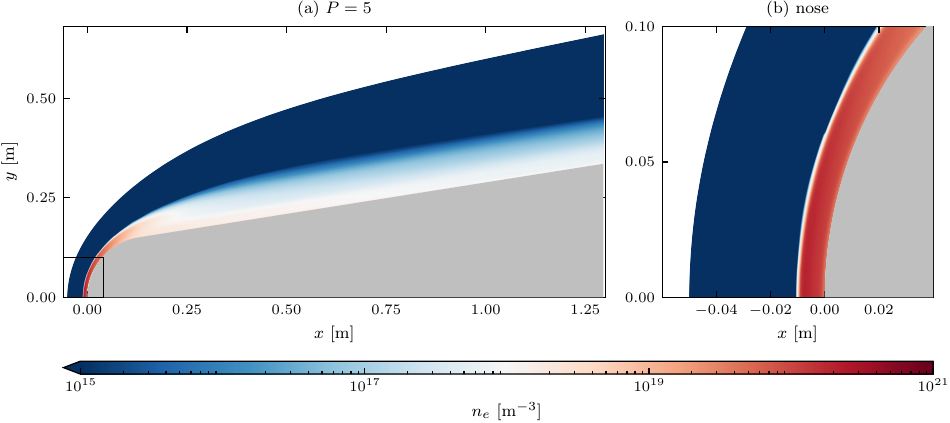}
  \caption{RAM-C~II at $P=5$: electron number density $n_e$, (a) along the body and (b) at the nose (frame in (a)).}
  \label{fig:ramc_ne_field}
\end{figure}

Figure~\ref{fig:ramc_alpha} shows the blending coefficient at $P=5$, which, as on the cylinder, is largest in the elements that contain the shock and zero near the wall. Behind the oblique part of the shock, along the cone, the flow remains supersonic, so the second-order subcells are used only in the subsonic region behind the shock at the nose, around the stagnation line (Fig.~\ref{fig:ramc_alpha}b). At the nose, the shock stands 9.8\,mm ahead of the wall, measured at the half rise of $T$ on the stagnation line (Fig.~\ref{fig:ramc_temperatures}b). Downstream of the nose, the shock layer carries a vibrational--electronic temperature $\Tve$ above $T$, which relaxes towards $T$ along the cone (Fig.~\ref{fig:ramc_temperatures}a). The electron number density peaks behind the shock at the nose, at $2\times10^{20}$\,m$^{-3}$, and decreases along the body to about $1.5\times10^{18}$\,m$^{-3}$ at the end of the domain (Fig.~\ref{fig:ramc_ne_field}).

\begin{figure}[pos=htbp]
  \centering
  \includegraphics[width=\textwidth]{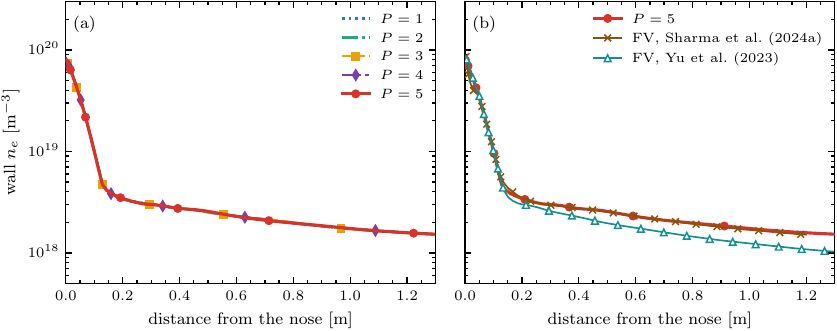}
  \caption{RAM-C~II: electron number density at the wall against the distance from the nose, (a) from $P=1$ to $P=5$ and (b) at $P=5$, with the FV computations of \citet[Fig.~9]{Sharma2024} and \citet[Fig.~12b]{Yu2023}, both with the same non-catalytic wall at 1500\,K.}
  \label{fig:ramc_ne}
\end{figure}

At the wall, Figure~\ref{fig:ramc_ne} compares the electron number density along the body with the FV computations of \citet{Sharma2024} and \citet{Yu2023}, both with a non-catalytic wall at 1500\,K. Beyond 2\,cm from the nose, the orders $P=1$ to $5$ agree within 2.5\,\%; at the stagnation point, the wall value still changes by 3.3\,\% from $P=4$ to $P=5$. The computation of \citet{Sharma2024}, with the same thermochemical library, is within 7\,\% of $P=5$ from 5\,cm to 1.2\,m from the nose, except around the junction of the nose and the cone, where it is up to 12\,\% higher; on the nose, between 1 and 3\,cm, it is up to 23\,\% lower. The computation of \citet{Yu2023}, with a different reaction set and four temperatures, agrees within 11\,\% on the nose and is 10 to 33\,\% lower along the cone beyond 20\,cm from the nose. The electron densities measured in flight on the cone are lower than these non-catalytic computations, a difference that \citet{Yu2023} attribute to wall catalysis, as recalled by \citet{Sharma2024}, so the comparison with the flight data is left to computations with a catalytic wall.

\begin{figure}[pos=htbp]
  \centering
  \includegraphics[width=0.5\textwidth]{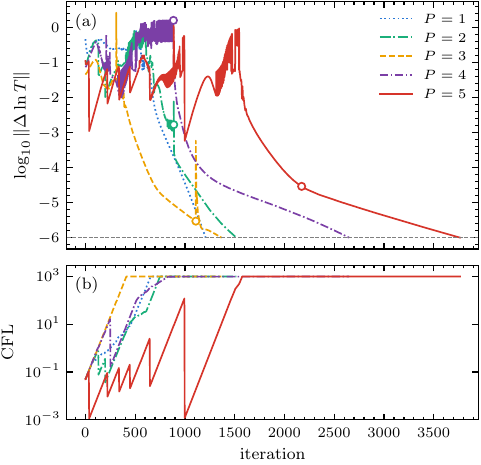}
  \caption{RAM-C~II: (a) norm of the $\ln T$ update and (b) CFL number against the iteration, for $P=1$ to $5$. The circles mark the automatic freeze of the blending coefficient; the dashed line is the stopping threshold.}
  \label{fig:ramc_conv}
\end{figure}

The parameters of the subcell order blending scheme used for this case are the same as those of the HEG cylinder, and each simulation of order $P$ again starts from the converged solution of the order below. The computations converge in 1209 to 3763 iterations (Fig.~\ref{fig:ramc_conv}). The growth of the CFL number is held back by limited updates at $P=1$, $2$, $4$ and $5$, where the maximum is reached at iterations 656, 748, 832 and 1575, and two updates are rejected at $P=2$, two at $P=4$ and twelve at $P=5$. At $P=4$, the update norm stalls at about $1$ until the blending coefficient is frozen at iteration 885, and then converges; at $P=2$ the freeze ends a stall at about $10^{-3}$, and at $P=3$ and $5$ it occurs late, below $10^{-4}$.

\section{Conclusions}
\label{sec:conclusions}

We presented an extended implicit high-order FR method able to tackle steady viscous hypersonic flows of neutral and ionised air in TCNEQ, implemented in COOLFluiD and using the PLATO library to compute all thermochemical/transport properties. The method rests on three main ingredients. (i) The scheme stores and updates the logarithms of the partial densities and temperatures, so that every state is positive without a positivity limiter, and, for ionised mixtures, the electron density is set from charge neutrality consistently with the implicit update. (ii) Shocks are captured by subcell order blending, extended to implicit computations of two-temperature ionised air and second-order subcells in subsonic elements to reduce first-order stagnation heat-flux error whilst the blending coefficient is frozen automatically once the convergence stalls. (iii) The steady state is reached with a pseudo-transient Newton method with a physicality-limited update and automatic CFL control, with the same settings at every order and for every case; all computations converged from the free stream in fewer than 3800 iterations.

The design order $P+1$ was verified from $P=1$ to $5$ on a manufactured solution for five-species air with finite-rate chemistry and vibrational relaxation. On the HEG cylinder, the stagnation heat flux changes by less than 1\,\% of its value from $P=4$ to $P=5$, and at $P=5$ is within 2\,\% of the FV solutions of SU2-NEMO and DPLR, and 6\,\% below and 3\,\% above the measurements in conditions I and III. Eleven-species air changes the stagnation values by less than 0.2\,\%. On RAM-C~II, the wall electron density agrees within 2.5\,\% between $P=1$ and $P=5$ beyond 2\,cm from the nose, and within 7\,\% of a finite-volume computation with the same thermochemical library over most of the body.

The HEG solution at $P=5$ uses 88\,560 solution points on 2460 elements, which is comparable in degrees of freedom to the second-order FV computations of SU2-NEMO \citep{Maier2021}, with 68\,000 cells for five-species air, and of \citet{Sharma2024b}, with 107\,256 cells for eleven-species air, while the mesh is between 28 and 44 times coarser in number of elements. It also uses about 3.5 times fewer solution points than the $P=3$ HDG computation of \citet{VanHeyningen2026}, although the latter relies on its own refinement and $r$-adaptation strategy, so that mesh sizes are not directly comparable. The resolution is therefore carried by the polynomial order rather than the number of elements, greatly simplifying mesh generation and alignment requirements, which are traditionally problematic for hypersonic flow simulations with state-of-the-art methods.

\section*{Acknowledgements}

The resources and services used in this work were provided by the VSC (Flemish Supercomputer Center), funded by
the Research Foundation - Flanders (FWO) and the Flemish Government. These results were also obtained in the framework of the projects
C16/24/010 (C1 project Internal Funds KU Leuven), G0B5823N and G002523N (WEAVE) (FWO-Vlaanderen), and 4000145223 (SIDC Data Exploitation (SIDEX2), ESA Prodex).  
FBA acknowledges support from the KAUST Global Fellowship Program (Award No.\ RFS-KGFP2023-6044).
SP is funded by the European Union (ERC-AdG agreement No 101141362, Open SESAME). Views and opinions expressed are, however, those of the author(s) only and do not necessarily reflect those of the European Union or the European Research Council. Neither the European Union nor the granting authority can be held responsible for them. 
The authors gratefully acknowledge the developers of the PLATO library, Dr A.\ Munafò and Prof.\ M.\ Panesi, for providing access to the code and allowing its interface with COOLFluiD.

\clearpage
\appendix

\section{Reaction mechanism of five-species air}
\label{app:air5}

Table~\ref{tab:air5} lists the 17 reactions of the five-species mechanism, with the forward rate coefficients of \citet{Park1993} as implemented in PLATO \citep{Munafo2025}. Each dissociation reaction is written once for every collision partner $\mathrm M$, and the partners with the same coefficients share a row. The forward coefficients follow the modified Arrhenius law~\eqref{eq:rxn_coeff} at the rate-controlling temperature $T_r$, and the backward coefficients follow from detailed balance at the temperature $T_b$ of the reverse process.

\begin{table}[pos=htbp]
  \centering
  \caption{Five-species air mechanism: coefficients of the forward rate law~\eqref{eq:rxn_coeff} \citep{Park1993}, with the rate-controlling temperatures $T_r$ of the forward and $T_b$ of the backward reaction.}
  \label{tab:air5}
  \begin{tabular}{rllrrrcc}
    \toprule
    $r$ & Reaction & $\mathrm M$ & $C_r$ (cm$^3$\,mol$^{-1}$\,s$^{-1}$) & $\eta_r$ & $E_{a,r}/k_B$ (K) & $T_r$ & $T_b$ \\
    \midrule
    1--2   & $\mathrm{N_2+M\rightleftharpoons N+N+M}$ & N, O             & $3.0\times10^{22}$ & $-1.6$ & 113\,200 & $\sqrt{T\Tve}$ & $T$ \\
    3--5   & $\mathrm{N_2+M\rightleftharpoons N+N+M}$ & N$_2$, O$_2$, NO & $7.0\times10^{21}$ & $-1.6$ & 113\,200 & $\sqrt{T\Tve}$ & $T$ \\
    6--7   & $\mathrm{O_2+M\rightleftharpoons O+O+M}$ & N, O             & $1.0\times10^{22}$ & $-1.5$ & 59\,500  & $\sqrt{T\Tve}$ & $T$ \\
    8--10  & $\mathrm{O_2+M\rightleftharpoons O+O+M}$ & N$_2$, O$_2$, NO & $2.0\times10^{21}$ & $-1.5$ & 59\,500  & $\sqrt{T\Tve}$ & $T$ \\
    11--13 & $\mathrm{NO+M\rightleftharpoons N+O+M}$ & N, O, NO         & $1.1\times10^{17}$ & $0$    & 75\,500  & $\sqrt{T\Tve}$ & $T$ \\
    14--15 & $\mathrm{NO+M\rightleftharpoons N+O+M}$ & N$_2$, O$_2$     & $5.0\times10^{15}$ & $0$    & 75\,500  & $\sqrt{T\Tve}$ & $T$ \\
    16     & $\mathrm{NO+O\rightleftharpoons O_2+N}$  & --               & $8.4\times10^{12}$ & $0$    & 19\,450  & $T$ & $T$ \\
    17     & $\mathrm{N_2+O\rightleftharpoons NO+N}$  & --               & $6.4\times10^{17}$ & $-1.0$ & 38\,400  & $T$ & $T$ \\
    \bottomrule
  \end{tabular}
\end{table}

\section{Subcell face normals on curved elements}
\label{app:subcell_metrics}

A basic requirement of any discretisation is free-stream preservation, that is, a zero residual when the state is the same at every solution point. FR preserves a uniform state on the curved Q2 meshes used in this work. This appendix shows that the subcell scheme of Sec.~\ref{sec:sob_fluxes} does so as well when its internal face normals are the telescoped normals~\eqref{eq:sob_telescope}, but it does not when the normals are evaluated from the mapping. The defects quoted below are computed from the geometry alone on the HEG mesh of Sec.~\ref{sec:val_heg}, for $P=1$ to $5$.

For a uniform state, the flux $\mathbf F^{C}$ is the same everywhere, and every numerical flux returns it exactly, so the subcell residual~\eqref{eq:sob_RSC} depends only on the face normals:
\begin{equation}
  \mathbf R^{C}_{\mathrm{SC},ij}=-\frac{\mathbf F^{C}\cdot\boldsymbol\Gamma_{ij}}{\varpi_i\varpi_j},\qquad
  \boldsymbol\Gamma_{ij}=\varpi_j(J\mathbf n)_{i+1/2,j}-\varpi_j(J\mathbf n)_{i-1/2,j}+\varpi_i(J\mathbf n)_{i,j+1/2}-\varpi_i(J\mathbf n)_{i,j-1/2}.
  \label{eq:sob_uniformSC}
\end{equation}
Each term of $\boldsymbol\Gamma_{ij}$ is the scaled normal of one face of the subcell $C_{ij}$, oriented outwards, times the reference length of that face, so the subcell scheme preserves a uniform state if and only if $\boldsymbol\Gamma_{ij}=\mathbf 0$ in every subcell. This condition holds for the exact face vectors, the outward normal of each face times its physical length,
\begin{equation}
  \mathbf A_{i+1/2,j}=\int_{\eta_{j-1/2}}^{\eta_{j+1/2}}(J\mathbf n)^{\xi}(\xi_{i+1/2},\eta)\,d\eta,
\end{equation}
and identically along $\eta$, since the four face vectors of the subcell, a closed curved quadrilateral in physical space, add up to zero. We measure the defect of the subcell scheme by $|\boldsymbol\Gamma_{ij}|/|\partial C_{ij}|$, where $|\partial C_{ij}|$, the same sum taken with the norms of the four terms, approximates the perimeter of the subcell.

\begin{figure}[pos=htbp]
  \centering
  \includegraphics[width=0.55\linewidth]{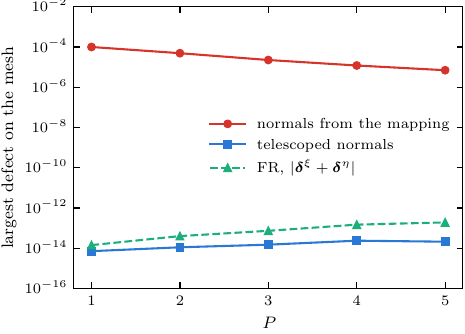}
  \caption{Free-stream preservation on the HEG mesh: largest defect against $P$, for the subcells with face normals from the mapping and with telescoped normals ($|\boldsymbol\Gamma_{ij}|/|\partial C_{ij}|$), and for FR ($|\boldsymbol\delta^{\xi}+\boldsymbol\delta^{\eta}|$, relative to the scaled normals).}
  \label{fig:sob_closure}
\end{figure}

Consider first internal face normals evaluated from the mapping, $(J\mathbf n)_{i+1/2,j}=(J\mathbf n)^{\xi}(\xi_{i+1/2},\eta_j)$ and $(J\mathbf n)_{i,j+1/2}=(J\mathbf n)^{\eta}(\xi_i,\eta_{j+1/2})$. Inside a Q2 element, $(J\mathbf n)^{\xi}$ varies linearly along any curve of constant $\xi$. Its integral along a subcell face is therefore $\varpi_j$ times its value at the middle of the face, $\bar\eta_j=\tfrac12(\eta_{j-1/2}+\eta_{j+1/2})$, and the evaluated normal gives
\begin{equation}
  \varpi_j\,(J\mathbf n)^{\xi}(\xi_{i+1/2},\eta_j)=\mathbf A_{i+1/2,j}+\varpi_j\,(\eta_j-\bar\eta_j)\,\frac{\partial(J\mathbf n)^{\xi}}{\partial\eta}(\xi_{i+1/2}),
\end{equation}
and identically along $\eta$. On elements with straight sides, whose mapping is bilinear, this derivative, $(\partial^2y/\partial\eta^2,\,-\partial^2x/\partial\eta^2)$, vanishes, so the evaluated normals are exact, and every subcell is closed. On curved elements, the normal turns along the subcell face. The solution point is not at the middle of the subcell face, $\eta_j\neq\bar\eta_j$, except for the central row when $P$ is even (Fig.~\ref{fig:sob_grid}), so the errors of the four faces do not cancel. On the HEG mesh the largest defect decreases from $10^{-4}$ at $P=1$ to $7\times10^{-6}$ at $P=5$, but it does not vanish (Fig.~\ref{fig:sob_closure}).

FR, in contrast, preserves a uniform state through its own derivative of the scaled normal. For a uniform state, the transformed flux is $\mathbf F^{C}$ times the scaled normal, and FR differentiates it like any flux. Along the row $j$, it builds the polynomial
\begin{equation}
  \mathbf p_j(\xi)=\sum_{m=0}^{P}l_m(\xi)\,(J\mathbf n)^{\xi}_{mj}
  +h_L(\xi)\Bigl((J\mathbf n)_{-1/2,j}-\sum_{m=0}^{P}l_m(-1)\,(J\mathbf n)^{\xi}_{mj}\Bigr)
  +h_R(\xi)\Bigl((J\mathbf n)_{P+1/2,j}-\sum_{m=0}^{P}l_m(1)\,(J\mathbf n)^{\xi}_{mj}\Bigr).
  \label{eq:sob_metricpoly}
\end{equation}
This polynomial is the interpolant of the scaled normal, corrected at both ends to the element-face values, $\mathbf p_j(\mp1)=(J\mathbf n)_{\mp1/2,j}$. Its derivative at the solution points is $\boldsymbol\delta^{\xi}_{ij}=\mathbf p_j'(\xi_i)$, Eq.~\eqref{eq:sob_metricderiv}. With $\boldsymbol\delta^{\eta}_{ij}$ built identically along the column $i$, the FR residual of a uniform state is
\begin{equation}
  \mathbf R^{C}_{\mathrm{FR},ij}=-\mathbf F^{C}\cdot\bigl(\boldsymbol\delta^{\xi}_{ij}+\boldsymbol\delta^{\eta}_{ij}\bigr).
  \label{eq:sob_uniformFR}
\end{equation}
FR therefore preserves a uniform state when $\boldsymbol\delta^{\xi}_{ij}+\boldsymbol\delta^{\eta}_{ij}=\mathbf 0$ at every solution point, the condition \citet{Abe2015} gives for free-stream preservation of FR. For the correction functions used in this work, $c_{\mathrm{VCJH}}=0$, the FR derivative is exact for polynomials of degree up to $P+1$, and the scaled normals of a Q2 element are of degree two along the direction in which they are differentiated, so the condition holds at every order $P\ge1$. On the HEG mesh $|\boldsymbol\delta^{\xi}_{ij}+\boldsymbol\delta^{\eta}_{ij}|$ stays at round-off, below $2\times10^{-13}$ relative to the scaled normals (Fig.~\ref{fig:sob_closure}).

\begin{figure}[pos=htbp]
  \centering
\begin{tikzpicture}[>=stealth,font=\small]
\draw[thick] (0,0) -- (0,1.0); \draw[thick] (8,0) -- (8,1.0);
\draw[dashed,gray] (2.222,0) -- (2.222,1.0); \draw[dashed,gray] (5.778,0) -- (5.778,1.0);
\draw (0,0.4) -- (8,0.4);
\fill (0.901,0.4) circle (2.2pt); \fill (4.0,0.4) circle (2.2pt); \fill (7.099,0.4) circle (2.2pt);
\node[above] at (0.901,0.45) {$\xi_0$}; \node[above] at (4.0,0.45) {$\xi_1$}; \node[above] at (7.099,0.45) {$\xi_2$};
\draw[->,blue!70!black,thick] (0.08,1.0) to[bend left=30] node[above,font=\footnotesize]{$+\,\varpi_0\boldsymbol\delta^{\xi}_{0j}$} (2.15,1.0);
\draw[->,blue!70!black,thick] (2.30,1.0) to[bend left=20] node[above,font=\footnotesize]{$+\,\varpi_1\boldsymbol\delta^{\xi}_{1j}$} (5.70,1.0);
\draw[->,blue!70!black,thick,dashed] (5.85,1.0) to[bend left=30] node[above,font=\footnotesize]{$+\,\varpi_2\boldsymbol\delta^{\xi}_{2j}$} (7.92,1.0);
\node[below] at (0,-0.05) {$-1$}; \node[below] at (2.222,-0.05) {$\xi_{1/2}$}; \node[below] at (5.778,-0.05) {$\xi_{3/2}$}; \node[below] at (8,-0.05) {$+1$};
\node[below,align=center,font=\footnotesize] at (0,-0.55) {$(J\mathbf n)_{-1/2,j}$\\$=-J_{f,q}\mathbf n_{f,q}$};
\node[below,font=\footnotesize] at (2.222,-0.55) {$(J\mathbf n)_{1/2,j}$};
\node[below,font=\footnotesize] at (5.778,-0.55) {$(J\mathbf n)_{3/2,j}$};
\node[below,align=center,font=\footnotesize] at (8,-0.55) {$(J\mathbf n)_{5/2,j}$\\$=J_{f,q}\mathbf n_{f,q}$};
\end{tikzpicture}
  \caption{Telescoped face normals along one row $j$ of a $P=2$ element. The recursion~\eqref{eq:sob_telescope} starts from the scaled normal of the element face at $\xi=-1$ and, across each subcell, adds its width times the FR derivative of the metric at its solution point. The internal normals $(J\mathbf n)_{1/2,j}$ and $(J\mathbf n)_{3/2,j}$ are used by the subcell fluxes; the last step (dashed) is not used but lands exactly on the element face at $\xi=+1$.}
  \label{fig:sob_telescope}
\end{figure}
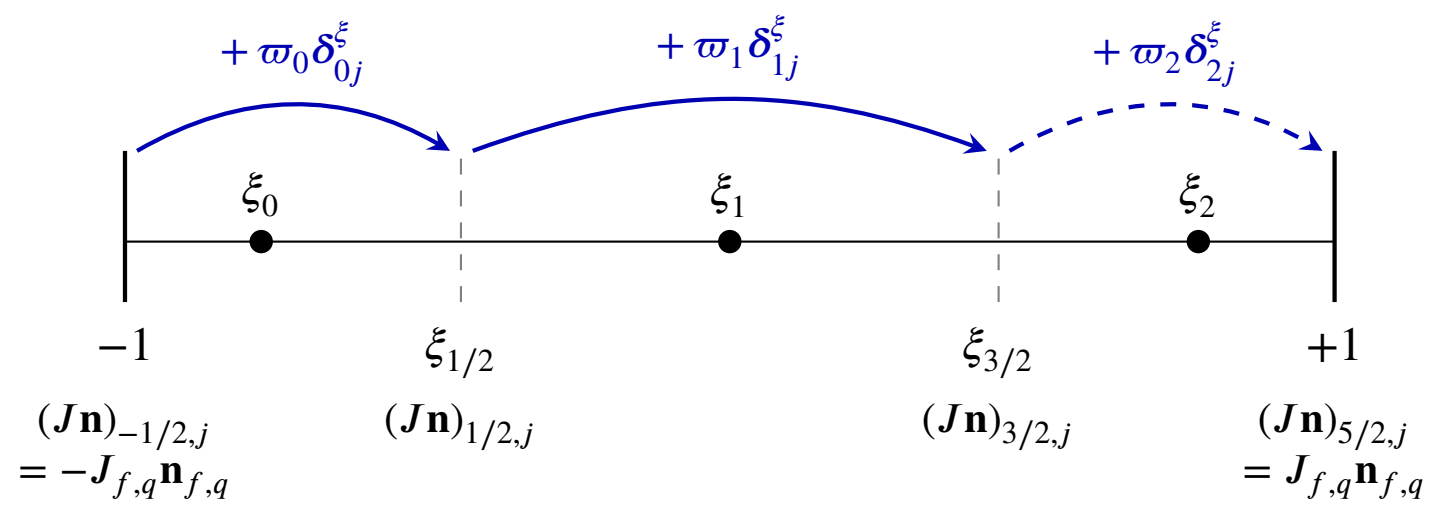

The telescoped normals give every subcell the change of normal that FR sees. The recursion~\eqref{eq:sob_telescope} (Fig.~\ref{fig:sob_telescope}) sets
\begin{equation}
  (J\mathbf n)_{i+1/2,j}-(J\mathbf n)_{i-1/2,j}=\varpi_i\,\boldsymbol\delta^{\xi}_{ij}
  \label{eq:sob_step}
\end{equation}
for the subcells $i=0,\dots,P-1$ of the row. The last subcell, $i=P$, is closed by the element-face normal $(J\mathbf n)_{P+1/2,j}$, and~\eqref{eq:sob_step} holds for it as well if the recursion lands on that normal, which it does because the GL quadrature with $P+1$ points integrates the derivative $\mathbf p_j'$, of degree $P$, exactly,
\begin{equation}
  \sum_{i=0}^{P}\varpi_i\,\boldsymbol\delta^{\xi}_{ij}=\int_{-1}^{1}\mathbf p_j'(\xi)\,d\xi=\mathbf p_j(1)-\mathbf p_j(-1)=(J\mathbf n)_{P+1/2,j}-(J\mathbf n)_{-1/2,j}.
\end{equation}
With~\eqref{eq:sob_step} along the rows and its counterpart along the columns, $\boldsymbol\Gamma_{ij}=\varpi_i\varpi_j\bigl(\boldsymbol\delta^{\xi}_{ij}+\boldsymbol\delta^{\eta}_{ij}\bigr)$, and~\eqref{eq:sob_uniformSC} becomes
\begin{equation}
  \mathbf R^{C}_{\mathrm{SC},ij}=\mathbf R^{C}_{\mathrm{FR},ij}.
\end{equation}
The subcells therefore preserve a uniform state exactly when FR does. On the HEG mesh, the last step lands on the element-face normal to $10^{-15}$, and the defect of the telescoped normals stays below $3\times10^{-14}$ for $P=1$ to $5$ (Fig.~\ref{fig:sob_closure}). The telescoped normals differ from the normals evaluated from the mapping by less than $6\times10^{-4}$ (relative).
The solver confirms this: for a uniform state on the HEG mesh, with $\alpha_K=1$ in every element, the norm of the density residual stays below $10^{-15}$ from $P=1$ to $5$, the round-off level of FR alone.

\section{Impact of first-order subcells on the wall heat flux}
\label{app:pointwise}

In an element with $\alpha_K>0$, the low-order residual $\mathbf R^{C}_{\mathrm{SC}}$ enters the blended scheme~\eqref{eq:sob_blend} with the weight $\alpha_K$, also where the flow is smooth, for instance behind a shock. This appendix measures the error of Option~1 for the simplest smooth field, a linear one, and its effect on the stagnation heat flux of a blunt body.

\subsection{Pointwise error}

The analysis is done in one dimension and extends directly to quadrilaterals, since the subcell scheme on a quadrilateral is the tensor product of the one-dimensional one. Consider a uniform mesh of one-dimensional elements, each with the reference coordinate $\xi\in[-1,1]$. The solution points $\xi_0<\dots<\xi_P$ of an element are the GL points, and each point $\xi_i$ owns a subcell $[\xi_{i-1/2},\xi_{i+1/2}]$ whose width is the GL weight $\varpi_i$ (Sec.~\ref{sec:subcell_ob}). Let $W$ be a scalar state and $F(W)$ its flux. The subcell residual~\eqref{eq:sob_RSC} at point $i$ is the difference of the fluxes through the two faces of its subcell, $R^{C}_{\mathrm{SC},i}=-\bigl(F^{*}_{i+1/2}-F^{*}_{i-1/2}\bigr)/\varpi_i$. With Option~1, the two states at a subcell face are the values at the two solution points around it, $W_i$ and $W_{i+1}$ (Eq.~\eqref{eq:sob_fo}). At the faces of the element, one of them is the closest solution point of the adjacent element, at $\xi_{-1}=-2-\xi_0$ or $\xi_{P+1}=2-\xi_P$ in the element's reference coordinate. The face flux is the local Lax--Friedrichs flux of Eq.~\eqref{eq:commonflux},
\begin{equation}
  F^{*}_{i+1/2}=\tfrac12\bigl(F(W_i)+F(W_{i+1})\bigr)-\tfrac12\lambda_{\max}\,(W_{i+1}-W_i),
\end{equation}
where $\lambda_{\max}\ge|\mathrm dF/\mathrm dW|$ is the largest wave speed.

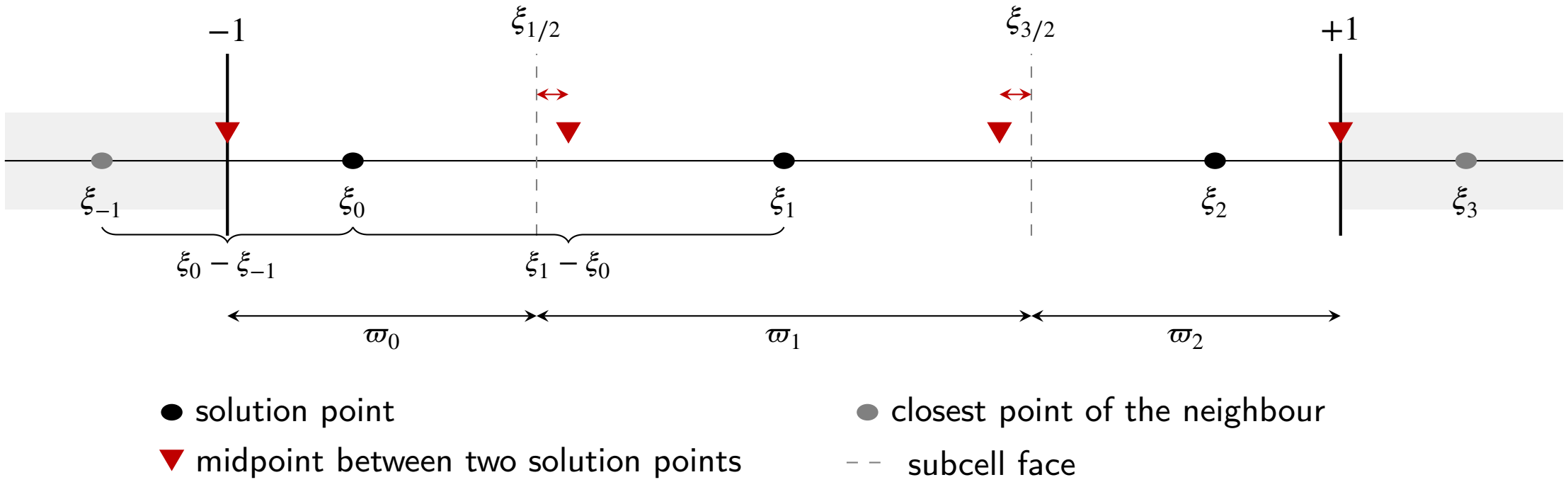
\begin{figure}[pos=htbp]
  \centering
\begin{tikzpicture}[>=stealth,font=\small,xscale=1.3]
\fill[gray!12] (-5.6,-0.45) rectangle (-4,0.45); \fill[gray!12] (4,-0.45) rectangle (5.6,0.45);
\draw (-5.6,0) -- (5.6,0);
\draw[thick] (-4,-0.7) -- (-4,1.0); \draw[thick] (4,-0.7) -- (4,1.0);
\draw[dashed,gray] (-1.778,-0.7) -- (-1.778,1.0); \draw[dashed,gray] (1.778,-0.7) -- (1.778,1.0);
\node[above] at (-4,1.0) {$-1$}; \node[above] at (4,1.0) {$+1$};
\node[above] at (-1.778,1.0) {$\xi_{1/2}$}; \node[above] at (1.778,1.0) {$\xi_{3/2}$};
\fill[gray] (-4.902,0) circle (2.2pt); \fill[gray] (4.902,0) circle (2.2pt);
\fill (-3.098,0) circle (2.2pt); \fill (0,0) circle (2.2pt); \fill (3.098,0) circle (2.2pt);
\node[below] at (-4.902,-0.12) {$\xi_{-1}$}; \node[below] at (-3.098,-0.12) {$\xi_{0}$};
\node[below] at (0,-0.12) {$\xi_{1}$}; \node[below] at (3.098,-0.12) {$\xi_{2}$}; \node[below] at (4.902,-0.12) {$\xi_{3}$};
\foreach \m in {-4,-1.549,1.549,4} \fill[red!75!black] (\m,0.16) -- ++(-0.09,0.2) -- ++(0.18,0) -- cycle;
\draw[<->,red!75!black] (-1.778,0.62) -- (-1.549,0.62); \draw[<->,red!75!black] (1.549,0.62) -- (1.778,0.62);
\draw[decorate,decoration={brace,amplitude=4pt,mirror}] (-4.902,-0.62) -- node[below=3pt,font=\footnotesize]{$\xi_0-\xi_{-1}$} (-3.098,-0.62);
\draw[decorate,decoration={brace,amplitude=4pt,mirror}] (-3.098,-0.62) -- node[below=3pt,font=\footnotesize]{$\xi_1-\xi_0$} (0,-0.62);
\draw[<->] (-4,-1.45) -- node[below,font=\footnotesize]{$\varpi_0$} (-1.778,-1.45);
\draw[<->] (-1.778,-1.45) -- node[below,font=\footnotesize]{$\varpi_1$} (1.778,-1.45);
\draw[<->] (1.778,-1.45) -- node[below,font=\footnotesize]{$\varpi_2$} (4,-1.45);
\fill (-4.4,-2.35) circle (2.2pt) node[right=3pt]{solution point};
\fill[gray] (0.6,-2.35) circle (2.2pt) node[right=3pt,black]{closest point of the neighbour};
\fill[red!75!black] (-4.4,-2.9) -- ++(-0.09,0.2) -- ++(0.18,0) -- cycle; \node[right=3pt] at (-4.4,-2.82) {midpoint between two solution points};
\draw[dashed,gray] (0.45,-2.82) -- (0.75,-2.82) node[right=2pt,black]{subcell face};
\end{tikzpicture}
  \caption{Option~1 along one row of a $P=2$ element on a uniform mesh, with the adjacent elements shaded. For a linear field, the average part of the face flux is the flux at the midpoint between the two solution points around the face (triangles). Inside the element, the midpoints are not on the subcell faces (red arrows), so $\theta^{\mathrm c}_i\neq1$. The dissipative part is proportional to the distance between the two points, which differs on the two sides of the end points, so $\theta^{\mathrm d}_0\neq0$ and $\theta^{\mathrm d}_2\neq0$.}
  \label{fig:sob_fo_error}
\end{figure}

Take a linear field, $W(\xi)=W(0)+W_\xi\,\xi$, and a flux with constant derivative $F_W=\mathrm dF/\mathrm dW$. The exact residual is $-F_W W_\xi$ at every point, and a consistent scheme should return this value at every solution point. For this field, the first term of the face flux is the exact flux at the midpoint $\tfrac12(\xi_i+\xi_{i+1})$ between the two solution points, and the second term is proportional to their distance $\xi_{i+1}-\xi_i$ (Fig.~\ref{fig:sob_fo_error}). The subcell residual at point $i$ is then
\begin{equation}
  R^{C}_{\mathrm{SC},i}=-\bigl(\theta^{\mathrm c}_i F_W-\theta^{\mathrm d}_i\lambda_{\max}\bigr)W_\xi,\qquad
  \theta^{\mathrm c}_i=\frac{\xi_{i+1}-\xi_{i-1}}{2\varpi_i},\qquad
  \theta^{\mathrm d}_i=\frac{\xi_{i+1}-2\xi_i+\xi_{i-1}}{2\varpi_i}.
  \label{eq:sob_fo_factors}
\end{equation}
It is exact at point $i$ only if $\theta^{\mathrm c}_i=1$ and $\theta^{\mathrm d}_i=0$. The first condition holds when the subcell faces lie halfway between the points, and the second when the point lies halfway between its two neighbours. Both factors depend only on the reference positions of the points, so the error does not decrease when the mesh is refined.

\begin{table}[pos=htbp]
  \centering
  \caption{Factors~\eqref{eq:sob_fo_factors} of Option~1 on GL points, for the points $i=0,\dots,P$ of a row.}
  \label{tab:sob_fo_factors}
  \begin{tabular}{lll}
    \toprule
    $P$ & $\theta^{\mathrm c}_i$ & $\theta^{\mathrm d}_i$ \\
    \midrule
    1 & $1,\ 1$ & $0.155,\ -0.155$ \\
    2 & $1.103,\ 0.871,\ 1.103$ & $0.291,\ 0,\ -0.291$ \\
    3 & $1.148,\ 0.921,\ 0.921,\ 1.148$ & $0.350,\ 0.122,\ -0.122,\ -0.350$ \\
    4 & $1.172,\ 0.947,\ 0.947,\ 0.947,\ 1.172$ & $0.380,\ 0.178,\ 0,\ -0.178,\ -0.380$ \\
    5 & $1.186,\ 0.962,\ \dots,\ 0.962,\ 1.186$ & $0.397,\ 0.210,\ 0.058,\ -0.058,\ -0.210,\ -0.397$ \\
    \bottomrule
  \end{tabular}
\end{table}

On GL points, neither condition holds in general (Table~\ref{tab:sob_fo_factors}): at $P=1$, the subcell face between the two points lies halfway between them, so $\theta^{\mathrm c}_i=1$, but each point is nearer to the closest point of the adjacent element than to the other point of its element, so $\theta^{\mathrm d}_i=\pm0.155$. From $P=2$ to $5$, $\theta^{\mathrm c}_i$ differs from one by 4 to 19\,\%, and $|\theta^{\mathrm d}_i|$ reaches 0.29 to 0.40. The second error is multiplied by $\lambda_{\max}$, which for a flow is $|\mathbf V|+a_f$ for every wave, while the entropy and shear waves travel at $|\mathbf V|$. Relative to their exact residual, the second error is therefore $|\theta^{\mathrm d}_i|(1+1/\mathit{Ma})$, with $\mathit{Ma}=|\mathbf V|/a_f$, which grows as the Mach number decreases, as near a stagnation point.

Both errors vanish in the mean over the element weighted by the subcell widths, so the element mean of the residual is exact, and the errors only redistribute it among the points. On a smooth flow, they leave an error that repeats in every element with $\alpha_K>0$. With Option~2, the reconstructed states on both sides of a subcell face equal the exact value at the face for a linear field (Sec.~\ref{sec:sob_subcell_options}). The flux is then exact, its dissipative part vanishes, and the residual is $-F_W W_\xi$ at every point.

\subsection{Stagnation heat flux on a blunt body}
\label{app:gnoffo}

To see how this error affects a viscous computation, we consider the hypersonic flow of a perfect gas over a circular cylinder studied by \citet{Ching2019}, where the blended elements behind the bow shock reach into the smooth subsonic flow ahead of the stagnation point.

A cylinder of radius 1\,m is placed in air at $\mathit{Ma}_\infty=17.605$ and $T_\infty=200$\,K, with a Reynolds number of $3.77\times10^{5}$ based on the radius and an isothermal wall at $T_w=500$\,K \citep[Table~2]{Ching2019}, which gives $\rho_\infty=1.00335\times10^{-3}$\,kg\,m$^{-3}$ and $V_\infty=4991.04$\,m\,s$^{-1}$. The gas is calorically perfect (ratio of specific heats 1.4, gas constant 287.046\,J\,kg$^{-1}$\,K$^{-1}$) with Sutherland's law and a Prandtl number of 0.72 (0.71 in \citet{Ching2019}). The wall heat flux $q_w$ is reported as the Stanton number
\begin{equation}
  \mathrm{St}=\frac{q_w}{c_p\,\rho_\infty V_\infty\,(T_{t,\infty}-T_w)},
\end{equation}
with $c_p$ the specific heat at constant pressure and $T_{t,\infty}$ the free-stream stagnation temperature \citep[Eq.~21]{Ching2019}. The reference is the DG solution of \citet[Fig.~6]{Ching2019} at $P=3$ on a mesh twice as fine as their finest configuration, with $\mathrm{St}_0=8.15\times10^{-3}$ at the stagnation point.

\begin{figure}[pos=htbp]
  \centering
  \includegraphics[width=0.5\linewidth]{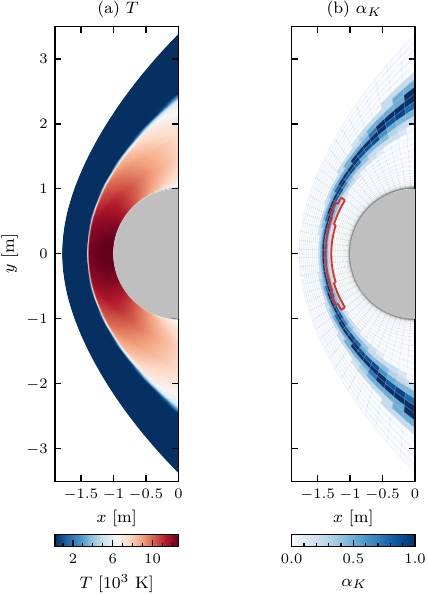}
  \caption{Cylinder on the medium mesh at $P=3$, selective reconstruction; the flow comes from the left. (a) Temperature. (b) Blending coefficient distribution; the elements with second-order subcells are outlined in red.}
  \label{fig:gn_field}
\end{figure}

The computations use the three meshes of Table~\ref{tab:gn_meshes}, with Q2 elements: from one level to the next, the number of elements about doubles in each direction and the wall spacing is halved, and the stagnation line runs through the middle of a row of elements. Figure~\ref{fig:gn_field} shows the medium mesh with the temperature field and the blending coefficient at $P=3$: on the stagnation line, the blended elements extend three elements behind the shock into the subsonic layer, where the selective reconstruction uses second-order subcells.

\begin{table}[pos=htbp]
  \centering
  \caption{Cylinder meshes: number of elements in the circumferential and radial directions, height $\Delta r_w$ of the first element row at the wall and growth ratio of the radial element size at the wall.}
  \label{tab:gn_meshes}
  \begin{tabular}{lrrrr}
    \toprule
    mesh & circumferential & radial & $\Delta r_w$ [m] & growth ratio \\
    \midrule
    coarse & 21 & 30  & $2\times10^{-5}$ & 1.563 \\
    medium & 41 & 60  & $1\times10^{-5}$ & 1.250 \\
    fine   & 81 & 120 & $5\times10^{-6}$ & 1.118 \\
    \bottomrule
  \end{tabular}
\end{table}

From $P=2$, the selective reconstruction of Sec.~\ref{sec:sob_selective}, with $\mathit{Ma}_{\mathrm{rec}}=1$, is compared with Option~1 in every element, each with and without the blending of the diffusive terms in the elements with first-order subcells, Eq.~\eqref{eq:sob_blend_fo}. Each order starts from the converged solution of the order below, $P=0\to1\to2\to3$, with first-order subcells at $P=1$. The CFL law of Sec.~\ref{sec:implicit} goes from $0.5$ to $\mathrm{CFL}_{\max}=1000$, and a computation is converged when the norm of the density update falls below $10^{-8}$. The stagnation value $\mathrm{St}_0$ is taken from the polynomial at the flux points on the wall face containing the stagnation point.

\begin{figure}[pos=htbp]
  \centering
  \includegraphics[width=0.75\linewidth]{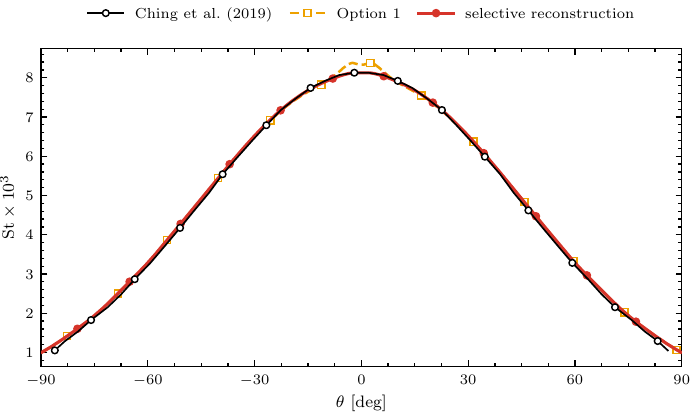}
  \caption{Wall Stanton number on the medium mesh at $P=3$, against the reference of \citet{Ching2019}. In both schemes, the diffusive terms are blended in the elements with first-order subcells (Sec.~\ref{sec:sob_option1}).}
  \label{fig:gn_wall}
\end{figure}

Figure~\ref{fig:gn_wall} shows the wall Stanton number on the medium mesh at $P=3$. Both schemes follow the reference over the whole wall and differ only within about $7^\circ$ of the stagnation point, where Option~1 gives a double hump up to 3.2\,\% above the selective reconstruction.

\begin{figure}[pos=htbp]
  \centering
  \includegraphics[width=\linewidth]{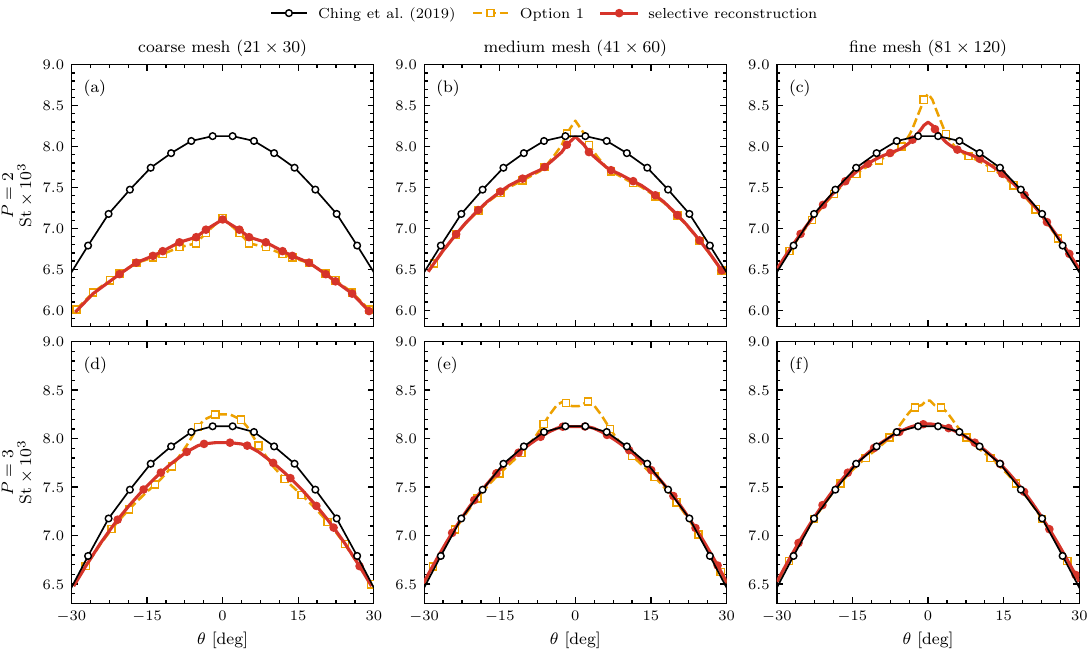}
  \caption{Wall Stanton number near the stagnation point: (a)--(c) $P=2$, (d)--(f) $P=3$, on the coarse, medium and fine meshes. Option~1 and the selective reconstruction, both with the diffusive terms blended in the elements with first-order subcells, against the reference of \citet{Ching2019}.}
  \label{fig:gn_zoom}
\end{figure}

Figure~\ref{fig:gn_zoom} compares the two schemes near the stagnation point on the three meshes, and Table~\ref{tab:gn_results} gives $\mathrm{St}_0$. With Option~1, the heat flux has a peak at the stagnation point, split into a double hump on the medium mesh at $P=3$ and confined to the elements next to the stagnation line, and this error does not decrease under mesh refinement: at $P=2$, $\mathrm{St}_0$ is 2.0\,\% above the reference on the medium mesh and 6.1\,\% above it on the fine mesh, and at $P=3$ it stays 2.5 to 3.8\,\% above the selective reconstruction on the three meshes. The selective reconstruction gives a smooth distribution at $P=3$, with $\mathrm{St}_0$ 0.3\,\% below the reference on the medium mesh and equal to it on the fine mesh. At $P=2$, it keeps a smaller peak, 1.8\,\% above the reference on the fine mesh. On the coarse mesh at $P=2$, both schemes are about 12.5\,\% below the reference.

\begin{figure}[pos=htbp]
  \centering
  \includegraphics[width=0.5\linewidth]{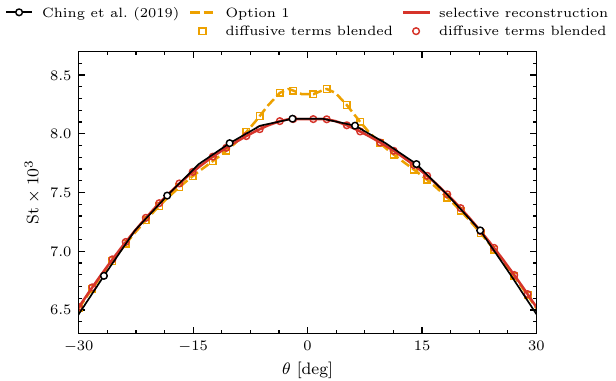}
  \caption{Wall Stanton number on the medium mesh at $P=3$, without (lines) and with (symbols, every second flux point) blending the diffusive terms in the elements with first-order subcells (Sec.~\ref{sec:sob_option1}).}
  \label{fig:gn_diffusion}
\end{figure}

Blending the diffusive terms in elements with first-order subcells changes the wall heat flux by less than 0.1\,\% (Fig.~\ref{fig:gn_diffusion}), so the error in Option~1 comes from its convective fluxes.

\begin{table}[pos=htbp]
  \centering
  \caption{Stagnation-point Stanton number $\mathrm{St}_0\times10^{3}$, with the diffusive terms blended into elements with first-order subcells; the reference value from \citet{Ching2019} is $8.15$.}
  \label{tab:gn_results}
  \begin{tabular}{llrr}
    \toprule
    $P$ & mesh & Option~1 & selective reconstruction \\
    \midrule
    2 & coarse & 7.13 & 7.11 \\
      & medium & 8.32 & 8.12 \\
      & fine   & 8.65 & 8.30 \\
    \addlinespace
    3 & coarse & 8.26 & 7.96 \\
      & medium & 8.33 & 8.12 \\
      & fine   & 8.40 & 8.15 \\
    \bottomrule
  \end{tabular}
\end{table}

\section{Subcell diffusive flux}
\label{app:subcell_diffusion}

In the elements with first-order subcells, the low-order diffusive residual $\mathbf R^{D}_{\mathrm{SC},ij}$ of Eq.~\eqref{eq:sob_blend_fo} is given by Eq.~\eqref{eq:sob_RSC} with the diffusive fluxes through the subcell faces. At the internal face between the solution points $\boldsymbol\xi_{ij}$ and $\boldsymbol\xi_{i+1,j}$,
\begin{equation}
  \hat{\mathbf F}^{D}_{i+1/2,j}=-J_{i+1/2,j}\,\mathbf F^{D}\bigl(\mathbf W_{i+1/2,j},\,\mathbf Q_{i+1/2,j}\bigr)\cdot\mathbf n_{i+1/2,j},
  \qquad
  \mathbf W_{i+1/2,j}=\tfrac12\bigl(\mathbf W_{ij}+\mathbf W_{i+1,j}\bigr),
  \label{eq:sob_diffflux}
\end{equation}
with the face Jacobian and normal of Sec.~\ref{sec:sob_normals}; the minus sign follows from $\mathbf F=\mathbf F^{C}-\mathbf F^{D}$. As in the common flux~\eqref{eq:commonflux}, the flux is evaluated at the average of the two states. The face gradient combines the average of the gradients at the two points with the derivative along the segment that joins them \citep{Haselbacher2000}. For each component $G_k$ of $\mathbf G$, with gradient $\mathbf Q_k$,
\begin{equation}
  \mathbf Q_{k,i+1/2,j}=\bar{\mathbf Q}_k+\bigl(g_k-\bar{\mathbf Q}_k\cdot\mathbf t\bigr)\,\mathbf t,
  \qquad
  \bar{\mathbf Q}_k=\tfrac12\bigl(\mathbf Q_{k,ij}+\mathbf Q_{k,i+1,j}\bigr),
  \label{eq:sob_diffgrad}
\end{equation}
where $\mathbf t=(\mathbf x_{i+1,j}-\mathbf x_{ij})/d$ is the unit vector from the physical position $\mathbf x_{ij}$ of the first point to that of the second, $d=|\mathbf x_{i+1,j}-\mathbf x_{ij}|$ their distance, and $g_k$ the derivative of $G_k$ along $\mathbf t$. For a component that is not a logarithm, $g_k=(G_{k,i+1,j}-G_{k,ij})/d$. For a logarithm, $G_k=\ln\phi$ with $\phi=y_s$, $T$ or $\Tve$, the difference is taken in $\phi$, $g_k=(\phi_{i+1,j}-\phi_{ij})/(\phi_{i+1/2,j}\,d)$, with $\phi_{i+1/2,j}$ computed from the face state $\mathbf W_{i+1/2,j}$. The derivative of $\phi$ along $\mathbf t$ given by Eq.~\eqref{eq:gradrecovery} at the face is then the difference $(\phi_{i+1,j}-\phi_{ij})/d$.

The gradients $\mathbf Q_{ij}$ at the solution points are computed on the subcell grid. Along row $j$, the derivative of $\mathbf G$ at $\boldsymbol\xi_{ij}$ is the difference between its values at the two neighbouring points divided by their distance in $\xi$, and similarly along column $i$. The two derivatives are transformed to physical space with the metric terms~\eqref{eq:sob_Jn} at the point, $\nabla G=\bigl((J\mathbf n)^{\xi}\,\partial_\xi G+(J\mathbf n)^{\eta}\,\partial_\eta G\bigr)/J_{K,ij}$. At the ends of a row or a column, the missing neighbour is replaced by a value on the element face: the average of the values at the solution points of the two elements closest to the flux point. At a physical boundary, the difference is one-sided.

At the element faces, the diffusive part of the common flux~\eqref{eq:commonflux} is blended like the convective part in Eq.~\eqref{eq:sob_faceflux}, with the flux~\eqref{eq:sob_diffflux} between the solution points of the two elements closest to the flux point, evaluated with $J_{f,q}$ and $\mathbf n_{f,q}$. In the weight $\alpha_f$, the coefficient for an element with second-order subcells is set to zero. If one of the two elements does not blend its diffusive terms, the gradient at its point is the FR gradient $\mathbf Q^{\delta D}_{ij}$ of Sec.~\ref{sec:fr_diff}. As for the convective terms, the blended value is used by both elements in the FR correction and, where the diffusive terms are blended, as the outer subcell flux, so the scheme remains conservative. At physical boundaries, the boundary flux of Sec.~\ref{sec:fr_bc} is used for both parts.

\section*{Declaration of generative AI and AI-assisted technologies in the writing process}
During the preparation of this work, the authors used Claude (Anthropic) in order to improve the language and readability of the manuscript. After using this tool, the authors reviewed and edited the content as needed and take full responsibility for the content of the publication.

\bibliographystyle{cas-model2-names}
\bibliography{refs}

\begin{thebibliography}{89}
\expandafter\ifx\csname natexlab\endcsname\relax\def\natexlab#1{#1}\fi
\providecommand{\url}[1]{\texttt{#1}}
\providecommand{\href}[2]{#2}
\providecommand{\path}[1]{#1}
\providecommand{\DOIprefix}{doi:}
\providecommand{\ArXivprefix}{arXiv:}
\providecommand{\URLprefix}{URL: }
\providecommand{\Pubmedprefix}{pmid:}
\providecommand{\doi}[1]{\href{http://dx.doi.org/#1}{\path{#1}}}
\providecommand{\Pubmed}[1]{\href{pmid:#1}{\path{#1}}}
\providecommand{\bibinfo}[2]{#2}
\ifx\xfnm\relax \def\xfnm[#1]{\unskip,\space#1}\fi
\bibitem[{Abe et~al.(2015)Abe, Haga, Nonomura and Fujii}]{Abe2015}
\bibinfo{author}{Abe, Y.}, \bibinfo{author}{Haga, T.},
  \bibinfo{author}{Nonomura, T.}, \bibinfo{author}{Fujii, K.},
  \bibinfo{year}{2015}.
\newblock \bibinfo{title}{On the freestream preservation of high-order
  conservative flux-reconstruction schemes}.
\newblock \bibinfo{journal}{Journal of Computational Physics}
  \bibinfo{volume}{281}, \bibinfo{pages}{28--54}.
\newblock \DOIprefix\doi{10.1016/j.jcp.2014.10.011}.
\bibitem[{Absillis et~al.(2025)Absillis, Luo, Greene, Nourgaliev and
  Goodson}]{Absillis2025}
\bibinfo{author}{Absillis, G.}, \bibinfo{author}{Luo, H.},
  \bibinfo{author}{Greene, P.}, \bibinfo{author}{Nourgaliev, R.},
  \bibinfo{author}{Goodson, M.}, \bibinfo{year}{2025}.
\newblock \bibinfo{title}{Direct discontinuous {G}alerkin methods for the
  reacting multi-component flow equations}.
\newblock \bibinfo{journal}{Computers \& Fluids} \bibinfo{volume}{301},
  \bibinfo{pages}{106774}.
\newblock \DOIprefix\doi{10.1016/j.compfluid.2025.106774}.
\bibitem[{van Albada et~al.(1982)van Albada, van Leer and
  Roberts}]{VanAlbada1982}
\bibinfo{author}{van Albada, G.D.}, \bibinfo{author}{van Leer, B.},
  \bibinfo{author}{Roberts, Jr., W.W.}, \bibinfo{year}{1982}.
\newblock \bibinfo{title}{A comparative study of computational methods in
  cosmic gas dynamics}.
\newblock \bibinfo{journal}{Astronomy and Astrophysics} \bibinfo{volume}{108},
  \bibinfo{pages}{76--84}.
\bibitem[{Alberti et~al.(2024)Alberti, Cammalleri, Carnevali and
  Nigro}]{Alberti2024}
\bibinfo{author}{Alberti, L.}, \bibinfo{author}{Cammalleri, E.},
  \bibinfo{author}{Carnevali, E.}, \bibinfo{author}{Nigro, A.},
  \bibinfo{year}{2024}.
\newblock \bibinfo{title}{On the use of different sets of variables for solving
  unsteady inviscid flows with an implicit discontinuous {G}alerkin method}.
\newblock \bibinfo{journal}{Fluids} \bibinfo{volume}{9}, \bibinfo{pages}{248}.
\newblock \DOIprefix\doi{10.3390/fluids9110248}.
\bibitem[{Babbar and Chandrashekar(2025)}]{Babbar2025}
\bibinfo{author}{Babbar, A.}, \bibinfo{author}{Chandrashekar, P.},
  \bibinfo{year}{2025}.
\newblock \bibinfo{title}{{L}ax--{W}endroff flux reconstruction on adaptive
  curvilinear meshes with error based time stepping for hyperbolic conservation
  laws}.
\newblock \bibinfo{journal}{Journal of Computational Physics}
  \bibinfo{volume}{522}, \bibinfo{pages}{113622}.
\newblock \DOIprefix\doi{10.1016/j.jcp.2024.113622}.
\bibitem[{Babbar and Chandrashekar(2026)}]{Babbar2026}
\bibinfo{author}{Babbar, A.}, \bibinfo{author}{Chandrashekar, P.},
  \bibinfo{year}{2026}.
\newblock \bibinfo{title}{Multiderivative {R}unge--{K}utta flux reconstruction
  for hyperbolic conservation laws}.
\newblock \bibinfo{journal}{Communications on Applied Mathematics and
  Computation} \bibinfo{volume}{8}, \bibinfo{pages}{664--704}.
\newblock \DOIprefix\doi{10.1007/s42967-024-00463-1}.
\bibitem[{Babbar et~al.(2024)Babbar, Kenettinkara and
  Chandrashekar}]{Babbar2024}
\bibinfo{author}{Babbar, A.}, \bibinfo{author}{Kenettinkara, S.K.},
  \bibinfo{author}{Chandrashekar, P.}, \bibinfo{year}{2024}.
\newblock \bibinfo{title}{Admissibility preserving subcell limiter for
  {L}ax--{W}endroff flux reconstruction}.
\newblock \bibinfo{journal}{Journal of Scientific Computing}
  \bibinfo{volume}{99}, \bibinfo{pages}{31}.
\newblock \DOIprefix\doi{10.1007/s10915-024-02482-9}.
\bibitem[{Balay et~al.(2018)Balay, Abhyankar, Adams, Brown, Brune, Buschelman,
  Dalcin, Eijkhout, Gropp, Kaushik, Knepley, May, McInnes, Mills, Munson, Rupp,
  Sanan, Smith, Zampini, Zhang and Zhang}]{Balay2018}
\bibinfo{author}{Balay, S.}, \bibinfo{author}{Abhyankar, S.},
  \bibinfo{author}{Adams, M.F.}, \bibinfo{author}{Brown, J.},
  \bibinfo{author}{Brune, P.}, \bibinfo{author}{Buschelman, K.},
  \bibinfo{author}{Dalcin, L.}, \bibinfo{author}{Eijkhout, V.},
  \bibinfo{author}{Gropp, W.D.}, \bibinfo{author}{Kaushik, D.},
  \bibinfo{author}{Knepley, M.G.}, \bibinfo{author}{May, D.A.},
  \bibinfo{author}{McInnes, L.C.}, \bibinfo{author}{Mills, R.T.},
  \bibinfo{author}{Munson, T.}, \bibinfo{author}{Rupp, K.},
  \bibinfo{author}{Sanan, P.}, \bibinfo{author}{Smith, B.F.},
  \bibinfo{author}{Zampini, S.}, \bibinfo{author}{Zhang, H.},
  \bibinfo{author}{Zhang, H.}, \bibinfo{year}{2018}.
\newblock \bibinfo{title}{{PETS}c Users Manual}.
\newblock \bibinfo{type}{Technical Report} \bibinfo{number}{ANL-95/11 -
  Revision 3.9}. Argonne National Laboratory.
\bibitem[{Bassi et~al.(2015)Bassi, Botti, Colombo, Ghidoni and
  Massa}]{Bassi2015}
\bibinfo{author}{Bassi, F.}, \bibinfo{author}{Botti, L.},
  \bibinfo{author}{Colombo, A.}, \bibinfo{author}{Ghidoni, A.},
  \bibinfo{author}{Massa, F.}, \bibinfo{year}{2015}.
\newblock \bibinfo{title}{Linearly implicit {R}osenbrock-type {R}unge--{K}utta
  schemes applied to the discontinuous {G}alerkin solution of compressible and
  incompressible unsteady flows}.
\newblock \bibinfo{journal}{Computers \& Fluids} \bibinfo{volume}{118},
  \bibinfo{pages}{305--320}.
\newblock \DOIprefix\doi{10.1016/j.compfluid.2015.06.007}.
\bibitem[{Bassi and Rebay(2000)}]{Bassi2000}
\bibinfo{author}{Bassi, F.}, \bibinfo{author}{Rebay, S.}, \bibinfo{year}{2000}.
\newblock \bibinfo{title}{{GMRES} discontinuous {G}alerkin solution of the
  compressible {N}avier--{S}tokes equations}, in: \bibinfo{editor}{Cockburn,
  B.}, \bibinfo{editor}{Karniadakis, G.E.}, \bibinfo{editor}{Shu, C.W.} (Eds.),
  \bibinfo{booktitle}{Discontinuous {G}alerkin Methods}.
  \bibinfo{publisher}{Springer}, \bibinfo{address}{Berlin, Heidelberg}.
  volume~\bibinfo{volume}{11} of \textit{\bibinfo{series}{Lecture Notes in
  Computational Science and Engineering}}, pp. \bibinfo{pages}{197--208}.
\newblock \DOIprefix\doi{10.1007/978-3-642-59721-3_14}.
\bibitem[{Bhatia(2014)}]{Bhatia2014}
\bibinfo{author}{Bhatia, A.}, \bibinfo{year}{2014}.
\newblock \bibinfo{title}{Application of Parallel Time-Implicit Discontinuous
  {G}alerkin Finite Element Methods to Hypersonic Nonequilibrium Flow
  Problems}.
\newblock Ph.D. thesis. University of Florida.
\bibitem[{Bhatia and Roy(2013)}]{Bhatia2013}
\bibinfo{author}{Bhatia, A.}, \bibinfo{author}{Roy, S.}, \bibinfo{year}{2013}.
\newblock \bibinfo{title}{2-{D} hypersonic non-equilibrium flow simulation
  using r-p adaptive time-implicit discontinuous {G}alerkin method}, in:
  \bibinfo{booktitle}{51st AIAA Aerospace Sciences Meeting},
  \bibinfo{address}{Grapevine, TX}.
\newblock \DOIprefix\doi{10.2514/6.2013-302}. \bibinfo{note}{aIAA Paper
  2013-0302}.
\bibitem[{Bose et~al.(2013)Bose, Brown, Prabhu, Gnoffo, Johnston and
  Hollis}]{Bose2013}
\bibinfo{author}{Bose, D.}, \bibinfo{author}{Brown, J.L.},
  \bibinfo{author}{Prabhu, D.K.}, \bibinfo{author}{Gnoffo, P.},
  \bibinfo{author}{Johnston, C.O.}, \bibinfo{author}{Hollis, B.},
  \bibinfo{year}{2013}.
\newblock \bibinfo{title}{Uncertainty assessment of hypersonic
  aerothermodynamics prediction capability}.
\newblock \bibinfo{journal}{Journal of Spacecraft and Rockets}
  \bibinfo{volume}{50}, \bibinfo{pages}{12--18}.
\bibitem[{Brezzi et~al.(2000)Brezzi, Manzini, Marini, Pietra and
  Russo}]{Brezzi2000}
\bibinfo{author}{Brezzi, F.}, \bibinfo{author}{Manzini, G.},
  \bibinfo{author}{Marini, D.}, \bibinfo{author}{Pietra, P.},
  \bibinfo{author}{Russo, A.}, \bibinfo{year}{2000}.
\newblock \bibinfo{title}{Discontinuous {G}alerkin approximations for elliptic
  problems}.
\newblock \bibinfo{journal}{Numerical Methods for Partial Differential
  Equations} \bibinfo{volume}{16}, \bibinfo{pages}{365--378}.
\newblock
  \DOIprefix\doi{10.1002/1098-2426(200007)16:4<365::AID-NUM2>3.0.CO;2-Y}.
\bibitem[{Candler(2019)}]{Candler2019}
\bibinfo{author}{Candler, G.V.}, \bibinfo{year}{2019}.
\newblock \bibinfo{title}{Rate effects in hypersonic flows}.
\newblock \bibinfo{journal}{Annual Review of Fluid Mechanics}
  \bibinfo{volume}{51}, \bibinfo{pages}{379--402}.
\newblock \DOIprefix\doi{10.1146/annurev-fluid-010518-040258}.
\bibitem[{Candler et~al.(2007)Candler, Barnhardt, Drayna, Nompelis, Peterson
  and Subbareddy}]{Candler2007}
\bibinfo{author}{Candler, G.V.}, \bibinfo{author}{Barnhardt, M.D.},
  \bibinfo{author}{Drayna, T.W.}, \bibinfo{author}{Nompelis, I.},
  \bibinfo{author}{Peterson, D.M.}, \bibinfo{author}{Subbareddy, P.K.},
  \bibinfo{year}{2007}.
\newblock \bibinfo{title}{Unstructured grid approaches for accurate aeroheating
  simulations}, in: \bibinfo{booktitle}{18th AIAA Computational Fluid Dynamics
  Conference}, \bibinfo{address}{Miami, FL}.
\newblock \DOIprefix\doi{10.2514/6.2007-3959}. \bibinfo{note}{aIAA Paper
  2007-3959}.
\bibitem[{Candler et~al.(2015)Candler, Johnson, Nompelis, Subbareddy, Drayna,
  Gidzak and Barnhardt}]{Candler2015}
\bibinfo{author}{Candler, G.V.}, \bibinfo{author}{Johnson, H.B.},
  \bibinfo{author}{Nompelis, I.}, \bibinfo{author}{Subbareddy, P.K.},
  \bibinfo{author}{Drayna, T.W.}, \bibinfo{author}{Gidzak, V.},
  \bibinfo{author}{Barnhardt, M.D.}, \bibinfo{year}{2015}.
\newblock \bibinfo{title}{Development of the {US3D} code for advanced
  compressible and reacting flow simulations}, in: \bibinfo{booktitle}{53rd
  AIAA Aerospace Sciences Meeting}.
\newblock \DOIprefix\doi{10.2514/6.2015-1893}. \bibinfo{note}{aIAA Paper
  2015-1893}.
\bibitem[{Candler et~al.(2009)Candler, Mavriplis and Trevino}]{Candler2009}
\bibinfo{author}{Candler, G.V.}, \bibinfo{author}{Mavriplis, D.J.},
  \bibinfo{author}{Trevino, L.}, \bibinfo{year}{2009}.
\newblock \bibinfo{title}{Current status and future prospects for the numerical
  simulation of hypersonic flows}, in: \bibinfo{booktitle}{47th AIAA Aerospace
  Sciences Meeting}, \bibinfo{address}{Orlando, FL}.
\newblock \DOIprefix\doi{10.2514/6.2009-153}. \bibinfo{note}{aIAA Paper
  2009-153}.
\bibitem[{Ceze and Fidkowski(2013)}]{Ceze2013}
\bibinfo{author}{Ceze, M.}, \bibinfo{author}{Fidkowski, K.J.},
  \bibinfo{year}{2013}.
\newblock \bibinfo{title}{Pseudo-transient continuation, solution update
  methods, and {CFL} strategies for {DG} discretizations of the {RANS-SA}
  equations}, in: \bibinfo{booktitle}{21st AIAA Computational Fluid Dynamics
  Conference}, \bibinfo{address}{San Diego, CA}.
\newblock \DOIprefix\doi{10.2514/6.2013-2686}. \bibinfo{note}{aIAA Paper
  2013-2686}.
\bibitem[{Ceze and Fidkowski(2015)}]{Ceze2015}
\bibinfo{author}{Ceze, M.}, \bibinfo{author}{Fidkowski, K.J.},
  \bibinfo{year}{2015}.
\newblock \bibinfo{title}{Constrained pseudo-transient continuation}.
\newblock \bibinfo{journal}{International Journal for Numerical Methods in
  Engineering} \bibinfo{volume}{102}, \bibinfo{pages}{1683--1703}.
\newblock \DOIprefix\doi{10.1002/nme.4858}.
\bibitem[{Ching et~al.(2025)Ching, Johnson, Burrows, Higgs and
  Kercher}]{Ching2025}
\bibinfo{author}{Ching, E.J.}, \bibinfo{author}{Johnson, R.F.},
  \bibinfo{author}{Burrows, S.}, \bibinfo{author}{Higgs, J.},
  \bibinfo{author}{Kercher, A.D.}, \bibinfo{year}{2025}.
\newblock \bibinfo{title}{Positivity-preserving and entropy-bounded
  discontinuous {G}alerkin method for the chemically reacting, compressible
  {N}avier-{S}tokes equations}.
\newblock \bibinfo{journal}{Journal of Computational Physics}
  \bibinfo{volume}{525}, \bibinfo{pages}{113746}.
\newblock \DOIprefix\doi{10.1016/j.jcp.2025.113746}.
\bibitem[{Ching et~al.(2019)Ching, Lv, Gnoffo, Barnhardt and Ihme}]{Ching2019}
\bibinfo{author}{Ching, E.J.}, \bibinfo{author}{Lv, Y.},
  \bibinfo{author}{Gnoffo, P.}, \bibinfo{author}{Barnhardt, M.},
  \bibinfo{author}{Ihme, M.}, \bibinfo{year}{2019}.
\newblock \bibinfo{title}{Shock capturing for discontinuous {G}alerkin methods
  with application to predicting heat transfer in hypersonic flows}.
\newblock \bibinfo{journal}{Journal of Computational Physics}
  \bibinfo{volume}{376}, \bibinfo{pages}{54--75}.
\newblock \DOIprefix\doi{10.1016/j.jcp.2018.09.016}.
\bibitem[{Dhib et~al.(2026)Dhib, {Ben Ameur}, Parsani, Lani and
  Poedts}]{Dhib2026}
\bibinfo{author}{Dhib, R.}, \bibinfo{author}{{Ben Ameur}, F.},
  \bibinfo{author}{Parsani, M.}, \bibinfo{author}{Lani, A.},
  \bibinfo{author}{Poedts, S.}, \bibinfo{year}{2026}.
\newblock \bibinfo{title}{Development of a high-order {F}lux {R}econstruction
  framework for solar and astrophysical magnetohydrodynamics: Methods and
  benchmarks}.
\newblock \bibinfo{journal}{The Astrophysical Journal Supplement Series}
  \bibinfo{volume}{283}, \bibinfo{pages}{75}.
\newblock \DOIprefix\doi{10.3847/1538-4365/ae4ec1}.
\bibitem[{Dhib et~al.(2025)Dhib, {Ben Ameur}, Sharma, Lani and
  Poedts}]{Dhib2025}
\bibinfo{author}{Dhib, R.}, \bibinfo{author}{{Ben Ameur}, F.},
  \bibinfo{author}{Sharma, V.}, \bibinfo{author}{Lani, A.},
  \bibinfo{author}{Poedts, S.}, \bibinfo{year}{2025}.
\newblock \bibinfo{title}{Toward high-order solar corona simulations: A
  high-order hyperbolized {P}oisson approach for magnetic field
  initialization}.
\newblock \bibinfo{journal}{The Astrophysical Journal} \bibinfo{volume}{980},
  \bibinfo{pages}{163}.
\newblock \DOIprefix\doi{10.3847/1538-4357/adace5}.
\bibitem[{Dhib et~al.(2024)Dhib, {Ben Ameur}, Vandenhoeck, Lani and
  Poedts}]{Dhib2024}
\bibinfo{author}{Dhib, R.}, \bibinfo{author}{{Ben Ameur}, F.},
  \bibinfo{author}{Vandenhoeck, R.}, \bibinfo{author}{Lani, A.},
  \bibinfo{author}{Poedts, S.}, \bibinfo{year}{2024}.
\newblock \bibinfo{title}{Development of an implicit high-order {F}lux
  {R}econstruction solver for high-speed flows on simplex elements}.
\newblock \bibinfo{journal}{Computer Physics Communications}
  \bibinfo{volume}{295}, \bibinfo{pages}{109006}.
\newblock \DOIprefix\doi{10.1016/j.cpc.2023.109006}.
\bibitem[{Du and Yang(2022)}]{Du2022}
\bibinfo{author}{Du, J.}, \bibinfo{author}{Yang, Y.}, \bibinfo{year}{2022}.
\newblock \bibinfo{title}{High-order bound-preserving discontinuous {G}alerkin
  methods for multicomponent chemically reacting flows}.
\newblock \bibinfo{journal}{Journal of Computational Physics}
  \bibinfo{volume}{469}, \bibinfo{pages}{111548}.
\newblock \DOIprefix\doi{10.1016/j.jcp.2022.111548}.
\bibitem[{Edquist et~al.(2013)Edquist, Hollis, Johnston, Bose, White and
  Mahzari}]{Edquist2013}
\bibinfo{author}{Edquist, K.T.}, \bibinfo{author}{Hollis, B.R.},
  \bibinfo{author}{Johnston, C.O.}, \bibinfo{author}{Bose, D.},
  \bibinfo{author}{White, T.R.}, \bibinfo{author}{Mahzari, M.},
  \bibinfo{year}{2013}.
\newblock \bibinfo{title}{{M}ars {S}cience {L}aboratory heatshield
  aerothermodynamics: Design and reconstruction}, in: \bibinfo{booktitle}{AIAA
  Thermophysics Conference}, \bibinfo{address}{San Diego, CA}.
\newblock \bibinfo{note}{NASA NTRS 20140000343}.
\bibitem[{Freno et~al.(2021)Freno, Carnes and Weirs}]{Freno2021}
\bibinfo{author}{Freno, B.A.}, \bibinfo{author}{Carnes, B.R.},
  \bibinfo{author}{Weirs, V.G.}, \bibinfo{year}{2021}.
\newblock \bibinfo{title}{Code-verification techniques for hypersonic reacting
  flows in thermochemical nonequilibrium}.
\newblock \bibinfo{journal}{Journal of Computational Physics}
  \bibinfo{volume}{425}, \bibinfo{pages}{109752}.
\newblock \DOIprefix\doi{10.1016/j.jcp.2020.109752}.
\bibitem[{Giangaspero et~al.(2023)Giangaspero, Sharma, Laur, Thoemel,
  Munaf{\`o}, Lani and Poedts}]{Giangaspero2023}
\bibinfo{author}{Giangaspero, V.F.}, \bibinfo{author}{Sharma, V.},
  \bibinfo{author}{Laur, J.}, \bibinfo{author}{Thoemel, J.},
  \bibinfo{author}{Munaf{\`o}, A.}, \bibinfo{author}{Lani, A.},
  \bibinfo{author}{Poedts, S.}, \bibinfo{year}{2023}.
\newblock \bibinfo{title}{{3D} ray tracing solver for communication blackout
  analysis in atmospheric entry missions}.
\newblock \bibinfo{journal}{Computer Physics Communications}
  \bibinfo{volume}{286}, \bibinfo{pages}{108663}.
\newblock \DOIprefix\doi{10.1016/j.cpc.2023.108663}.
\bibitem[{Giangaspero et~al.(2026)Giangaspero, Sharma, Poedts and
  Lani}]{Giangaspero2026}
\bibinfo{author}{Giangaspero, V.F.}, \bibinfo{author}{Sharma, V.},
  \bibinfo{author}{Poedts, S.}, \bibinfo{author}{Lani, A.},
  \bibinfo{year}{2026}.
\newblock \bibinfo{title}{{BORAT-EM}: High frequency electromagnetic solver for
  atmospheric entry blackout analysis}.
\newblock \bibinfo{journal}{Computer Physics Communications}
  \bibinfo{volume}{323}, \bibinfo{pages}{110076}.
\newblock \DOIprefix\doi{10.1016/j.cpc.2026.110076}.
\bibitem[{Gnoffo(2007)}]{Gnoffo2007}
\bibinfo{author}{Gnoffo, P.A.}, \bibinfo{year}{2007}.
\newblock \bibinfo{title}{Simulation of stagnation region heating in hypersonic
  flow on tetrahedral grids}, in: \bibinfo{booktitle}{18th AIAA Computational
  Fluid Dynamics Conference}, \bibinfo{address}{Miami, FL}.
\newblock \DOIprefix\doi{10.2514/6.2007-3960}. \bibinfo{note}{aIAA Paper
  2007-3960}.
\bibitem[{Gnoffo(2009)}]{Gnoffo2009}
\bibinfo{author}{Gnoffo, P.A.}, \bibinfo{year}{2009}.
\newblock \bibinfo{title}{Multi-dimensional, inviscid flux reconstruction for
  simulation of hypersonic heating on tetrahedral grids}, in:
  \bibinfo{booktitle}{47th AIAA Aerospace Sciences Meeting},
  \bibinfo{address}{Orlando, FL}.
\newblock \DOIprefix\doi{10.2514/6.2009-599}. \bibinfo{note}{aIAA Paper
  2009-599}.
\bibitem[{Gnoffo et~al.(1989)Gnoffo, Gupta and Shinn}]{Gnoffo1989}
\bibinfo{author}{Gnoffo, P.A.}, \bibinfo{author}{Gupta, R.N.},
  \bibinfo{author}{Shinn, J.L.}, \bibinfo{year}{1989}.
\newblock \bibinfo{title}{Conservation Equations and Physical Models for
  Hypersonic Air Flows in Thermal and Chemical Nonequilibrium}.
\newblock \bibinfo{type}{NASA Technical Paper} \bibinfo{number}{TP-2867}.
  National Aeronautics and Space Administration. \bibinfo{address}{Washington,
  D.C.}
\bibitem[{Gordon and McBride(1994)}]{Gordon1994}
\bibinfo{author}{Gordon, S.}, \bibinfo{author}{McBride, B.J.},
  \bibinfo{year}{1994}.
\newblock \bibinfo{title}{Computer Program for Calculation of Complex Chemical
  Equilibrium Compositions and Applications. {I}. Analysis}.
\newblock \bibinfo{type}{NASA Reference Publication} \bibinfo{number}{1311}.
  National Aeronautics and Space Administration, Lewis Research Center.
  \bibinfo{address}{Cleveland, OH}.
\bibitem[{Grantham(1970)}]{Grantham1970}
\bibinfo{author}{Grantham, W.L.}, \bibinfo{year}{1970}.
\newblock \bibinfo{title}{Flight Results of a 25\,000-Foot-per-Second Reentry
  Experiment Using Microwave Reflectometers to Measure Plasma Electron Density
  and Standoff Distance}.
\newblock \bibinfo{type}{NASA Technical Note} \bibinfo{number}{TN D-6062}.
  National Aeronautics and Space Administration. \bibinfo{address}{Washington,
  D.C.}
\bibitem[{Grover et~al.(2025)Grover, Valentini and Bisek}]{Grover2025}
\bibinfo{author}{Grover, M.S.}, \bibinfo{author}{Valentini, P.},
  \bibinfo{author}{Bisek, N.}, \bibinfo{year}{2025}.
\newblock \bibinfo{title}{Ab initio investigation of a hypersonic double cone
  experiment}.
\newblock \bibinfo{journal}{Science Advances} \bibinfo{volume}{11},
  \bibinfo{pages}{eads2147}.
\newblock \DOIprefix\doi{10.1126/sciadv.ads2147}.
\bibitem[{Hartmann(2008)}]{Hartmann2008}
\bibinfo{author}{Hartmann, R.}, \bibinfo{year}{2008}.
\newblock \bibinfo{title}{Numerical analysis of higher order discontinuous
  {G}alerkin finite element methods}, in: \bibinfo{editor}{Deconinck, H.}
  (Ed.), \bibinfo{booktitle}{VKI Lecture Series}. \bibinfo{publisher}{von
  Karman Institute for Fluid Dynamics},
  \bibinfo{address}{Rhode-Saint-Gen{\`e}se, Belgium}, pp.
  \bibinfo{pages}{1--107}.
\bibitem[{Haselbacher and Blazek(2000)}]{Haselbacher2000}
\bibinfo{author}{Haselbacher, A.}, \bibinfo{author}{Blazek, J.},
  \bibinfo{year}{2000}.
\newblock \bibinfo{title}{Accurate and efficient discretization of
  {N}avier--{S}tokes equations on mixed grids}.
\newblock \bibinfo{journal}{AIAA Journal} \bibinfo{volume}{38},
  \bibinfo{pages}{2094--2102}.
\newblock \DOIprefix\doi{10.2514/2.871}.
\bibitem[{Hash et~al.(2007)Hash, Olejniczak, Wright, Prabhu, Pulsonetti,
  Hollis, Gnoffo, Barnhardt, Nompelis and Candler}]{Hash2007}
\bibinfo{author}{Hash, D.}, \bibinfo{author}{Olejniczak, J.},
  \bibinfo{author}{Wright, M.}, \bibinfo{author}{Prabhu, D.},
  \bibinfo{author}{Pulsonetti, M.}, \bibinfo{author}{Hollis, B.},
  \bibinfo{author}{Gnoffo, P.}, \bibinfo{author}{Barnhardt, M.},
  \bibinfo{author}{Nompelis, I.}, \bibinfo{author}{Candler, G.},
  \bibinfo{year}{2007}.
\newblock \bibinfo{title}{{FIRE II} calculations for hypersonic nonequilibrium
  aerothermodynamics code verification: {DPLR}, {LAURA}, and {US3D}}, in:
  \bibinfo{booktitle}{45th AIAA Aerospace Sciences Meeting and Exhibit},
  \bibinfo{address}{Reno, NV}.
\newblock \DOIprefix\doi{10.2514/6.2007-605}. \bibinfo{note}{aIAA Paper
  2007-605}.
\bibitem[{Hennemann et~al.(2021)Hennemann, Rueda-Ram{\'\i}rez, Hindenlang and
  Gassner}]{Hennemann2021}
\bibinfo{author}{Hennemann, S.}, \bibinfo{author}{Rueda-Ram{\'\i}rez, A.M.},
  \bibinfo{author}{Hindenlang, F.J.}, \bibinfo{author}{Gassner, G.J.},
  \bibinfo{year}{2021}.
\newblock \bibinfo{title}{A provably entropy stable subcell shock capturing
  approach for high order split form {DG} for the compressible {E}uler
  equations}.
\newblock \bibinfo{journal}{Journal of Computational Physics}
  \bibinfo{volume}{426}, \bibinfo{pages}{109935}.
\newblock \DOIprefix\doi{10.1016/j.jcp.2020.109935}.
\bibitem[{Holloway et~al.(2022)Holloway, Chaudhry and Boyd}]{Holloway2022}
\bibinfo{author}{Holloway, M.E.}, \bibinfo{author}{Chaudhry, R.S.},
  \bibinfo{author}{Boyd, I.D.}, \bibinfo{year}{2022}.
\newblock \bibinfo{title}{Assessment of hypersonic double-cone experiments for
  validation of thermochemistry models}.
\newblock \bibinfo{journal}{Journal of Spacecraft and Rockets}
  \bibinfo{volume}{59}, \bibinfo{pages}{389--400}.
\newblock \DOIprefix\doi{10.2514/1.A35052}.
\bibitem[{Holloway et~al.(2020)Holloway, Hanquist and Boyd}]{Holloway2020}
\bibinfo{author}{Holloway, M.E.}, \bibinfo{author}{Hanquist, K.M.},
  \bibinfo{author}{Boyd, I.D.}, \bibinfo{year}{2020}.
\newblock \bibinfo{title}{Assessment of thermochemistry modeling for hypersonic
  flow over a double cone}.
\newblock \bibinfo{journal}{Journal of Thermophysics and Heat Transfer}
  \bibinfo{volume}{34}, \bibinfo{pages}{538--547}.
\newblock \DOIprefix\doi{10.2514/1.T5792}.
\bibitem[{Hoskin et~al.(2024)Hoskin, Van~Heyningen, Nguyen, Vila-P{\'e}rez,
  Harris and Peraire}]{Hoskin2024}
\bibinfo{author}{Hoskin, D.S.}, \bibinfo{author}{Van~Heyningen, R.L.},
  \bibinfo{author}{Nguyen, N.C.}, \bibinfo{author}{Vila-P{\'e}rez, J.},
  \bibinfo{author}{Harris, W.L.}, \bibinfo{author}{Peraire, J.},
  \bibinfo{year}{2024}.
\newblock \bibinfo{title}{Discontinuous {G}alerkin methods for hypersonic
  flows}.
\newblock \bibinfo{journal}{Progress in Aerospace Sciences}
  \bibinfo{volume}{146}, \bibinfo{pages}{100999}.
\newblock \DOIprefix\doi{10.1016/j.paerosci.2024.100999}.
\bibitem[{Huynh(2007)}]{Huynh2007}
\bibinfo{author}{Huynh, H.T.}, \bibinfo{year}{2007}.
\newblock \bibinfo{title}{A flux reconstruction approach to high-order schemes
  including discontinuous {G}alerkin methods}, in: \bibinfo{booktitle}{18th
  AIAA Computational Fluid Dynamics Conference}, \bibinfo{address}{Miami, FL}.
\newblock \DOIprefix\doi{10.2514/6.2007-4079}. \bibinfo{note}{aIAA Paper
  2007-4079}.
\bibitem[{Jones and Cross(1972)}]{Jones1972}
\bibinfo{author}{Jones, Jr., W.L.}, \bibinfo{author}{Cross, A.E.},
  \bibinfo{year}{1972}.
\newblock \bibinfo{title}{Electrostatic-Probe Measurements of Plasma Parameters
  for Two Reentry Flight Experiments at 25\,000 Feet per Second}.
\newblock \bibinfo{type}{NASA Technical Note} \bibinfo{number}{TN D-6617}.
  National Aeronautics and Space Administration. \bibinfo{address}{Washington,
  D.C.}
\bibitem[{Karl et~al.(2003)Karl, Martinez~Schramm and Hannemann}]{Karl2003}
\bibinfo{author}{Karl, S.}, \bibinfo{author}{Martinez~Schramm, J.},
  \bibinfo{author}{Hannemann, K.}, \bibinfo{year}{2003}.
\newblock \bibinfo{title}{High enthalpy cylinder flow in {HEG}: A basis for
  {CFD} validation}, in: \bibinfo{booktitle}{33rd AIAA Fluid Dynamics
  Conference and Exhibit}, \bibinfo{address}{Orlando, FL}.
\newblock \DOIprefix\doi{10.2514/6.2003-4252}. \bibinfo{note}{aIAA Paper
  2003-4252}.
\bibitem[{Kimpe et~al.(2005)Kimpe, Lani, Quintino, Poedts and
  Vandewalle}]{kimpe2}
\bibinfo{author}{Kimpe, D.}, \bibinfo{author}{Lani, A.},
  \bibinfo{author}{Quintino, T.}, \bibinfo{author}{Poedts, S.},
  \bibinfo{author}{Vandewalle, S.}, \bibinfo{year}{2005}.
\newblock \bibinfo{title}{The {COOLFluiD} parallel architecture}, in:
  \bibinfo{editor}{B.~Di~Martino, D.K.}, \bibinfo{editor}{Dongarra, J.J.}
  (Eds.), \bibinfo{booktitle}{Proc. 12th European Parallel Virtual Machine and
  Message Passing Interface Conference}, \bibinfo{publisher}{Springer},
  \bibinfo{address}{Sorrento}. pp. \bibinfo{pages}{520--527}.
\newblock \DOIprefix\doi{10.1007/11557265_66}.
\bibitem[{Knight et~al.(2017)Knight, Chazot, Austin, Badr, Candler, Celik,
  de~Rosa, Donelli, Komives, Lani, Levin, Nompelis, Panesi, Pezzella, Reimann,
  Tumuklu and Yuceil}]{Knight2017}
\bibinfo{author}{Knight, D.}, \bibinfo{author}{Chazot, O.},
  \bibinfo{author}{Austin, J.}, \bibinfo{author}{Badr, M.A.},
  \bibinfo{author}{Candler, G.}, \bibinfo{author}{Celik, B.},
  \bibinfo{author}{de~Rosa, D.}, \bibinfo{author}{Donelli, R.},
  \bibinfo{author}{Komives, J.}, \bibinfo{author}{Lani, A.},
  \bibinfo{author}{Levin, D.}, \bibinfo{author}{Nompelis, I.},
  \bibinfo{author}{Panesi, M.}, \bibinfo{author}{Pezzella, G.},
  \bibinfo{author}{Reimann, B.}, \bibinfo{author}{Tumuklu, O.},
  \bibinfo{author}{Yuceil, K.}, \bibinfo{year}{2017}.
\newblock \bibinfo{title}{Assessment of predictive capabilities for aerodynamic
  heating in hypersonic flow}.
\newblock \bibinfo{journal}{Progress in Aerospace Sciences}
  \bibinfo{volume}{90}, \bibinfo{pages}{39--53}.
\newblock \DOIprefix\doi{10.1016/j.paerosci.2017.02.001}.
\bibitem[{Knight et~al.(2012)Knight, Longo, Drikakis, Gaitonde, Lani, Nompelis,
  Reimann and Walpot}]{Knight2012}
\bibinfo{author}{Knight, D.}, \bibinfo{author}{Longo, J.},
  \bibinfo{author}{Drikakis, D.}, \bibinfo{author}{Gaitonde, D.},
  \bibinfo{author}{Lani, A.}, \bibinfo{author}{Nompelis, I.},
  \bibinfo{author}{Reimann, B.}, \bibinfo{author}{Walpot, L.},
  \bibinfo{year}{2012}.
\newblock \bibinfo{title}{Assessment of {CFD} capability for prediction of
  hypersonic shock interactions}.
\newblock \bibinfo{journal}{Progress in Aerospace Sciences}
  \bibinfo{volume}{48--49}, \bibinfo{pages}{8--26}.
\newblock \DOIprefix\doi{10.1016/j.paerosci.2011.10.001}.
\bibitem[{Krais et~al.(2021)Krais, Beck, Bolemann, Frank, Flad, Gassner,
  Hindenlang, Hoffmann, Kuhn, Sonntag and Munz}]{Krais2021}
\bibinfo{author}{Krais, N.}, \bibinfo{author}{Beck, A.},
  \bibinfo{author}{Bolemann, T.}, \bibinfo{author}{Frank, H.},
  \bibinfo{author}{Flad, D.}, \bibinfo{author}{Gassner, G.},
  \bibinfo{author}{Hindenlang, F.}, \bibinfo{author}{Hoffmann, M.},
  \bibinfo{author}{Kuhn, T.}, \bibinfo{author}{Sonntag, M.},
  \bibinfo{author}{Munz, C.D.}, \bibinfo{year}{2021}.
\newblock \bibinfo{title}{{FLEXI}: A high order discontinuous {G}alerkin
  framework for hyperbolic--parabolic conservation laws}.
\newblock \bibinfo{journal}{Computers \& Mathematics with Applications}
  \bibinfo{volume}{81}, \bibinfo{pages}{186--219}.
\newblock \DOIprefix\doi{10.1016/j.camwa.2020.05.004}.
\bibitem[{Lani(2009)}]{Lani2009}
\bibinfo{author}{Lani, A.}, \bibinfo{year}{2009}.
\newblock \bibinfo{title}{An Object Oriented and High Performance Platform for
  Aerothermodynamics Simulation}.
\newblock Ph.D. thesis. Universit{\'e} Libre de Bruxelles and von Karman
  Institute for Fluid Dynamics.
\bibitem[{Lani et~al.(2005)Lani, Quintino, Kimpe, Deconinck, Vandewalle and
  Poedts}]{lani1}
\bibinfo{author}{Lani, A.}, \bibinfo{author}{Quintino, T.},
  \bibinfo{author}{Kimpe, D.}, \bibinfo{author}{Deconinck, H.},
  \bibinfo{author}{Vandewalle, S.}, \bibinfo{author}{Poedts, S.},
  \bibinfo{year}{2005}.
\newblock \bibinfo{title}{The {COOLFluiD} framework: Design solutions for
  high-performance object oriented scientific computing software}, in:
  \bibinfo{editor}{Sunderan, V.S.}, \bibinfo{editor}{van Albada, G.D.},
  \bibinfo{editor}{Sloot, P.M.A.}, \bibinfo{editor}{Dongarra, J.J.} (Eds.),
  \bibinfo{booktitle}{Computational Science ICCS 2005},
  \bibinfo{organization}{Emory University}. \bibinfo{publisher}{Springer},
  \bibinfo{address}{Atlanta, GA, USA}. pp. \bibinfo{pages}{281--286}.
\newblock \DOIprefix\doi{10.1007/11428831_35}.
\bibitem[{Lani et~al.(2013)Lani, Villedieu, Bensassi, Kapa, Vymazal, Yalim and
  Panesi}]{lani13}
\bibinfo{author}{Lani, A.}, \bibinfo{author}{Villedieu, N.},
  \bibinfo{author}{Bensassi, K.}, \bibinfo{author}{Kapa, L.},
  \bibinfo{author}{Vymazal, M.}, \bibinfo{author}{Yalim, M.S.},
  \bibinfo{author}{Panesi, M.}, \bibinfo{year}{2013}.
\newblock \bibinfo{title}{{COOLFluiD}: an open computational platform for
  multi-physics simulation and research}, in: \bibinfo{booktitle}{AIAA
  2013-2589}, \bibinfo{organization}{21th AIAA CFD Conference},
  \bibinfo{address}{San Diego (CA)}.
\newblock \DOIprefix\doi{10.2514/6.2013-2589}.
\bibitem[{Lin et~al.(2023)Lin, Chan and Tomas}]{Lin2023}
\bibinfo{author}{Lin, Y.}, \bibinfo{author}{Chan, J.}, \bibinfo{author}{Tomas,
  I.}, \bibinfo{year}{2023}.
\newblock \bibinfo{title}{A positivity preserving strategy for entropy stable
  discontinuous {G}alerkin discretizations of the compressible {E}uler and
  {N}avier--{S}tokes equations}.
\newblock \bibinfo{journal}{Journal of Computational Physics}
  \bibinfo{volume}{475}, \bibinfo{pages}{111850}.
\newblock \DOIprefix\doi{10.1016/j.jcp.2022.111850}.
\bibitem[{Liu and Yuan(2023)}]{LiuYuan2023}
\bibinfo{author}{Liu, S.}, \bibinfo{author}{Yuan, L.}, \bibinfo{year}{2023}.
\newblock \bibinfo{title}{A modified a posteriori subcell limiter for high
  order flux reconstruction scheme for one-dimensional detonation simulation}.
\newblock \bibinfo{journal}{Journal of Scientific Computing}
  \bibinfo{volume}{97}, \bibinfo{pages}{31}.
\newblock \DOIprefix\doi{10.1007/s10915-023-02347-7}.
\bibitem[{Luo et~al.(2026)Luo, Absillis, Goodson, Nielsen and
  Salazar}]{Luo2026}
\bibinfo{author}{Luo, H.}, \bibinfo{author}{Absillis, G.},
  \bibinfo{author}{Goodson, M.}, \bibinfo{author}{Nielsen, T.},
  \bibinfo{author}{Salazar, G.}, \bibinfo{year}{2026}.
\newblock \bibinfo{title}{A $r$-adaptive discontinuous {G}alerkin method based
  on interface conservation for reacting hypersonic viscous flows}, in:
  \bibinfo{booktitle}{AIAA SciTech 2026 Forum}, \bibinfo{address}{Orlando, FL}.
\newblock \DOIprefix\doi{10.2514/6.2026-1544}. \bibinfo{note}{aIAA Paper
  2026-1544}.
\bibitem[{Maier et~al.(2021)Maier, Needels, Garbacz, Morgado, Alonso and
  Fossati}]{Maier2021}
\bibinfo{author}{Maier, W.T.}, \bibinfo{author}{Needels, J.T.},
  \bibinfo{author}{Garbacz, C.}, \bibinfo{author}{Morgado, F.},
  \bibinfo{author}{Alonso, J.J.}, \bibinfo{author}{Fossati, M.},
  \bibinfo{year}{2021}.
\newblock \bibinfo{title}{{SU2-NEMO}: An open-source framework for high-{M}ach
  nonequilibrium multi-species flows}.
\newblock \bibinfo{journal}{Aerospace} \bibinfo{volume}{8},
  \bibinfo{pages}{193}.
\newblock \DOIprefix\doi{10.3390/aerospace8070193}.
\bibitem[{May et~al.(2021)May, Devesse, Rangarajan and Magin}]{May2021}
\bibinfo{author}{May, G.}, \bibinfo{author}{Devesse, K.},
  \bibinfo{author}{Rangarajan, A.}, \bibinfo{author}{Magin, T.},
  \bibinfo{year}{2021}.
\newblock \bibinfo{title}{A hybridized discontinuous {G}alerkin solver for
  high-speed compressible flow}.
\newblock \bibinfo{journal}{Aerospace} \bibinfo{volume}{8},
  \bibinfo{pages}{322}.
\newblock \DOIprefix\doi{10.3390/aerospace8110322}.
\bibitem[{Mazaheri et~al.(2011)Mazaheri, Gnoffo, Johnston and
  Kleb}]{Mazaheri2011}
\bibinfo{author}{Mazaheri, A.}, \bibinfo{author}{Gnoffo, P.A.},
  \bibinfo{author}{Johnston, C.O.}, \bibinfo{author}{Kleb, B.},
  \bibinfo{year}{2011}.
\newblock \bibinfo{title}{{LAURA} Users Manual: 5.4-54166}.
\newblock \bibinfo{type}{NASA Technical Memorandum}
  \bibinfo{number}{NASA/TM-2011-217092}. NASA Langley Research Center.
\bibitem[{Millikan and White(1963)}]{Millikan1963}
\bibinfo{author}{Millikan, R.C.}, \bibinfo{author}{White, D.R.},
  \bibinfo{year}{1963}.
\newblock \bibinfo{title}{Systematics of vibrational relaxation}.
\newblock \bibinfo{journal}{The Journal of Chemical Physics}
  \bibinfo{volume}{39}, \bibinfo{pages}{3209--3213}.
\newblock \DOIprefix\doi{10.1063/1.1734182}.
\bibitem[{Mondal and Jagtap(2026)}]{Mondal2026}
\bibinfo{author}{Mondal, T.}, \bibinfo{author}{Jagtap, A.D.},
  \bibinfo{year}{2026}.
\newblock \bibinfo{title}{{MARUT}: An exascale-ready, {GPU}-accelerated
  high-order {CFD} framework with {AMR} for high-speed flows and finite-rate
  chemistry}.
\newblock \bibinfo{note}{ArXiv:2605.26388}.
\bibitem[{Munaf{\`o} and Panesi(2025)}]{Munafo2025}
\bibinfo{author}{Munaf{\`o}, A.}, \bibinfo{author}{Panesi, M.},
  \bibinfo{year}{2025}.
\newblock \bibinfo{title}{{PLATO}: A high-fidelity library for multicomponent
  gases and plasmas}.
\newblock \bibinfo{journal}{Journal of Thermophysics and Heat Transfer}
  \bibinfo{volume}{39}, \bibinfo{pages}{850--870}.
\newblock \DOIprefix\doi{10.2514/1.T7097}.
\bibitem[{Panesi et~al.(2007)Panesi, Lani, Magin, Pinna, Chazot and
  Deconinck}]{panesi07}
\bibinfo{author}{Panesi, M.}, \bibinfo{author}{Lani, A.},
  \bibinfo{author}{Magin, T.}, \bibinfo{author}{Pinna, F.},
  \bibinfo{author}{Chazot, O.}, \bibinfo{author}{Deconinck, H.},
  \bibinfo{year}{2007}.
\newblock \bibinfo{title}{Numerical investigation of the non equilibrium
  shock-layer around the expert vehicle}, in: \bibinfo{booktitle}{AIAA Paper
  2007-4317}, \bibinfo{organization}{38th AIAA Plasmadynamics and Lasers
  Conference}, \bibinfo{address}{Miami (Florida)}.
\bibitem[{Papoutsakis et~al.(2014)Papoutsakis, Panourgias and
  Ekaterinaris}]{Papoutsakis2014}
\bibinfo{author}{Papoutsakis, A.E.}, \bibinfo{author}{Panourgias, K.T.},
  \bibinfo{author}{Ekaterinaris, J.A.}, \bibinfo{year}{2014}.
\newblock \bibinfo{title}{Discontinuous {G}alerkin discretization of chemically
  reacting flows}, in: \bibinfo{booktitle}{52nd AIAA Aerospace Sciences
  Meeting}, \bibinfo{address}{National Harbor, MD}.
\newblock \DOIprefix\doi{10.2514/6.2014-0068}. \bibinfo{note}{aIAA Paper
  2014-0068}.
\bibitem[{Park(1989)}]{Park1989}
\bibinfo{author}{Park, C.}, \bibinfo{year}{1989}.
\newblock \bibinfo{title}{Assessment of two-temperature kinetic model for
  ionizing air}.
\newblock \bibinfo{journal}{Journal of Thermophysics and Heat Transfer}
  \bibinfo{volume}{3}, \bibinfo{pages}{233--244}.
\newblock \DOIprefix\doi{10.2514/3.28771}.
\bibitem[{Park(1990)}]{Park1990}
\bibinfo{author}{Park, C.}, \bibinfo{year}{1990}.
\newblock \bibinfo{title}{Nonequilibrium Hypersonic Aerothermodynamics}.
\newblock \bibinfo{publisher}{Wiley}, \bibinfo{address}{New York}.
\bibitem[{Park(1993)}]{Park1993}
\bibinfo{author}{Park, C.}, \bibinfo{year}{1993}.
\newblock \bibinfo{title}{Review of chemical-kinetic problems of future {NASA}
  missions. {I} - {E}arth entries}.
\newblock \bibinfo{journal}{Journal of Thermophysics and Heat Transfer}
  \bibinfo{volume}{7}, \bibinfo{pages}{385--398}.
\newblock \DOIprefix\doi{10.2514/3.431}.
\bibitem[{Persson and Peraire(2006)}]{Persson2006}
\bibinfo{author}{Persson, P.O.}, \bibinfo{author}{Peraire, J.},
  \bibinfo{year}{2006}.
\newblock \bibinfo{title}{Sub-cell shock capturing for discontinuous {G}alerkin
  methods}, in: \bibinfo{booktitle}{44th AIAA Aerospace Sciences Meeting and
  Exhibit}, \bibinfo{address}{Reno, NV}.
\newblock \DOIprefix\doi{10.2514/6.2006-112}. \bibinfo{note}{aIAA Paper
  2006-112}.
\bibitem[{Peyvan et~al.(2023)Peyvan, Shukla, Chan and Karniadakis}]{Peyvan2023}
\bibinfo{author}{Peyvan, A.}, \bibinfo{author}{Shukla, K.},
  \bibinfo{author}{Chan, J.}, \bibinfo{author}{Karniadakis, G.},
  \bibinfo{year}{2023}.
\newblock \bibinfo{title}{High-order methods for hypersonic flows with strong
  shocks and real chemistry}.
\newblock \bibinfo{journal}{Journal of Computational Physics}
  \bibinfo{volume}{490}, \bibinfo{pages}{112310}.
\newblock \DOIprefix\doi{10.1016/j.jcp.2023.112310}.
\bibitem[{Peyvan et~al.(2027)Peyvan, Shukla and Karniadakis}]{Peyvan2027}
\bibinfo{author}{Peyvan, A.}, \bibinfo{author}{Shukla, K.},
  \bibinfo{author}{Karniadakis, G.E.}, \bibinfo{year}{2027}.
\newblock \bibinfo{title}{{H$^3$PC}: Hypersonic, high-order, high-performance
  code with adaptive mesh refinement and real chemistry}.
\newblock \bibinfo{journal}{Journal of Computational Physics}
  \bibinfo{volume}{568}, \bibinfo{pages}{115394}.
\newblock \DOIprefix\doi{10.1016/j.jcp.2026.115394}.
\bibitem[{Rueda-Ram{\'\i}rez et~al.(2022)Rueda-Ram{\'\i}rez, Pazner and
  Gassner}]{RuedaRamirez2022}
\bibinfo{author}{Rueda-Ram{\'\i}rez, A.M.}, \bibinfo{author}{Pazner, W.},
  \bibinfo{author}{Gassner, G.J.}, \bibinfo{year}{2022}.
\newblock \bibinfo{title}{Subcell limiting strategies for discontinuous
  {G}alerkin spectral element methods}.
\newblock \bibinfo{journal}{Computers \& Fluids} \bibinfo{volume}{247},
  \bibinfo{pages}{105627}.
\newblock \DOIprefix\doi{10.1016/j.compfluid.2022.105627}.
\bibitem[{Rusanov(1962)}]{Rusanov1962}
\bibinfo{author}{Rusanov, V.V.}, \bibinfo{year}{1962}.
\newblock \bibinfo{title}{The calculation of the interaction of non-stationary
  shock waves and obstacles}.
\newblock \bibinfo{journal}{USSR Computational Mathematics and Mathematical
  Physics} \bibinfo{volume}{1}, \bibinfo{pages}{304--320}.
\newblock \DOIprefix\doi{10.1016/0041-5553(62)90062-9}.
\bibitem[{Saad and Schultz(1986)}]{Saad1986}
\bibinfo{author}{Saad, Y.}, \bibinfo{author}{Schultz, M.H.},
  \bibinfo{year}{1986}.
\newblock \bibinfo{title}{{GMRES}: A generalized minimal residual algorithm for
  solving nonsymmetric linear systems}.
\newblock \bibinfo{journal}{SIAM Journal on Scientific and Statistical
  Computing} \bibinfo{volume}{7}, \bibinfo{pages}{856--869}.
\newblock \DOIprefix\doi{10.1137/0907058}.
\bibitem[{Sabo(2022)}]{Sabo2022}
\bibinfo{author}{Sabo, K.M.}, \bibinfo{year}{2022}.
\newblock \bibinfo{title}{Application of Ab-initio Quantum Chemistry Techniques
  to Hypersonic Flows for Plasma Blackout Alleviation}.
\newblock Ph.D. thesis. Massachusetts Institute of Technology.
\bibitem[{Scalabrin and Boyd(2006)}]{Scalabrin2006}
\bibinfo{author}{Scalabrin, L.}, \bibinfo{author}{Boyd, I.},
  \bibinfo{year}{2006}.
\newblock \bibinfo{title}{Numerical simulation of weakly ionized hypersonic
  flow for reentry configurations}, in: \bibinfo{booktitle}{9th AIAA/ASME Joint
  Thermophysics and Heat Transfer Conference}, \bibinfo{address}{San Francisco,
  CA}.
\newblock \DOIprefix\doi{10.2514/6.2006-3773}. \bibinfo{note}{aIAA Paper
  2006-3773}.
\bibitem[{Sharma et~al.(2024a)Sharma, Giangaspero, Munaf{\`o}, Poedts and
  Lani}]{Sharma2024}
\bibinfo{author}{Sharma, V.}, \bibinfo{author}{Giangaspero, V.F.},
  \bibinfo{author}{Munaf{\`o}, A.}, \bibinfo{author}{Poedts, S.},
  \bibinfo{author}{Lani, A.}, \bibinfo{year}{2024}a.
\newblock \bibinfo{title}{Validation and verification of the implicit
  thermo-chemical non-equilibrium {CFD} solver in {COOLFluiD} with {PLATO}},
  in: \bibinfo{booktitle}{AIAA SciTech 2024 Forum}, \bibinfo{address}{Orlando,
  FL}.
\newblock \DOIprefix\doi{10.2514/6.2024-2728}. \bibinfo{note}{aIAA Paper
  2024-2728}.
\bibitem[{Sharma et~al.(2024b)Sharma, Giangaspero, Poedts and
  Lani}]{Sharma2024b}
\bibinfo{author}{Sharma, V.}, \bibinfo{author}{Giangaspero, V.F.},
  \bibinfo{author}{Poedts, S.}, \bibinfo{author}{Lani, A.},
  \bibinfo{year}{2024}b.
\newblock \bibinfo{title}{Influence of magnetohydrodynamics configuration on
  aerothermodynamics during {M}artian reentry}.
\newblock \bibinfo{journal}{Physics of Fluids} \bibinfo{volume}{36},
  \bibinfo{pages}{036103}.
\newblock \DOIprefix\doi{10.1063/5.0191101}.
\bibitem[{Sonntag and Munz(2017)}]{Sonntag2017}
\bibinfo{author}{Sonntag, M.}, \bibinfo{author}{Munz, C.D.},
  \bibinfo{year}{2017}.
\newblock \bibinfo{title}{Efficient parallelization of a shock capturing for
  discontinuous {G}alerkin methods using finite volume sub-cells}.
\newblock \bibinfo{journal}{Journal of Scientific Computing}
  \bibinfo{volume}{70}, \bibinfo{pages}{1262--1289}.
\newblock \DOIprefix\doi{10.1007/s10915-016-0287-5}.
\bibitem[{Takahashi et~al.(2014)Takahashi, Yamada and Abe}]{Takahashi2014}
\bibinfo{author}{Takahashi, Y.}, \bibinfo{author}{Yamada, K.},
  \bibinfo{author}{Abe, T.}, \bibinfo{year}{2014}.
\newblock \bibinfo{title}{Prediction performance of blackout and plasma
  attenuation in atmospheric reentry demonstrator mission}.
\newblock \bibinfo{journal}{Journal of Spacecraft and Rockets}
  \bibinfo{volume}{51}, \bibinfo{pages}{1954--1964}.
\newblock \DOIprefix\doi{10.2514/1.A32880}.
\bibitem[{Van~Heyningen et~al.(2026)Van~Heyningen, Nguyen and
  Peraire}]{VanHeyningen2026}
\bibinfo{author}{Van~Heyningen, R.L.}, \bibinfo{author}{Nguyen, N.C.},
  \bibinfo{author}{Peraire, J.}, \bibinfo{year}{2026}.
\newblock \bibinfo{title}{An $r$-adaptive hybridizable discontinuous {G}alerkin
  method for nonequilibrium hypersonic flows}, in: \bibinfo{booktitle}{AIAA
  SciTech 2026 Forum}, \bibinfo{address}{Orlando, FL}.
\newblock \DOIprefix\doi{10.2514/6.2026-1543}. \bibinfo{note}{aIAA Paper
  2026-1543}.
\bibitem[{Vandenhoeck and Lani(2019)}]{Vandenhoeck2019}
\bibinfo{author}{Vandenhoeck, R.}, \bibinfo{author}{Lani, A.},
  \bibinfo{year}{2019}.
\newblock \bibinfo{title}{Implicit high-order flux reconstruction solver for
  high-speed compressible flows}.
\newblock \bibinfo{journal}{Computer Physics Communications}
  \bibinfo{volume}{242}, \bibinfo{pages}{1--24}.
\newblock \DOIprefix\doi{10.1016/j.cpc.2019.04.015}.
\bibitem[{van~der Vegt et~al.(2019)van~der Vegt, Xia and Xu}]{vanderVegt2019}
\bibinfo{author}{van~der Vegt, J.J.W.}, \bibinfo{author}{Xia, Y.},
  \bibinfo{author}{Xu, Y.}, \bibinfo{year}{2019}.
\newblock \bibinfo{title}{Positivity preserving limiters for time-implicit
  higher order accurate discontinuous {G}alerkin discretizations}.
\newblock \bibinfo{journal}{SIAM Journal on Scientific Computing}
  \bibinfo{volume}{41}, \bibinfo{pages}{A2037--A2063}.
\newblock \DOIprefix\doi{10.1137/18M1227998}.
\bibitem[{Venkatakrishnan(1995)}]{Venkatakrishnan1995}
\bibinfo{author}{Venkatakrishnan, V.}, \bibinfo{year}{1995}.
\newblock \bibinfo{title}{Convergence to steady state solutions of the {E}uler
  equations on unstructured grids with limiters}.
\newblock \bibinfo{journal}{Journal of Computational Physics}
  \bibinfo{volume}{118}, \bibinfo{pages}{120--130}.
\newblock \DOIprefix\doi{10.1006/jcph.1995.1084}.
\bibitem[{Vincent et~al.(2011)Vincent, Castonguay and Jameson}]{Vincent2011}
\bibinfo{author}{Vincent, P.E.}, \bibinfo{author}{Castonguay, P.},
  \bibinfo{author}{Jameson, A.}, \bibinfo{year}{2011}.
\newblock \bibinfo{title}{A new class of high-order energy stable flux
  reconstruction schemes}.
\newblock \bibinfo{journal}{Journal of Scientific Computing}
  \bibinfo{volume}{47}, \bibinfo{pages}{50--72}.
\newblock \DOIprefix\doi{10.1007/s10915-010-9420-z}.
\bibitem[{Wright et~al.(1998)Wright, Candler and Bose}]{Wright1998}
\bibinfo{author}{Wright, M.J.}, \bibinfo{author}{Candler, G.V.},
  \bibinfo{author}{Bose, D.}, \bibinfo{year}{1998}.
\newblock \bibinfo{title}{Data-parallel line relaxation method for the
  {N}avier--{S}tokes equations}.
\newblock \bibinfo{journal}{AIAA Journal} \bibinfo{volume}{36},
  \bibinfo{pages}{1603--1609}.
\newblock \DOIprefix\doi{10.2514/2.586}.
\bibitem[{Yu et~al.(2023)Yu, Qiu and Takahashi}]{Yu2023}
\bibinfo{author}{Yu, M.}, \bibinfo{author}{Qiu, Z.},
  \bibinfo{author}{Takahashi, Y.}, \bibinfo{year}{2023}.
\newblock \bibinfo{title}{Numerical investigation of surface catalytic effect
  on the plasma sheath of a hypersonic re-entry capsule}.
\newblock \bibinfo{journal}{Physics of Fluids} \bibinfo{volume}{35},
  \bibinfo{pages}{056106}.
\newblock \DOIprefix\doi{10.1063/5.0149660}.
\bibitem[{Zhang and Shu(2010)}]{Zhang2010}
\bibinfo{author}{Zhang, X.}, \bibinfo{author}{Shu, C.W.}, \bibinfo{year}{2010}.
\newblock \bibinfo{title}{On positivity-preserving high order discontinuous
  {G}alerkin schemes for compressible {E}uler equations on rectangular meshes}.
\newblock \bibinfo{journal}{Journal of Computational Physics}
  \bibinfo{volume}{229}, \bibinfo{pages}{8918--8934}.
\newblock \DOIprefix\doi{10.1016/j.jcp.2010.08.016}.
\bibitem[{Zhang and Shu(2011)}]{Zhang2011}
\bibinfo{author}{Zhang, X.}, \bibinfo{author}{Shu, C.W.}, \bibinfo{year}{2011}.
\newblock \bibinfo{title}{Positivity-preserving high order discontinuous
  {G}alerkin schemes for compressible {E}uler equations with source terms}.
\newblock \bibinfo{journal}{Journal of Computational Physics}
  \bibinfo{volume}{230}, \bibinfo{pages}{1238--1248}.
\newblock \DOIprefix\doi{10.1016/j.jcp.2010.10.036}.
\bibitem[{Zuo and Hu(2021)}]{Zuo2021}
\bibinfo{author}{Zuo, F.Y.}, \bibinfo{author}{Hu, S.L.}, \bibinfo{year}{2021}.
\newblock \bibinfo{title}{Thermochemical non-equilibrium effects on
  aerothermodynamic prediction of laminar double-cone flow}.
\newblock \bibinfo{journal}{Acta Astronautica} \bibinfo{volume}{182},
  \bibinfo{pages}{179--188}.
\newblock \DOIprefix\doi{10.1016/j.actaastro.2021.01.058}.

\end{thebibliography}

\end{document}